# Geomorphometric modeling and mapping of Antarctic oases

## I.V. Florinsky[*]

Institute of Mathematical Problems of Biology, Keldysh Institute of Applied Mathematics,
Russian Academy of Sciences, Pushchino, Moscow Region, 142290, Russia

### Abstract

Geomorphometric modeling is widely used in geosciences. However, geomorphometric modeling and mapping of Antarctic oases has not been performed so far. This article presents the first results of our work on geomorphometric modeling and mapping of several Antarctic oases including the Larsemann Hills, Thala Hills, Schirmacher Oasis, and Fildes Peninsula. As input data, we used fragments of the Reference Elevation Model of Antarctica. For each territory, we derived digital models of the following 17 morphometric variables from the extracted and edited digital elevation models: slope, aspect, horizontal curvature, vertical curvature, mean curvature, Gaussian curvature, minimal curvature, maximal curvature, unsphericity curvature, difference curvature, vertical excess curvature, horizontal excess curvature, ring curvature, accumulation curvature, catchment area, topographic index, and stream power index. Derived geomorphometric maps can be useful for structural geological and process-oriented hydrological studies. The ultimate goal of the ongoing work is to create a digital large-scale geomorphometric atlas of Antarctic oases and other ice-free Antarctic territories.

**Keywords:** topography, digital elevation model, geomorphometry, surface.

## 1. Introduction

Antarctic oases are commonly referred to glacier-free areas of the Antarctic coastal zone with an area of several tens to several thousands square kilometers, which are characterized by: (1) a local climate largely determined by the surrounding ice sheet; (2) the presence of unfrozen water in the form of a system of seasonal streams and unfrozen lakes; and (3) the presence of primitive soils and biota (Simonov, 1971; Sokratova, 2010).

Despite their relatively small size, Antarctic oases are of scientific and practical importance for Antarctic research. Because of their relative accessibility, these territories are convenient for construction of year-round polar stations and seasonal bases, and also attract scientists by the possibility of obtaining a significant amount of scientific field material in a short period of austral summer (Sokratova, 2010).

Geomorphometry is a scientific discipline lying at the intersection of geoinformatics, remote sensing, photogrammetry, and computational mathematics, dealing with mathematical modeling and analysis of topography, as well as the relationships between topography and other components of geosystems (Evans, 1972; Shary, 1995; Shary et al., 2002; Hengl and Reuter, 2009; Florinsky, 2016, 2017). Modern geomorphometry has a developed physical and mathematical theory and a powerful toolkit of computational methods. The initial data for modeling are digital elevation models (DEMs) (Guth et al., 2021). The objects of modeling are usually land surface, submarine topography, glacier surface, subglacial topography, and surfaces of extraterrestrial territories (Florinsky, 2017). These non-smooth objects are usually approximated by a topographic surface – a closed oriented infinitely differentiable two-dimensional manifold in the three-dimensional Euclidean space (Shary, 1995; Florinsky, 2016).

---

[*] Correspondence to: iflor@mail.ru

Geomorphometric methods are widely used to solve various multi-scale problems of geomorphology, hydrology, soil science, geobotany, forestry, geology, oceanology, climatology, planetology, and other geosciences. DEM spatial resolution may lie in a wide range: from centimeters and decimeters to tens and hundreds of meters and tens of kilometers. International experience of research in the field of geomorphometry is summarized in a series of analytical reviews and books (Moore et al., 1991; Wilson and Gallant, 2000; Li et al., 2005; Hengl and Reuter, 2009; Wilson, 2018; Minár et al., 2016; Lv et al., 2017; Florinsky, 2016, 2017, 2021).

Prospects of geomorphometric modeling and mapping of Antarctic oases are quite obvious both for obtaining new knowledge about quantitative characteristics of the topography of these unique natural objects, and for further use of the obtained geomorphometric information for solving fundamental and applied problems of geomorphology, geology, geophysics, glaciology, climatology, etc. Nevertheless, geomorphometric modeling and mapping of Antarctic oases has not been performed so far.

This article presents the first results of our works on geomorphometric modeling and mapping of several Antarctic oases including the Larsemann Hills, Thala Hills, Schirmacher Oasis, and Fildes Peninsula. The ultimate goal of the ongoing work under this multiyear program is to create a digital large-scale geomorphometric atlas of Antarctic oases and other ice-free Antarctic territories.

## 2. Study areas
### 2.1. Larsemann Hills

The Larsemann Hills, East Antarctica (Figure 1) is an Antarctic coastal oasis including ice-free, low rounded hills with an area of about 40 km$^2$ located on the south-eastern shore of the Prydz Bay of the Commonwealth Sea, Ingrid Christensen Coast, Princess Elizabeth Land. The oasis consists of two large peninsulas, Stornes and Broknes, three small peninsulas, Grovnes, Brattnevet, and Stinear, and about 130 islands including three relatively large, Fisher, Manning, and McLeod Islands (Figure 1). The peninsulas and islands are separated by a system of fjords and bays, such as, Thala, Nella, and Clemence Fjords, as well as Dålkoy, Wilcock, and Quilty Bays. The oasis is bounded by the Dålk outlet glacier in the east, by the glacial plateau in the south, and by the Polararboken outlet glacier in the west.

The Larsemann Hills are composed mainly of the Proterozoic garnet-biotite gneisses and garnet-sillimanite schists; intrusions of pegmatites and granites are also present (Stüwe et al., 1989; Carson and Grew, 2007; ATCM, 2021). Unique borosilicate and phosphate complexes are developed on the Stornes Peninsula (ATCM, 2014).

The Larsemann Hills were deglaciated between 20,000 and 10,000 years ago (Burgess et al., 1994; Hodgson et al., 2001; Verleyen et al., 2004; Kiernan et al., 2009). The oasis topography is a structurally controlled, exarated rocky hill terrain (Bolshiyanov, 2011). The elongated shape of the large landform elements of the Larsemann Hills is generally related to compositional layering and the presence of folds and faults in the metamorphic bedrock (ATCM, 2021). Nivation dominates among exogenous topography-forming processes. The landscape is rugged with large, structurally controlled fjords and valleys with steep slopes, which are rarely deeper than 100 m on land and have a maximum length of 3 km. Elevations range from 30 m to 120 m. The maximum elevation above sea level is 158 m (Blundell Peak on the Stornes Peninsula).

There are about 150 lakes in the oasis. Drainage network is not developed. Snowfields are widespread. There is a small passive glacial dome on the Stornes Peninsula.

The Larsemann Hills is the Antarctic Specially Managed Area No. 6 (ATCM, 2021), while its largest portion, the Stornes Peninsula, is the Antarctic Specially Protected Area (ASPA) No. 174, some sort of a geological and mineralogical reserve (ATCM, 2014).

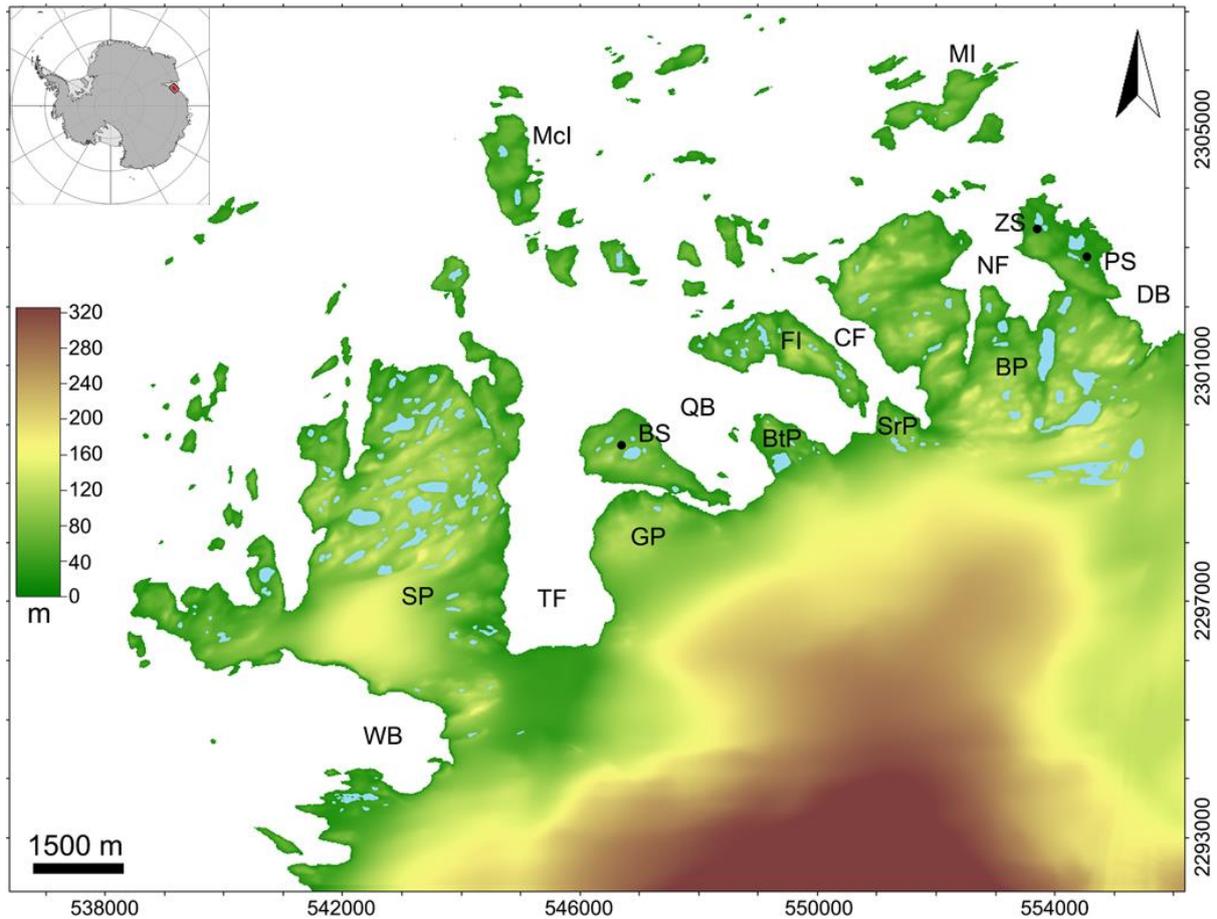

Figure 1. The geographic location and elevations of the Larsemann Hills extracted from the REMA DEM (Howat et al., 2019; REMA, 2018–2022). Peninsulas: BP – Broknes, BtP – Brattnevet, GP – Grovnes, SP – Stornes, SrP – Stinear; islands: FI – Fisher, MI – Manning, McI – McLeod; fjords: CF – Clemence, NF – Nella, TF – Thala; bays: DB – Dålkoy, WB – Wilcock, QB – Quilty; stations: PS – Progress (Russia), ZS – Zhogshan (China), BS – Bharati (India).

There are three year-round operated polar stations in the oasis: Progress (Russia), Zhongshan (China), and Bharati (India) (Figure 1). There is the snow-ice airfield Zenith (Russia) on the glacial plateau, to the south of the Progress Station.

### 2.2. Thala Hills

The Thala Hills (Figure 2) include two ice-separated coastal oases, Molodezhny and Vecherny. These are ice-free hills with an area of about 20 km² located on the shore of the Alasheev Gulf of the Cosmonauts Sea (Enderby Land, East Antarctica). The Thala Hills are bounded by the Hays outlet glacier in the east, by the glacial plateau in the south, and by the sea in the west.

The Thala Hills are predominantly composed of the Proterozoic charnockite series (amphibole-pyroxene-plagioclase-quartz feldspar charnockitized enderbites, amphibole-feldspar-quartz-plagioclase charnockites, and ultrametamorphic feldspar-quartz-plagioclase-amphibole-biotite gneissic-rapakivite-like charnockites) and Archean biotite-duroxene plagiogneisses (Grew, 1978; Myasnikov, 2011; Myasnikov et al., 2021).

Topographically, the Thala Hills is exarctic rocky hummocky terrain consisting of several ridges stretched parallel to the coast and separated by snow-covered terraced valleys with glaciers, lakes, and channels of temporary watercourses (Kakareka et al., 2015). The elevations range from 0 m to 103 m in the Molodezhny oasis, and from 0 m to 272 m (Mount

Vecheryaya) in the Vecherny oasis.

There are two seasonal bases, Molodezhnaya (Russia) and Gora Vechernyaya (Belarus) at a distance of 12 km from each other, in the oases Molodezhny and Vecherny, respectively. There are two snow-ice airfields (Russia and Belarus) on the glacial plateau, to the south of the Molodezhnaya and Gora Vechernyaya Bases.

### 2.3. Schirmacher Oasis

The Schirmacher Oasis (Figure 3) is located on the Princess Astrid Coast (Queen Maud Land, East Antarctica). This is an ice-free area 1.5–3.5 km wide stretched for about 20 km in the sublatitudinal direction. In the north, the Lazarev Ice Shelf with a width of about 80 km separates the oasis from the Lazarev Sea. In the south, the ice plateau, separated from the inland ice sheet by the Wohlthat Mountains, bound the oasis. In the east and west the oasis is bounded by the lateral branches of outlet glaciers enveloping those mountains (Bormann and Fritzsche, 1995; Dharwadkar et al, 2018; Shrivastava et al., 2019).

The Schirmacher Oasis is composed of metamorphosed rocks of the Precambrian crystalline basement including banded, augen, and garnet-biotite gneisses, alaskite, chondalites with migmatites; which in places are intersected by basalt, lamprophyre, pegmatite, dolerite, and apatite dikes (Bormann and Fritzsche, 1995; Dharwadkar et al., 2018; Shrivastava et al., 2019).

The topography of the Schirmacher oasis is an exarctic rocky hummocky terrain including roche moutonees and glacial valleys, with elevations ranging from 0 to 228 m and average elevations of about 100 m. There are up to 180 lakes within the oasis.

There are two year-round operated polar stations in the oasis: Novolazarevskaya (Russia) and Maitri (India) (Figure 3). There is the snow-ice airfield (Russia) on the glacial plateau, to the south of the Novolazarevskaya Station.

There is a glaciological ASPA No. 163 'Dakshin Gangotri Glacier' in the region.

### 2.4. Fildes Peninsula

The Fildes Peninsula is the ice-free southwestern tip of the King George Island, part of the South Shetland Islands (West Antarctica) (Figure 4). The peninsula measures about 15 km × 5 km, extends from southwest to northeast, and is bounded by the Maxwell Bay of the Bransfield Strait in the southeast, by the Fildes Strait in the southwest, by the Drake Passage in the northwest, and by the Arctowski Icefield in the northeast.

The Fildes Peninsula is composed of the late Cretaceous volcanogenic rocks (mainly basalts, andesites, and tuffs), including interlayers of volcaniclastic deposits between the andesite rocks with overlying layers of limestone, tuff conglomerates, sandstone, and clay of the early and middle Eocene (ATCM, 2009a; Schmid et al., 2017).

The topography of the peninsula is mainly an exarctic rocky hummocky terrain with elevations from 0 m to ~160 m and includes a rocky coastline and bays, an exarctic-denudation plateau, ancient and modern abrasion surfaces, denudation-accumulative marine terraces, rocky lowlands, and valleys (ATCM, 2009a; Schmid et al., 2017). The northeastern part of the peninsula, between the ice-free area and the Arctowsky Icefield, is occupied by the Bellingshausen Glacial Dome with elevations ranging from 0 m to 250 m.

At the Fildes Peninsula, there are small snowfields and small lakes. The territory is drained by dozens of minor streams. There are relatively small areas of marshy ground everywhere. The free of ice and snow territory of the peninsula is subantarctic tundra.

The Fildes Peninsula is one of the most densely populated areas of Antarctica. Four year-round stations are located there (Figure 4): Bellingshausen (Russia), Presidente Eduardo Montalva (Chile; territorially coinciding with the village of Villa las Estrellas), Artigas (Uruguay), and Chancheng (China). Also, there are the Chilean Rodolfo Marsh airport.

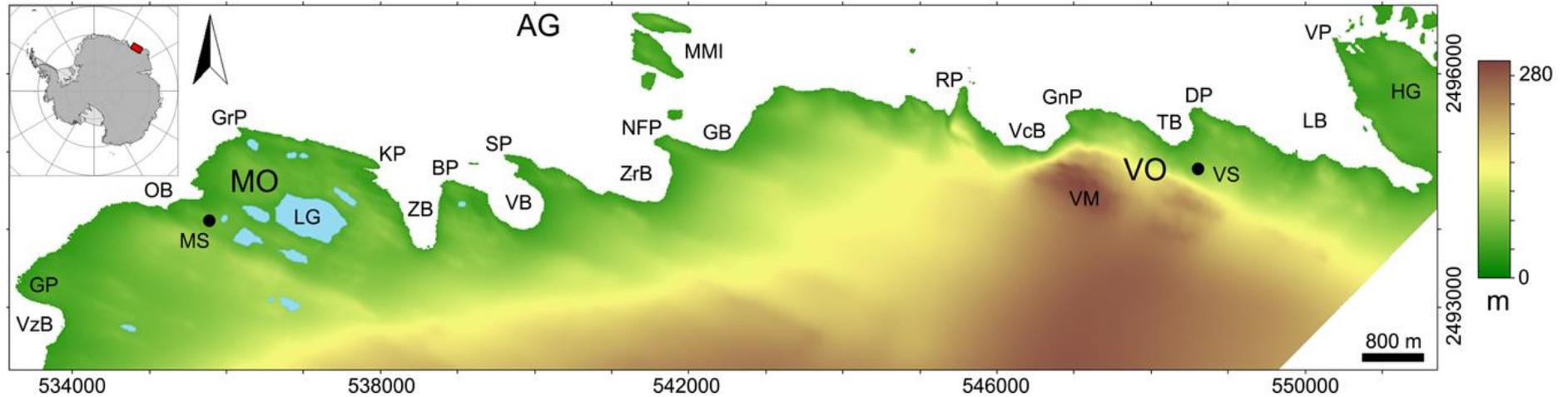

Figure 2. The geographic location and elevations of the Thala Hills extracted from the REMA DEM (Howat et al., 2019; REMA, 2018–2022). Oases: MO – Molodezhny, VO – Vecherny; AG – Alasheev Gulf; VM – Mount Vechernyaya; MMI – McMahon Islands; HG – Hays Glacier; bays: GB – Groznaya, LB – Lazurnaya, OB – Opasnaya, TB – Terpeniya, VB – Voskhod, VcB – Vechernyaya, VzB – Vozrozhdeniya, ZB – Zarya, ZrB – Zerkalnaya; capes: BP – Bliznetsov, DP – Dostupny, GP – Gaudisa, GnP – Gnezdovoi, GrP – Granat, KP – Kogot, NFP – Nikolay Feoktistov, RP – Rog, SP – Steregushchy, VP – Vyvodnoy; LG – Lake Glubokoe; seasonal bases: MS – Molodezhnaya, VS – Gora Vechernyaya.

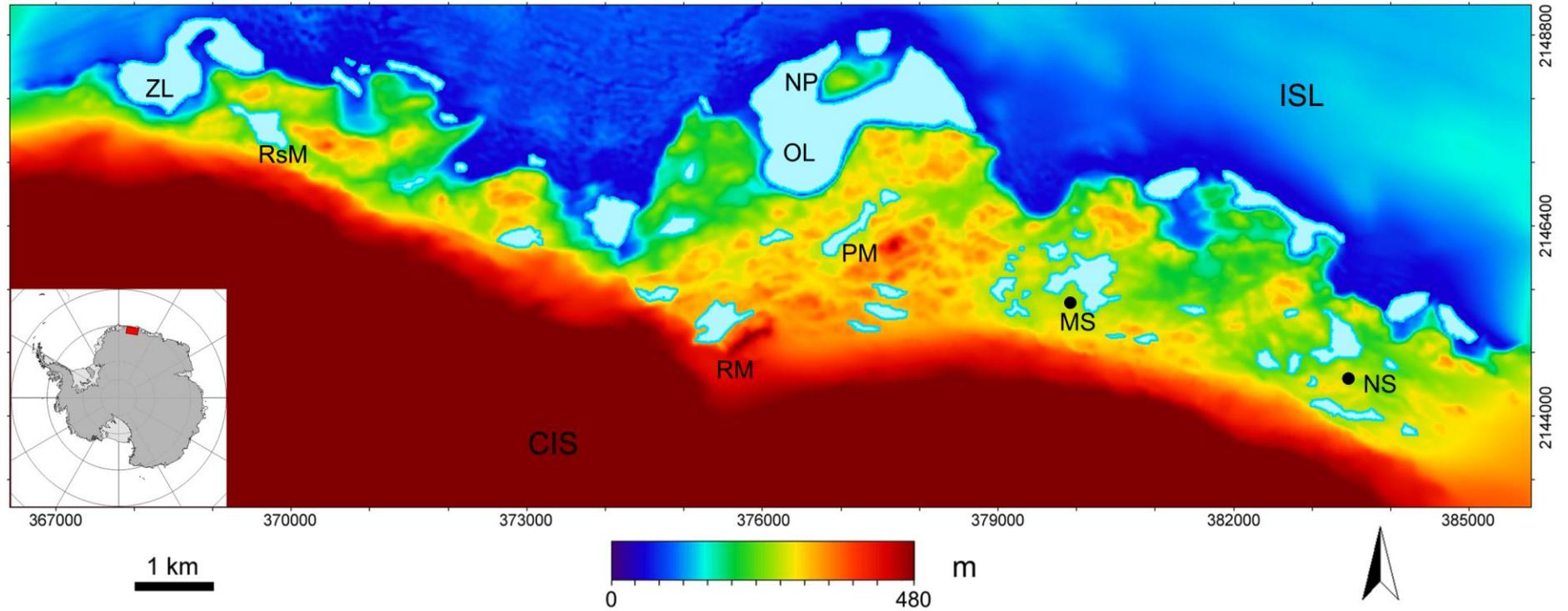

Figure 3. The geographic location and elevations of the Schirmacher Oasis extracted from the REMA DEM (Howat et al., 2019; REMA, 2018–2022). CIS – continental ice sheet; ISL – the Lazarev ice shelf; polar stations: NS – Novolazarevskaya (Russia), MS – Maitri (India); lakes: OL – Ozhidaniya, ZL – Zigzag; mountains: PM – Primetnaya, RM – Rebristaya, RsM – Ryskalina; NP – Nadezhda Peninsula

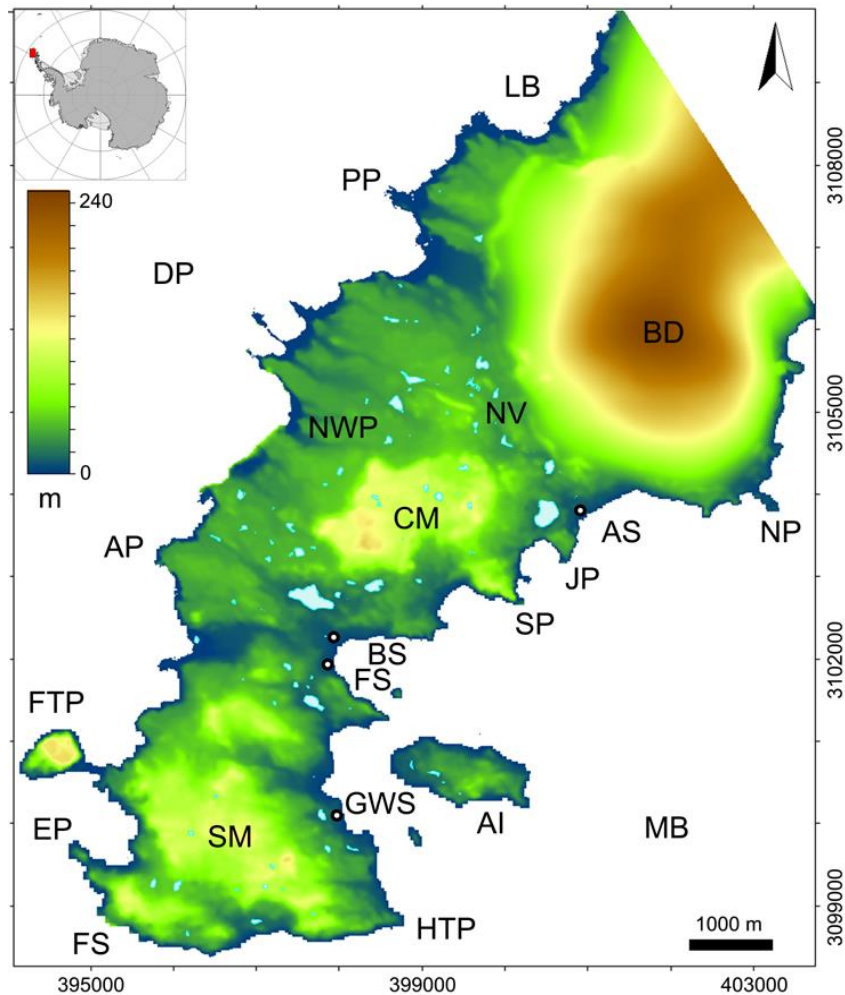

Figure 4. The geographic location and elevations of the Fildes Peninsula extracted from the REMA DEM (Howat et al., 2019; REMA, 2018–2022). DP – Drake Passage, FS – Fildes Strait; MB – Maxwell Bay; AI – Ardley Island; BD – Bellingshausen Ice Dome, NV – North Valley, NWP – North-western Platform, CM – Central Massif, SM – South Mountains; capes: AP – Aerodromic, EP – Exotic, HTP – Half Three, NP – Nebles Point, PP – Priroda, SP – Suffield; LB – Lednikovaya Bay; polar stations: AS – Artigas (Uruguay), BS – Bellingshausen (Russia), FS – Frei (Chile), GWS – Chancheng (China)

In the region, there are two ASPAs of paleontological and ornithological profiles: No. 125 'Fildes Peninsula' and No. 150 'Ardley Island' (ATCM, 2009a, 2009b).

### 3. Materials and methods

We used fragments of the Reference Elevation Model of Antarctica (REMA) ver. 1.1 DEM (Howat et al., 2019; REMA, 2018–2022) as input data for geomorphometric calculations, modeling, and mapping. To date, this DEM obtained photogrammetrically from WorldView-1, 2, and 3 satellite images is the most complete and accurate DEM of Antarctica.

Our preliminary assessment of the accuracy and quality of the REMA DEM showed that within the almost all territories under study, they generally meet the requirements of geomorphometric modeling, that is, low level of high-frequency noise, low number of obvious artifacts, and sufficient surface smoothness not requiring additional smoothing or filtering).

However, for the Fildes Peninsula of the King George Island (and generally for all the South Shetland Islands), the accuracy and quality of the REMA DEM are lower comparing with the most territory of Antarctica. In particular, this territory is marked by a high number of artifacts along island shores. This is connected with peculiarities of photogrammetric processing of water surface images. At the same time, such artifacts can be identified and edited (removed).

It is expected that description of the South Shetland Islands will be improved in next versions of the REMA DEM.

From the REMA DEM, we extracted the following four fragments:

- For the Larsemann Hills: a fragment of 18,168 m × 14,968 m, which includes the Larsemann Hills oasis and adjacent areas of the ice sheet and outlet glaciers. The extracted DEM contains 4,253,184 points (the elevation matrix of 2,272 × 1,872).
- For the Thala Hills: a fragment of 4,688 m × 20,920 m, which includes the Molodezhny and Vecherny Oases and adjacent areas of the ice sheet and outlet glaciers. The extracted DEM contains 1,535,592 points (the elevation matrix of 587 × 2,616).
- For the Schirmacher Oasis: a fragment of 6,328 m × 20,544 m, which includes the oasis and adjacent areas of the ice sheet and ice shelf. The extracted DEM contains 2,034,648 points (the elevation matrix of 792 × 2,569).
- For the Fildes Peninsula: a fragment of 9,672 m × 11,616 m, which includes the peninsula, adjacent islands, and a portion of the Bellingshausen Ice Dome. The extracted DEM contains 1,758,130 points (the elevation matrix of 1,210 × 1,453).

Since the REMA DEM does not describe the submarine topography, the extracted DEMs includes 2,068,674, 783,812, and 680,592 cells with elevation values for the Larsemann Hills, Thala Hills, and Fildes Peninsula, correspondingly.

For all extracted DEM, the DEM grid spacing is 8 m.

Before further processing, the extracted DEMs presented in the polar stereographic projection were reprojected into the UTM one, zone 43S, 38S, 33S, and 21S for the Larsemann Hills, Thala Hills, Schirmacher Oasis, and Fildes Peninsula, correspondingly (Figures 1–4). Reprojected DEMs also have the grid spacing of 8 m.

At the pre-processing stage, we manually edited the extracted DEMs in order to remove obvious artifacts, in particular, false islands, which are actually images of icebergs as well as well as a consequence of incorrect photogrammetric processing of satellite scenes containing images of the sea surface. To edit the DEMs, we used the available topographic maps and orthomosaic (Ministry of Merchant Marine of the USSR, 1972; Instituto Geográfico Militar de Chile, 1996; Australian Antarctic Division, 2005; Florinsky and Skrypitsyna, 2022).

From the edited DEMs, we derived digital models of 17 morphometric characteristics, namely: 14 local morphometric variables, such as: slope ($G$), aspect ($A$), horizontal curvature ($k_h$), vertical curvature ($k_v$), difference curvature ($E$), horizontal excess curvature ($k_{he}$), vertical excess curvature ($k_{ve}$), accumulation curvature ($K_a$), ring curvature ($K_r$), minimal curvature ($k_{min}$), maximal curvature ($k_{max}$), mean curvature ($H$), Gaussian curvature ($K$), unsphericity curvature ($M$); one nonlocal morphometric variable, catchment area ($CA$), as well as two combined morphometric variables: topographic index ($TI$) and stream power index ($SI$). The definitions, formulas, and brief interpretations of these morphometric variables are presented in Table 1.

Then we produced maps of the morphometric variables from the calculated digital models. Morphometric maps for the Larsemann Hills, Thala Hills, Schirmacher Oasis, and Fildes Peninsula are presented on Figures 5–21, 22–38, 39–55, and 56–72, correspondingly.

For the pre-processing and editing of DEM, geomorphometric calculations, and mapping, we used the System for Automated Geoscientific Analyses (SAGA) Version 7.8.2 software (Conrad et al., 2015). It should be noted that the capabilities of SAGA allow the direct computation of digital models for only a part of the morphometric variables mentioned above. In particular, we directly calculated digital models of six local morphometric variables using the Evans method (Evans, 1972): $G$, $A$, $k_h$, $k_v$, $k_{min}$, and $k_{max}$.

Digital models of the remaining eight local morphometric characteristics − $E$, $k_{he}$, $k_{ve}$, $K_a$, $K_r$, $H$, $K$, and $M$ − were sequentially calculated using their formulas (Table 1) and the SAGA option allowing arithmetic operations with raster layers. Also, we similarly derived digital models of $CA$, $TI$, and $SI$ using a digital model of 'total catchment area' directly computed in SAGA. In calculating $TI$ and $SI$, we also used the previously derived digital models of $G$.

Table 1. Formulas, definitions, and interpretations of calculated morphometric variables (Florinsky, 2016, 2017).

| Variable, notation, and unit | Formula, definition and interpretation |
|---|---|
| Slope, $G$ (°) | $$G = \arctan\sqrt{p^2 + q^2}$$ An angle between the tangential and horizontal planes at a given point of the topographic surface. Relates to the velocity of gravity-driven flows. |
| Aspect, $A$ (°) | $$A = -90\left[1 - \text{sign}(q)\right]\left(1 - \left|\text{sign}(p)\right|\right) + 180\left[1 + \text{sign}(p)\right] - \frac{180}{\pi}\text{sign}(p)\arccos\left(\frac{-q}{\sqrt{p^2+q^2}}\right)$$ An angle between the northern direction and the horizontal projection of the two-dimensional vector of gradient counted clockwise at a given point of the topographic surface. A measure of the direction of gravity-driven flows. |
| Horizontal (tangential) curvature, $k_h$ (m$^{-1}$) | $$k_h = -\frac{q^2 r - 2pqs + p^2 t}{(p^2+q^2)\sqrt{1+p^2+q^2}}$$ A curvature of a normal section tangential to a contour line at a given point of the surface. $k_h$ is a measure of flow convergence and divergence. Gravity-driven lateral flows converge where $k_h < 0$, and diverge where $k_h > 0$. $k_h$ mapping reveals ridge and valley spurs. |
| Vertical (profile) curvature, $k_v$ (m$^{-1}$) | $$k_v = -\frac{p^2 r + 2pqs + q^2 t}{(p^2+q^2)\sqrt{(1+p^2+q^2)^3}}$$ A curvature of a normal section having a common tangent line with a slope line at a given point of the surface. $k_v$ is a measure of relative deceleration and acceleration of gravity-driven flows. They are decelerated where $k_v < 0$, and are accelerated where $k_v > 0$. $k_v$ mapping allows revealing terraces and scarps. |
| Difference curvature, $E$ (m$^{-1}$) | $$E = \frac{1}{2}(k_v - k_h) = \frac{q^2 r - 2pqs + p^2 t}{(p^2+q^2)\sqrt{1+p^2+q^2}} - \frac{(1+q^2)r - 2pqs + (1+p^2)t}{2\sqrt{(1+p^2+q^2)^3}}$$ A half-difference of vertical and horizontal curvatures. Shows to what extent the relative deceleration of flows (measured by $k_v$) is higher than flow convergence (measured by $k_h$) at a given point of the topographic surface. |
| Minimal curvature, $k_{min}$ (m$^{-1}$) | $$k_{\min} = H - M = H - \sqrt{H^2 - K}$$ A curvature of a principal section with the lowest value of curvature at a given point of the surface. $k_{min} > 0$ corresponds to local convex landforms, while $k_{min} < 0$ relates to elongated concave landforms (e.g., hills and troughs, correspondingly). |
| Maximal curvature, $k_{max}$ (m$^{-1}$) | $$k_{\max} = H + M = H + \sqrt{H^2 - K}$$ A curvature of a principal section with the highest value of curvature at a given point of the surface. $k_{max} > 0$ corresponds to elongated convex landforms, while $k_{max} < 0$ relate to local concave landforms (e.g., crests and holes, correspondingly). |
| Horizontal excess curvature, $k_{he}$ (m$^{-1}$) | $$k_{he} = k_h - k_{\min} = M - E$$ A difference of horizontal and minimal curvatures. Shows to what extent the bending of a normal section tangential to a contour line is larger than the minimal bending at a given point of the topographic surface. |
| Vertical excess curvature, $k_{ve}$ (m$^{-1}$) | $$k_{ve} = k_v - k_{\min} = M + E$$ A difference of vertical and minimal curvatures. Shows to what extent the bending of a normal section having a common tangent line with a slope line is larger than the minimal bending at a given point of the topographic surface. |
| Accumulation curvature, $K_a$ (m$^{-2}$) | $K_a = k_h k_v = \dfrac{\left(q^2 r - 2pqs + p^2 t\right)\left(p^2 r + 2pqs + q^2 t\right)}{\left[\left(p^2 + q^2\right)\left(1 + p^2 + q^2\right)\right]^2}$ |

A product of vertical and horizontal curvatures. A measure of the extent of local accumulation of flows at a given point of the topographic surface.

| | |
|---|---|
| Ring curvature, $K_r$ (m$^{-2}$) | $K_r = k_{he} k_{ve} = M^2 - E^2 = \left[\dfrac{\left(p^2 - q^2\right)s - pq(r - t)}{\left(p^2 + q^2\right)\left(1 + p^2 + q^2\right)}\right]^2$ |

A product of horizontal excess and vertical excess curvatures. Describes flow line twisting.

| | |
|---|---|
| Mean curvature, $H$ (m$^{-1}$) | $H = \dfrac{1}{2}\left(k_{\min} + k_{\max}\right) = \dfrac{1}{2}\left(k_h + k_v\right) = -\dfrac{\left(1 + q^2\right)r - 2pqs + \left(1 + p^2\right)t}{2\sqrt{\left(1 + p^2 + q^2\right)^3}}$ |

A half-sum of curvatures of any two orthogonal normal sections at a given point of the topographic surface. Represents two accumulation mechanisms of gravity-driven substances — convergence and relative deceleration of flows — with equal weights.

| | |
|---|---|
| Gaussian curvature, $K$ (m$^{-2}$) | $K = k_{\min} k_{\max} = \dfrac{rt - s^2}{\left(1 + p^2 + q^2\right)^2}$ |

A product of maximal and minimal curvatures. Gaussian curvature retains values in each point of the surface after its bending without breaking, stretching, and compressing.

| | |
|---|---|
| Unsphericity curvature, $M$ (m$^{-1}$) | $M = \dfrac{1}{2}\left(k_{\max} - k_{\min}\right) = \sqrt{H^2 - K}$ |

A half-difference of maximal and minimal curvatures. Shows the extent to which the shape of the surface is non-spherical at a given point.

| | |
|---|---|
| Catchment area, $CA$ (m$^2$) | An area of a closed figure formed by a contour segment at a given point of the surface and two flow lines coming from upslope to the contour segment ends. CA is a measure of the contributing area. |
| Topographic index, $TI$ | $TI = \ln\left[1 + CA/\left(10^{-3} + \text{tg}\,G\right)\right]$ |

A ratio of catchment area to slope at a given point of the topographic surface. A measure of the extent of flow accumulation.

| | |
|---|---|
| Stream power index, $SI$ | $SI = \ln\left(1 + CA \cdot \text{tg}\,G\right)$ |

A product of catchment area and slope at a given point of the topographic surface. A measure of potential flow erosion and related landscape processes.

$p = \dfrac{\partial z}{\partial x}$, $q = \dfrac{\partial z}{\partial y}$, $r = \dfrac{\partial^2 z}{\partial x^2}$, $t = \dfrac{\partial^2 z}{\partial y^2}$, $s = \dfrac{\partial^2 z}{\partial x \partial y}$ for $z = f(x, y)$, where z is elevation, x and y are Cartesian coordinates.

The REMA DEM does not include the lake bathymetry. However, DEM cells relating lakes are not indicated as 'no data value' like submarine topography cells. Lake cells include 'elevation' values obtained by interpolation of lakeshore elevations, that is, these cells are artifacts of data processing. Due to this fact, lakes on all derived maps should be masked.

To create lake masks for maps of the Larsemann Hills (Figures 1, 5–21), Thala Hills (Figures 2, 22–38), Schirmacher Oasis (Figures 3, 39–55), and Fildes Peninsula (Figures 4, 56–72), we utilized the topographic maps (Australian Antarctic Division, 2005; Ministry of Merchant Marine of the USSR, 1972; Instituto Geográfico Militar de Chile, 1996) and the orthomozaic (Florinsky and Skrypitsyna, 2022).

Morphometric maps are presented in the UTM projection; zone 43S for the Larsemann Hills (Figures 5–21), zone 38S for the Thala Hills (Figures 22–38), zone 33S for the Schirmacher Oasis (Figures 39– 55), and zone 21S for the Fildes Peninsula (Figures 56–72).

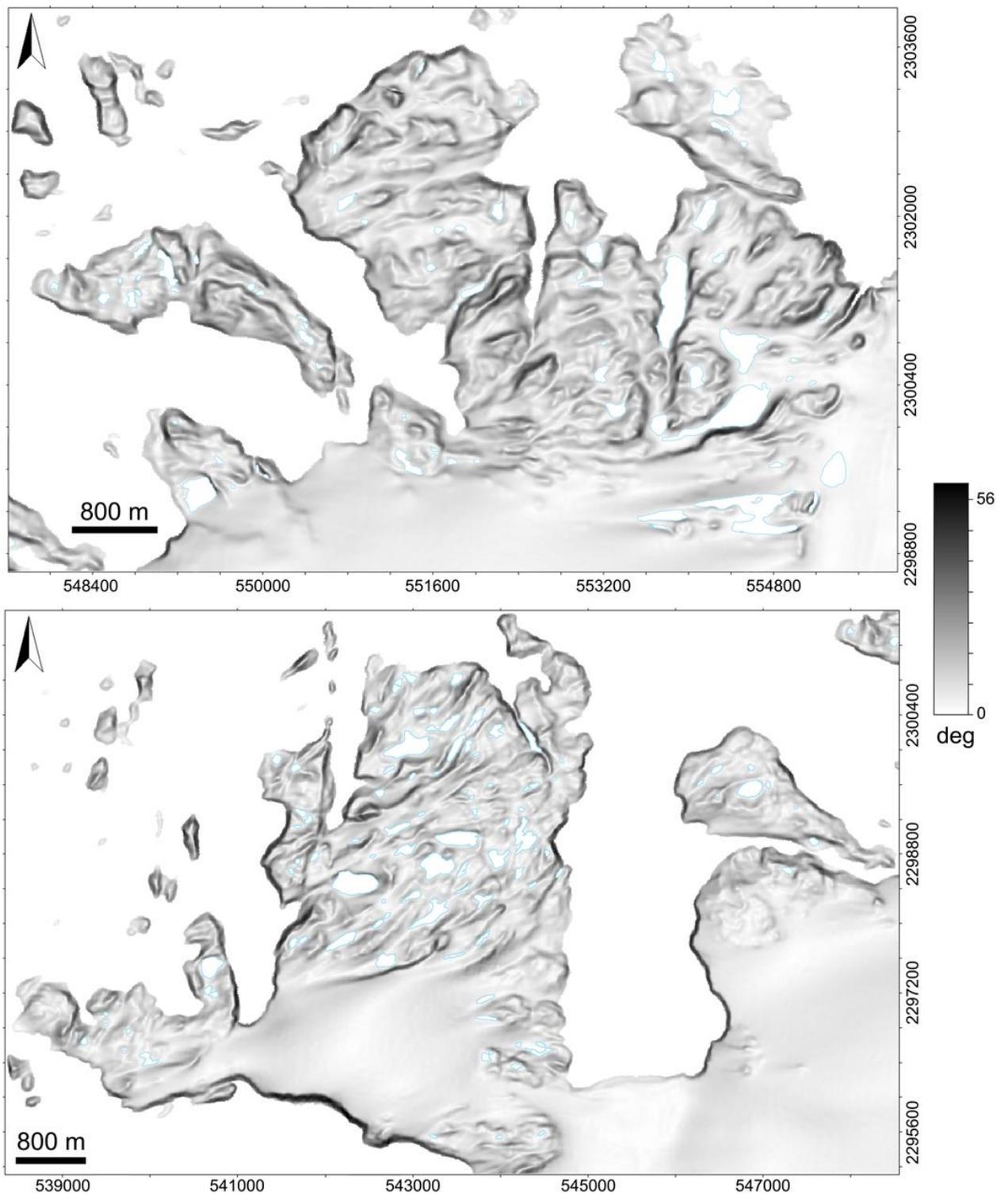

Figure 5. Larsemann Hills, slope.
Upper: Broknes Peninsula. Lower: Stornes Peninsula.

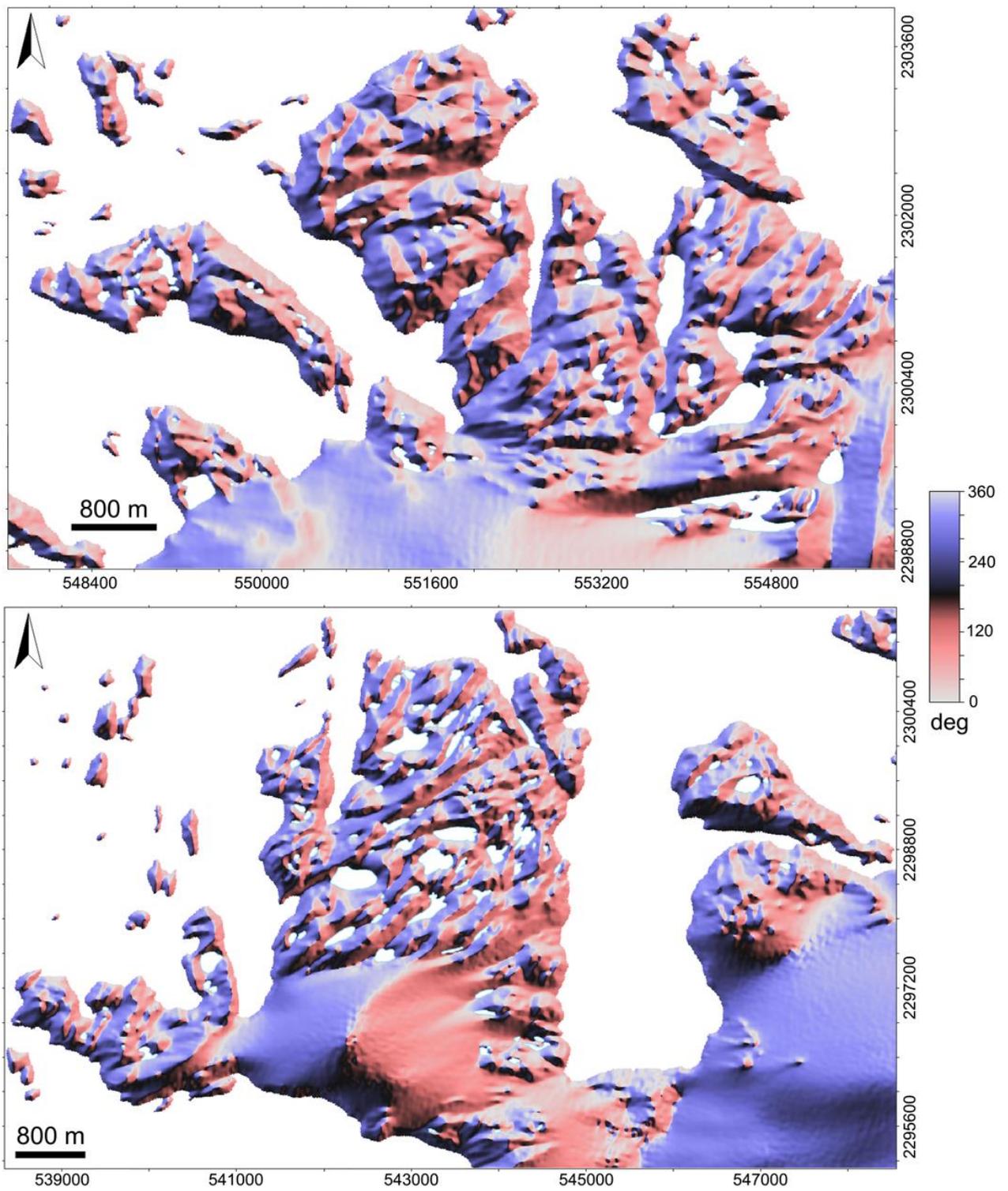

Figure 6. Larsemann Hills, aspect.
Upper: Broknes Peninsula. Lower: Stornes Peninsula.

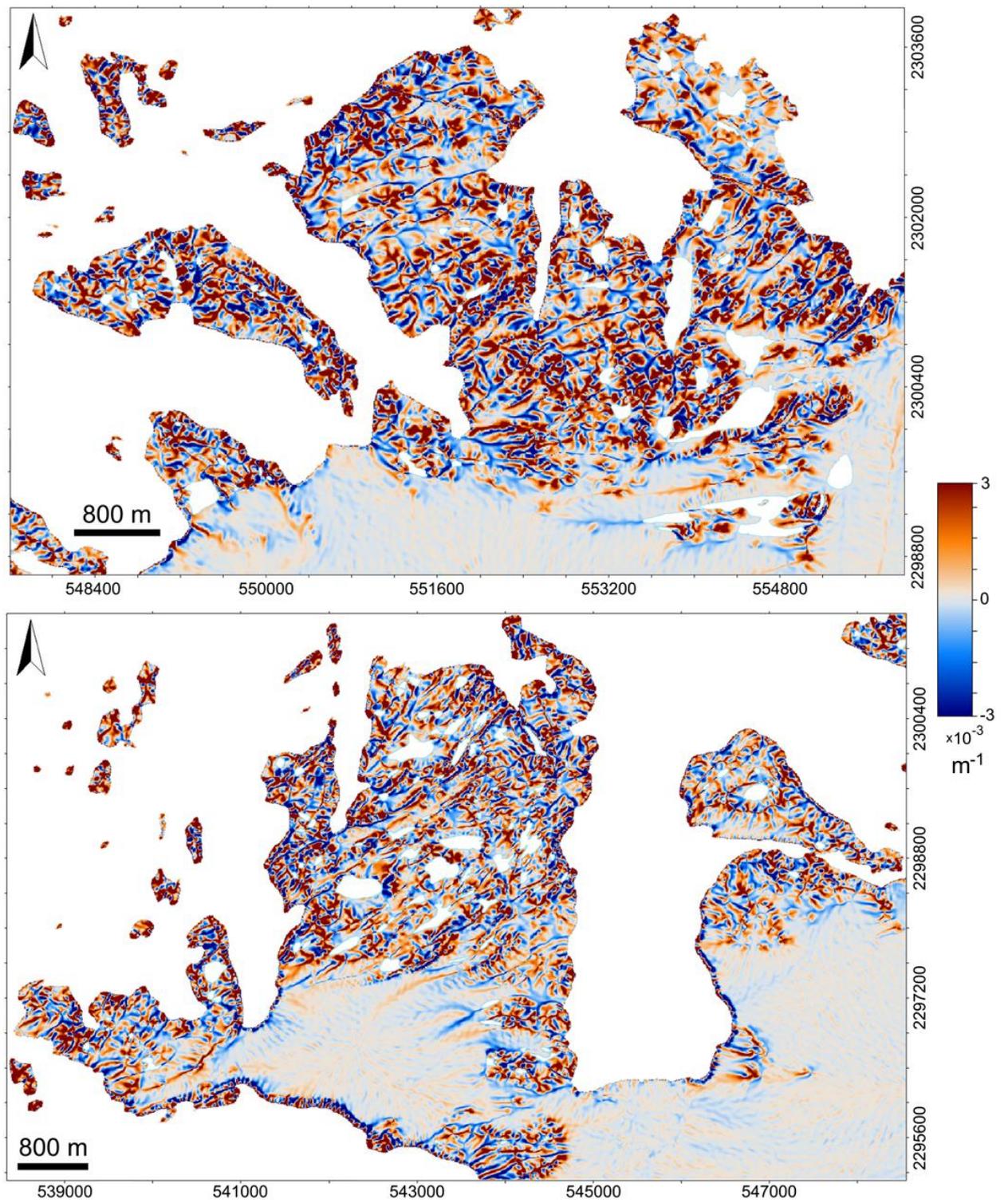

Figure 7. Larsemann Hills, horizontal curvature.
Upper: Broknes Peninsula. Lower: Stornes Peninsula.

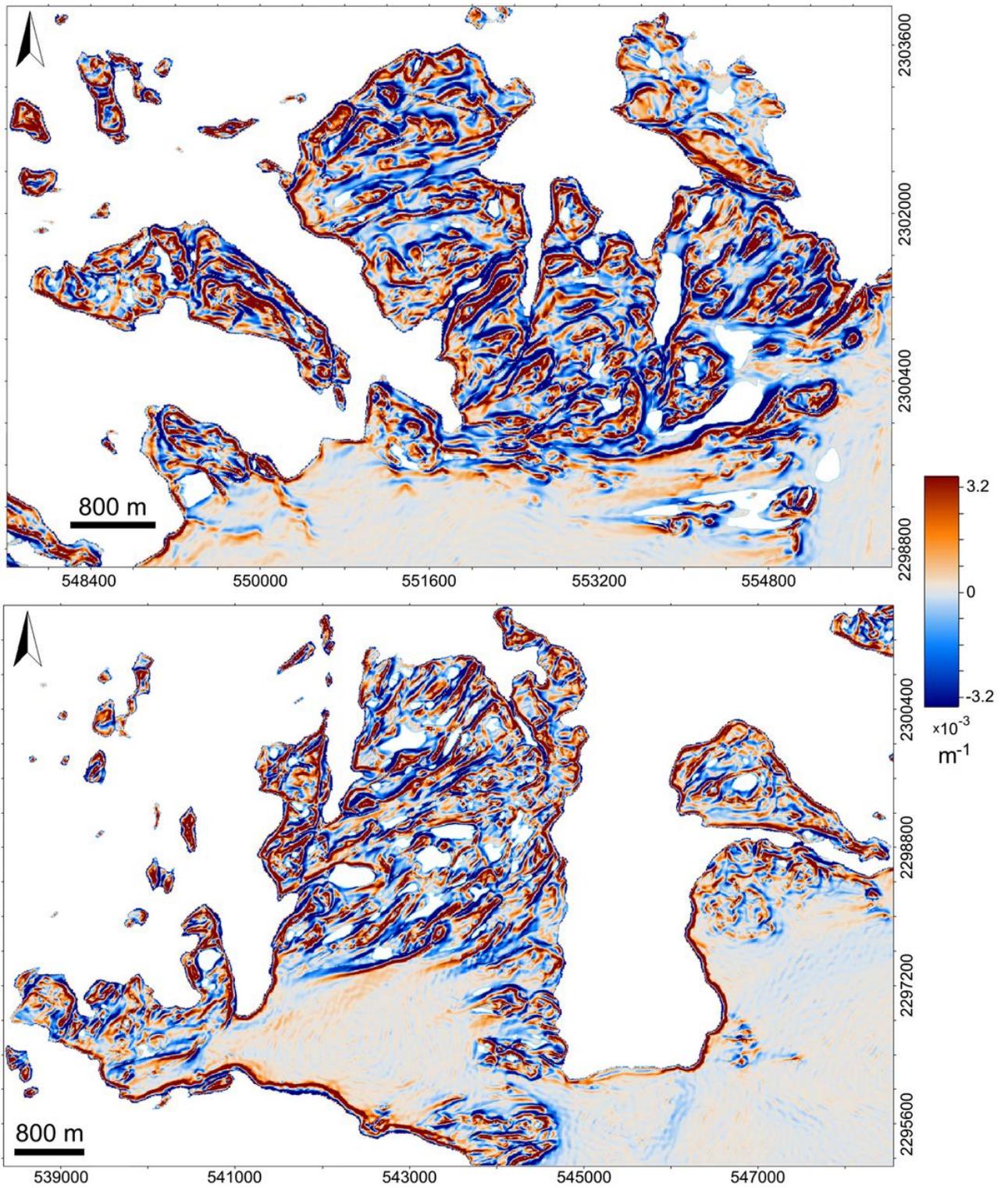

Figure 8. Larsemann Hills, vertical curvature.
Upper: Broknes Peninsula. Lower: Stornes Peninsula.

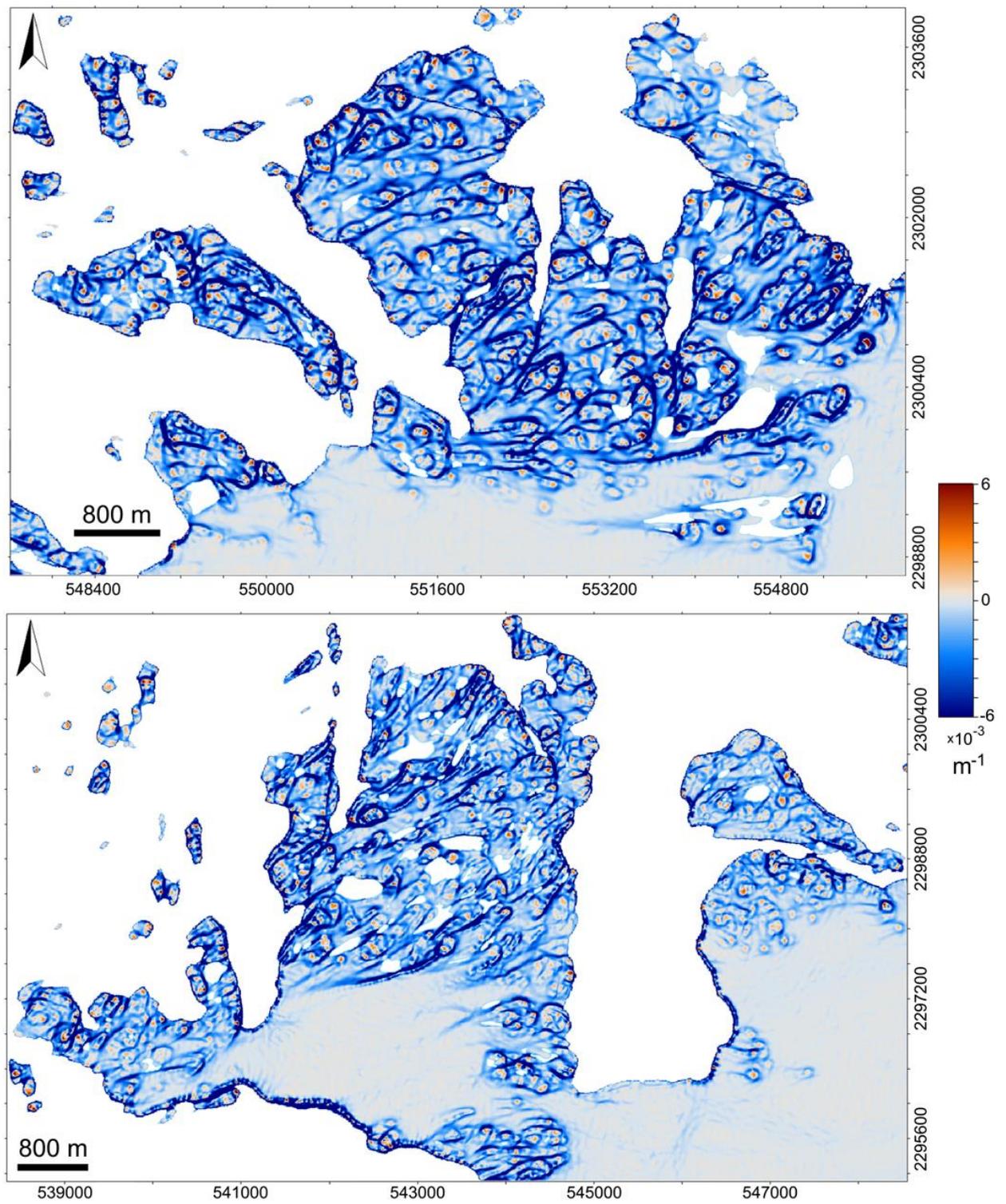

Figure 9. Larsemann Hills, minimal curvature.
Upper: Broknes Peninsula. Lower: Stornes Peninsula.

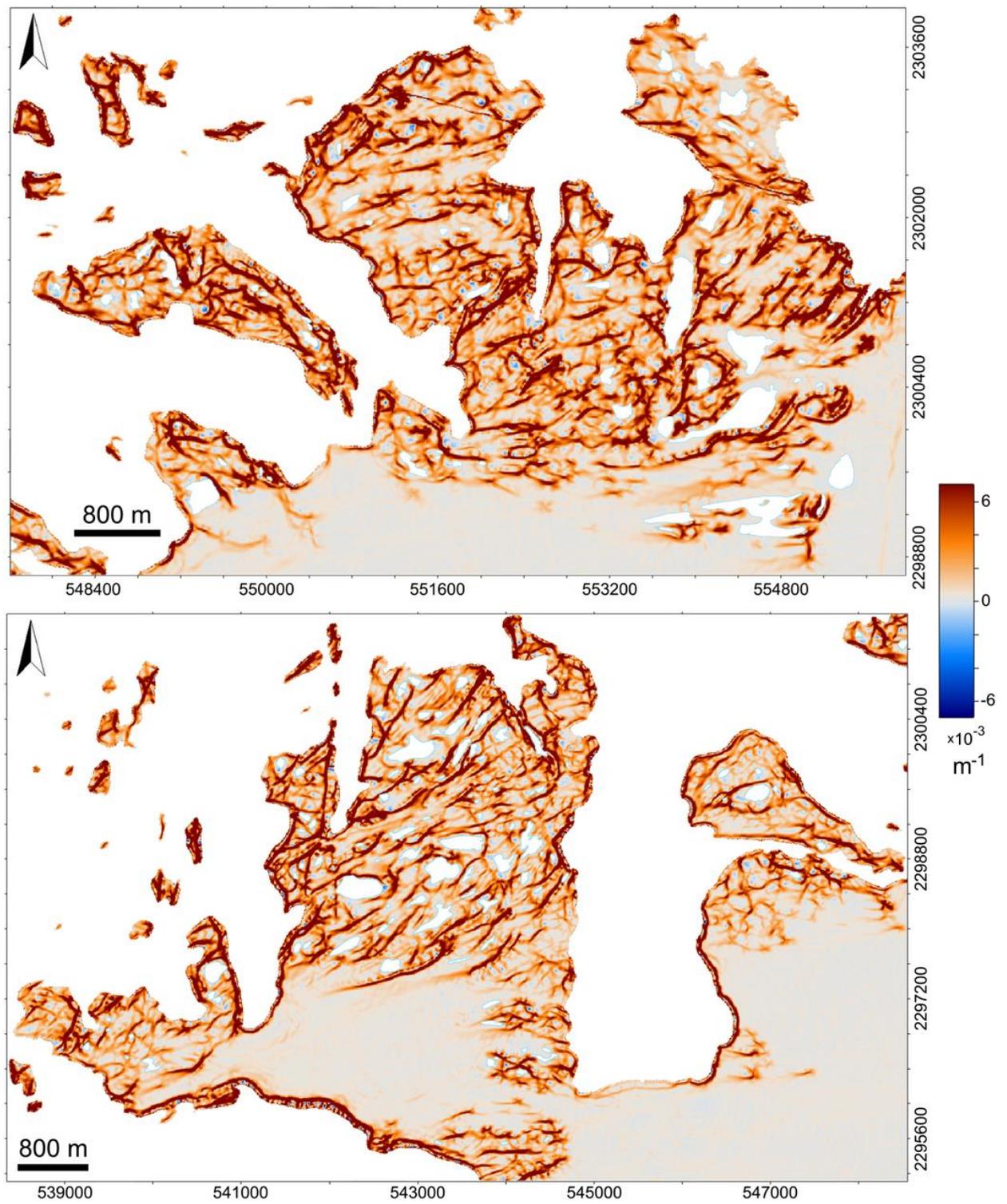

Figure 10. Larsemann Hills, maximal curvature.
Upper: Broknes Peninsula. Lower: Stornes Peninsula.

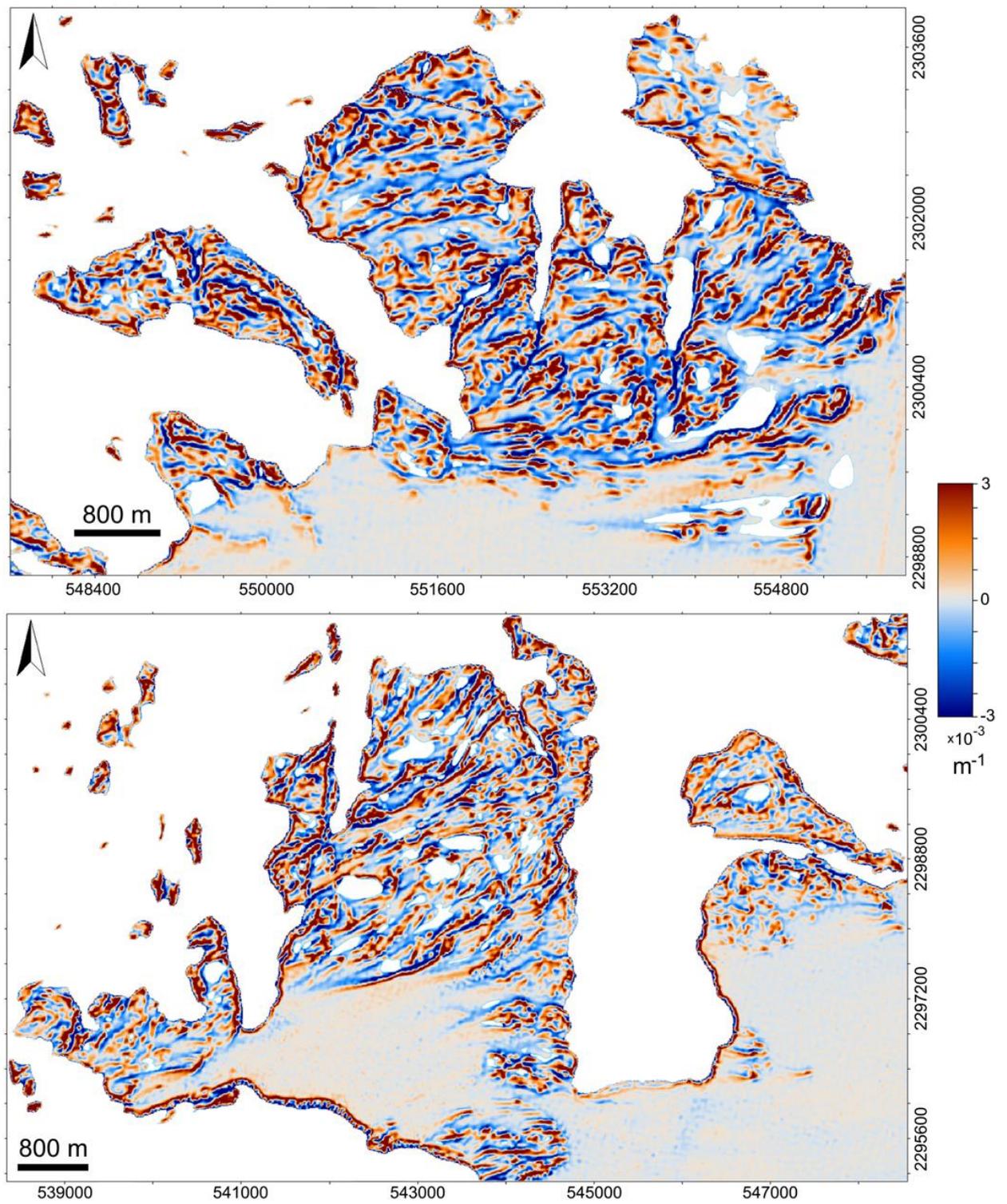

Figure 11. Larsemann Hills, mean curvature.
Upper: Broknes Peninsula. Lower: Stornes Peninsula.

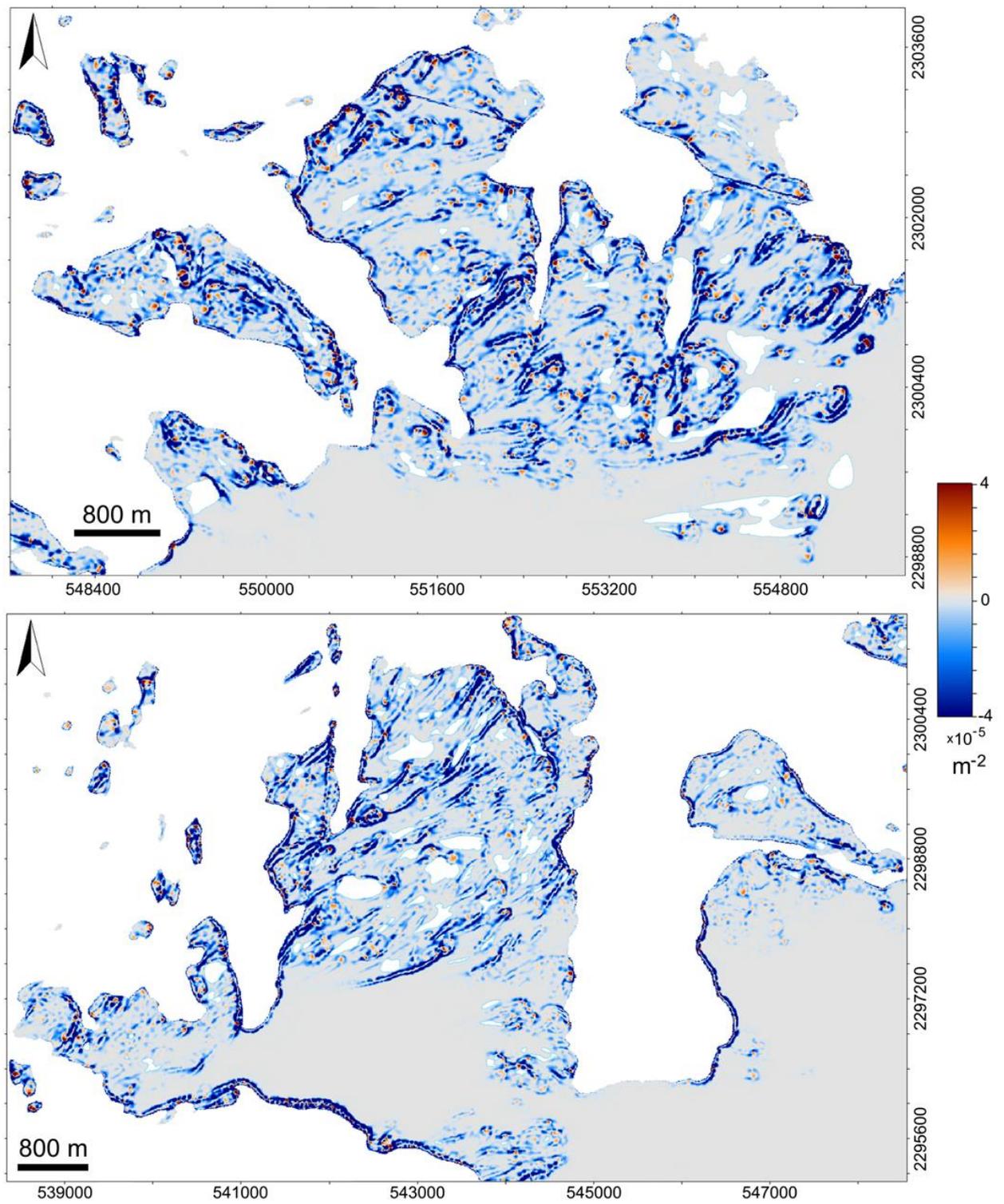

Figure 12. Larsemann Hills, Gaussian curvature.
Upper: Broknes Peninsula. Lower: Stornes Peninsula.

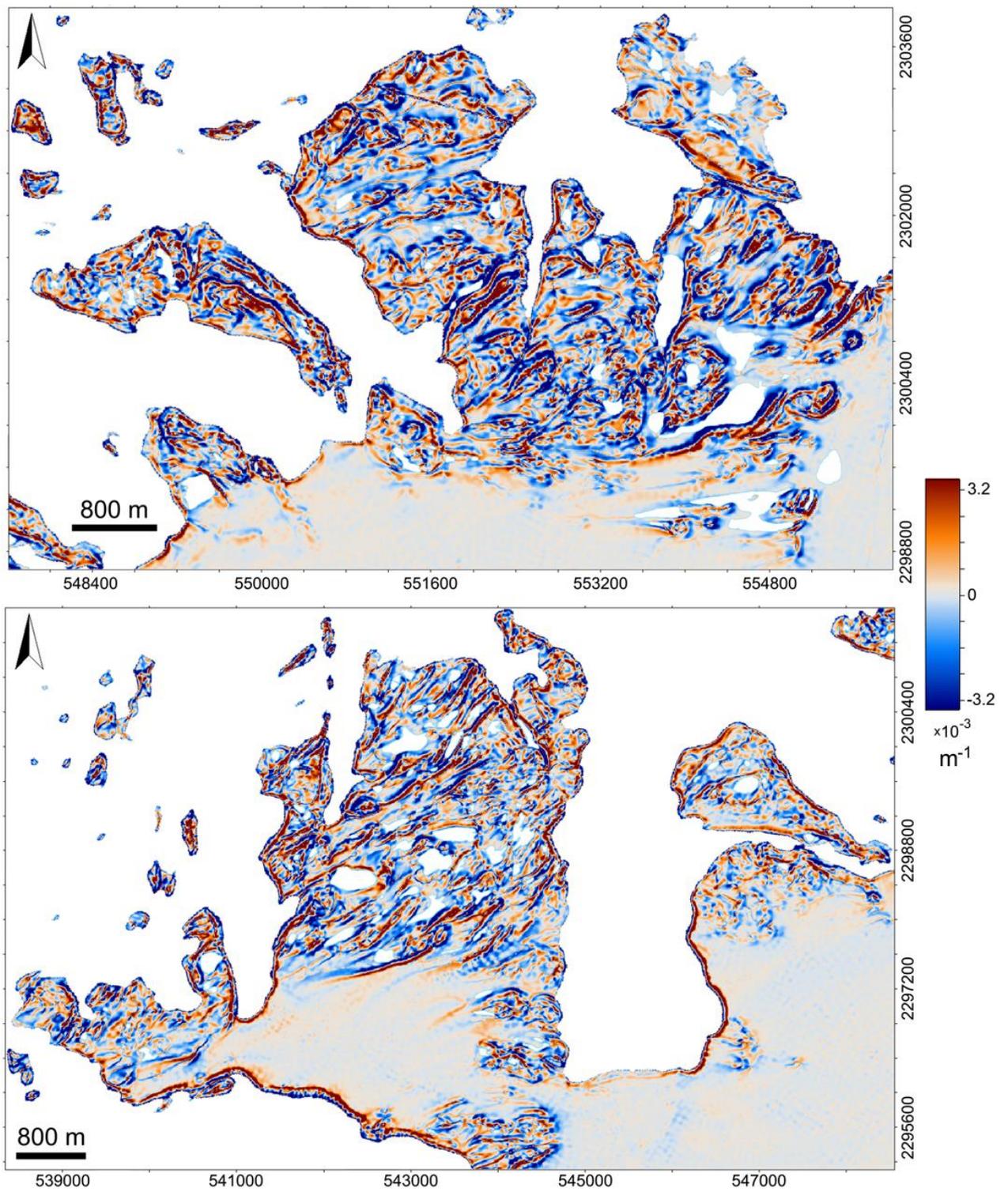

Figure 13. Larsemann Hills, difference curvature.
Upper: Broknes Peninsula. Lower: Stornes Peninsula.

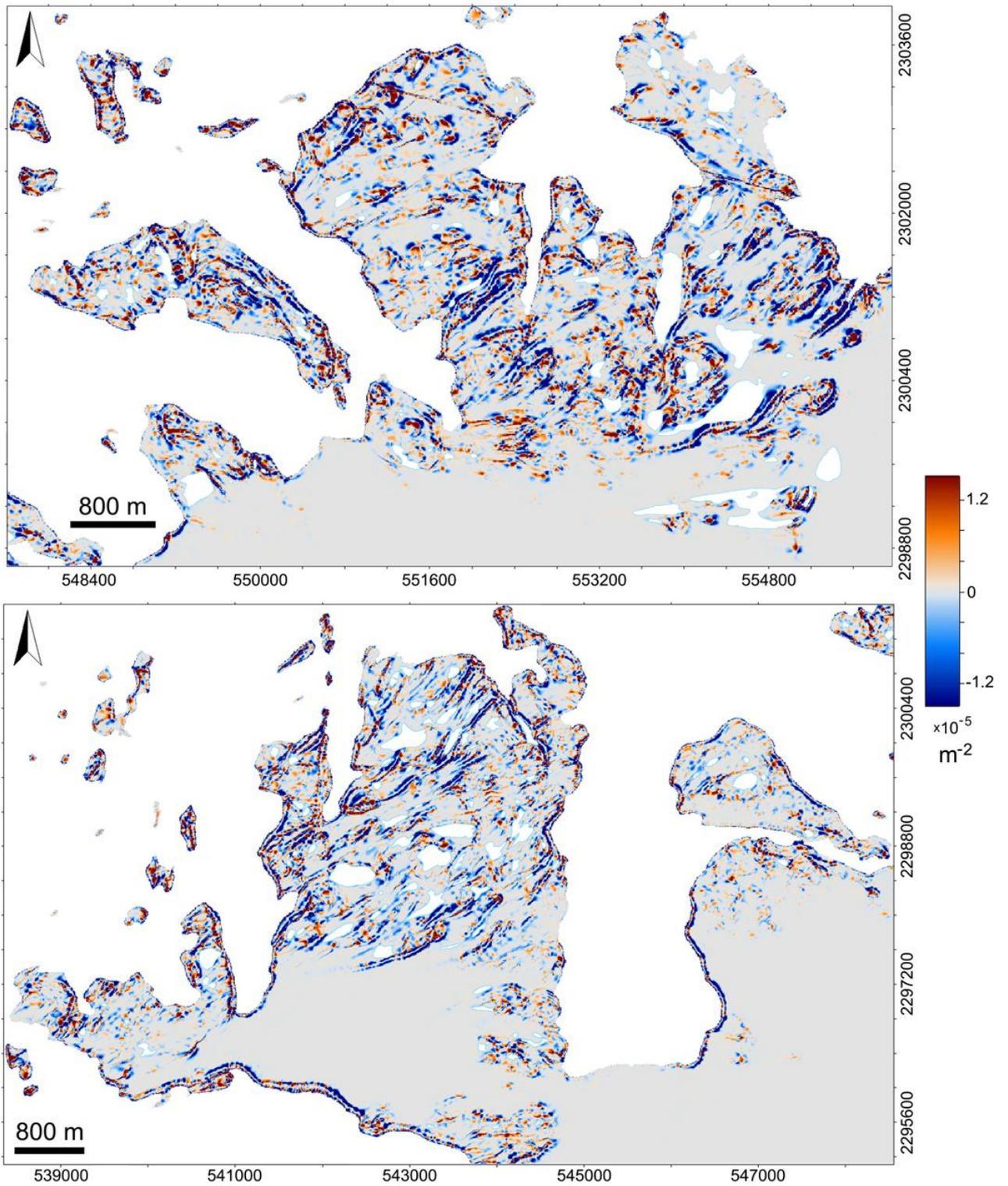

Figure 14. Larsemann Hills, accumulation curvature.
Upper: Brokness Peninsula. Lower: Stornes Peninsula.

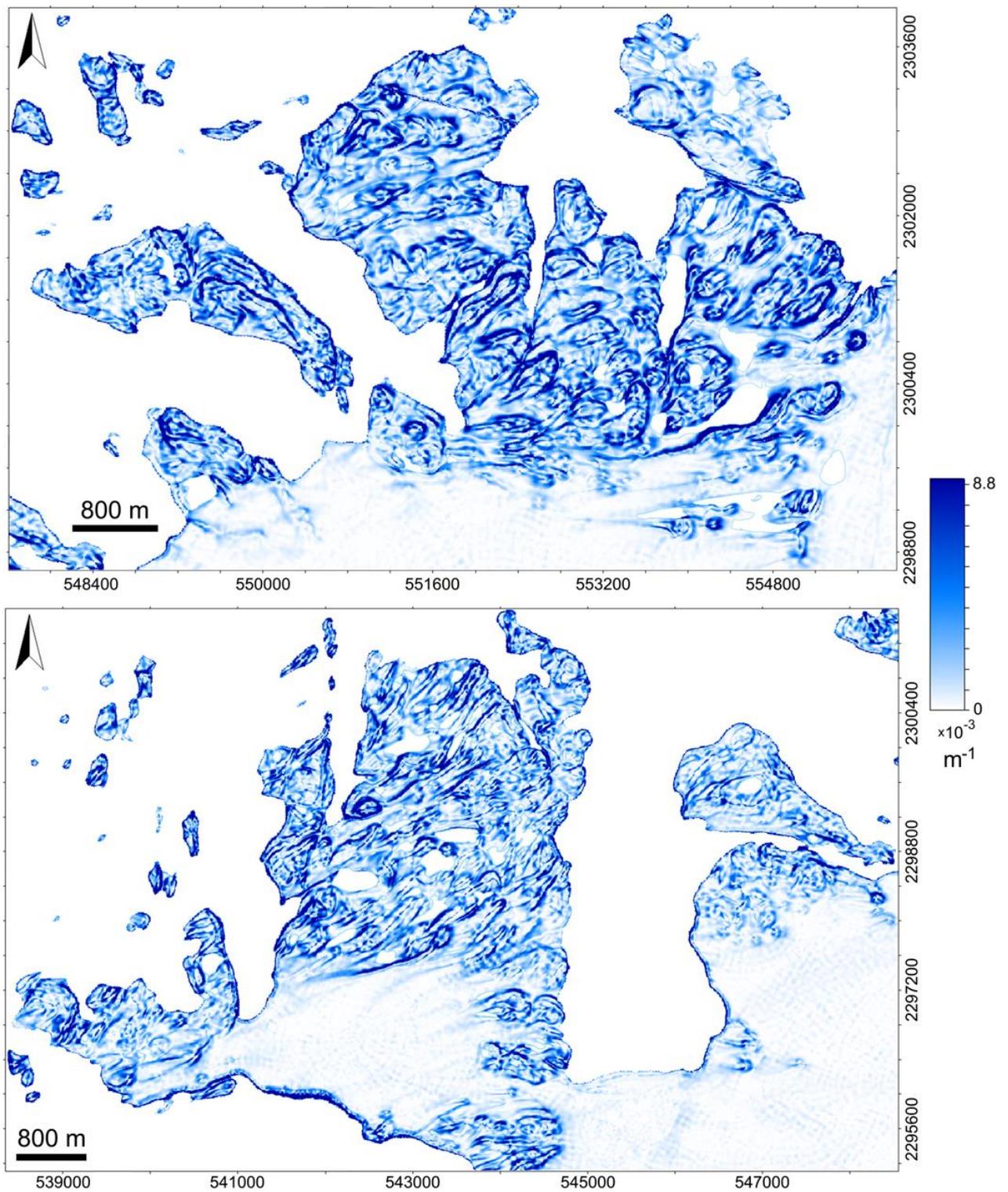

Figure 15. Larsemann Hills, horizontal excess curvature.
Upper: Brokness Peninsula. Lower: Stornes Peninsula.

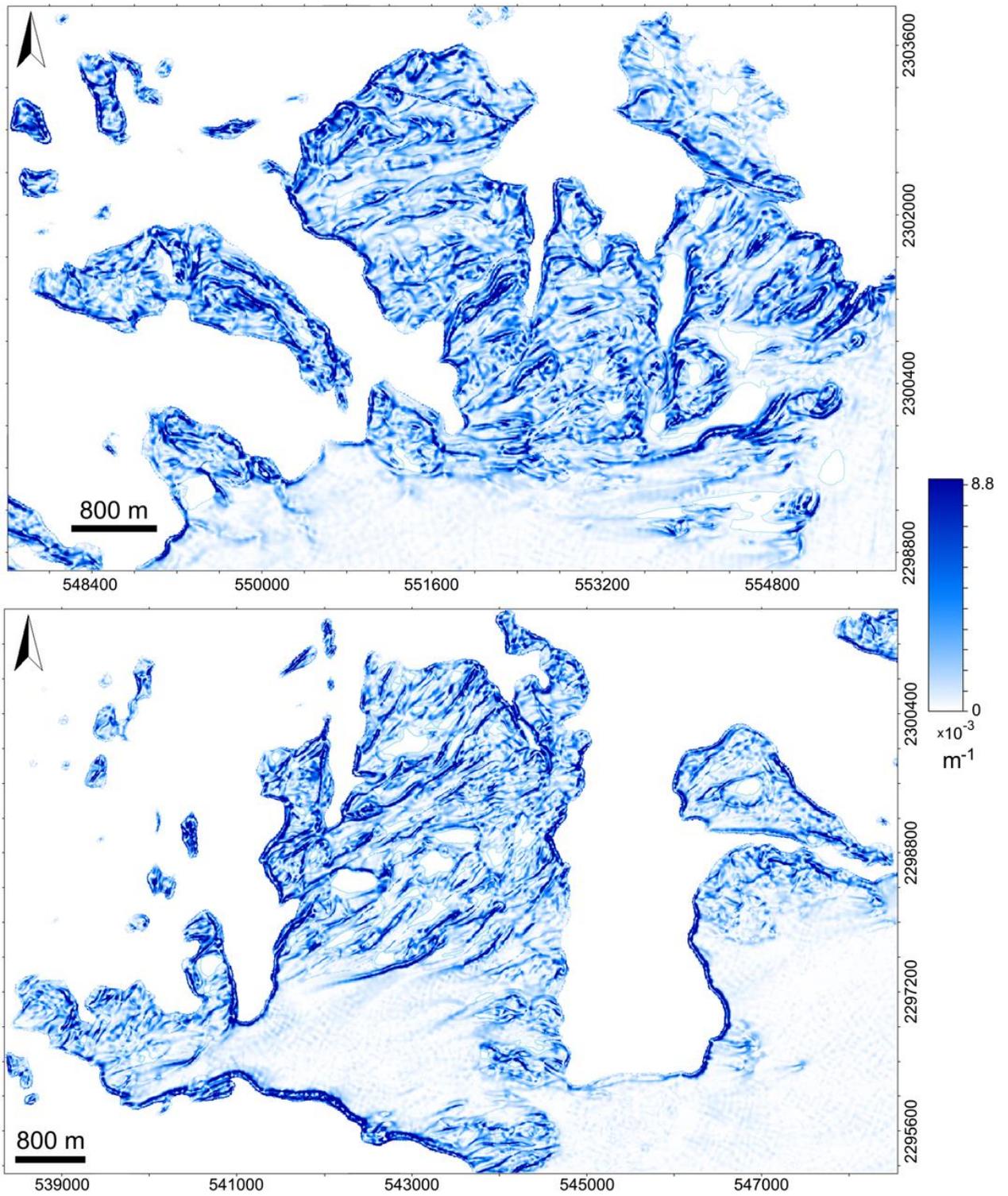

Figure 16. Larsemann Hills, vertical excess curvature.
Upper: Broknes Peninsula. Lower: Stornes Peninsula.

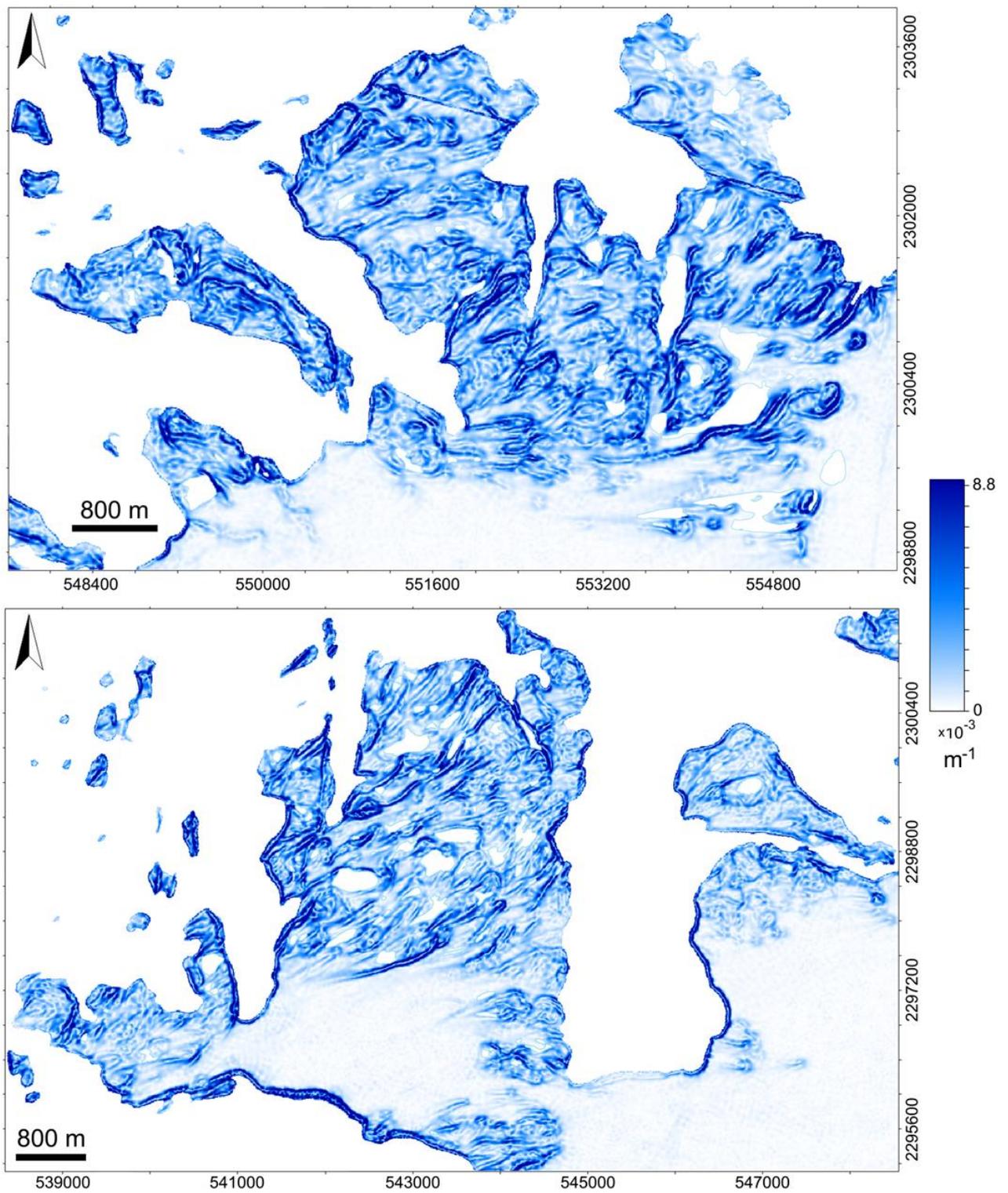

Figure 17. Larsemann Hills, unsphericity curvature.
Upper: Broknes Peninsula. Lower: Stornes Peninsula.

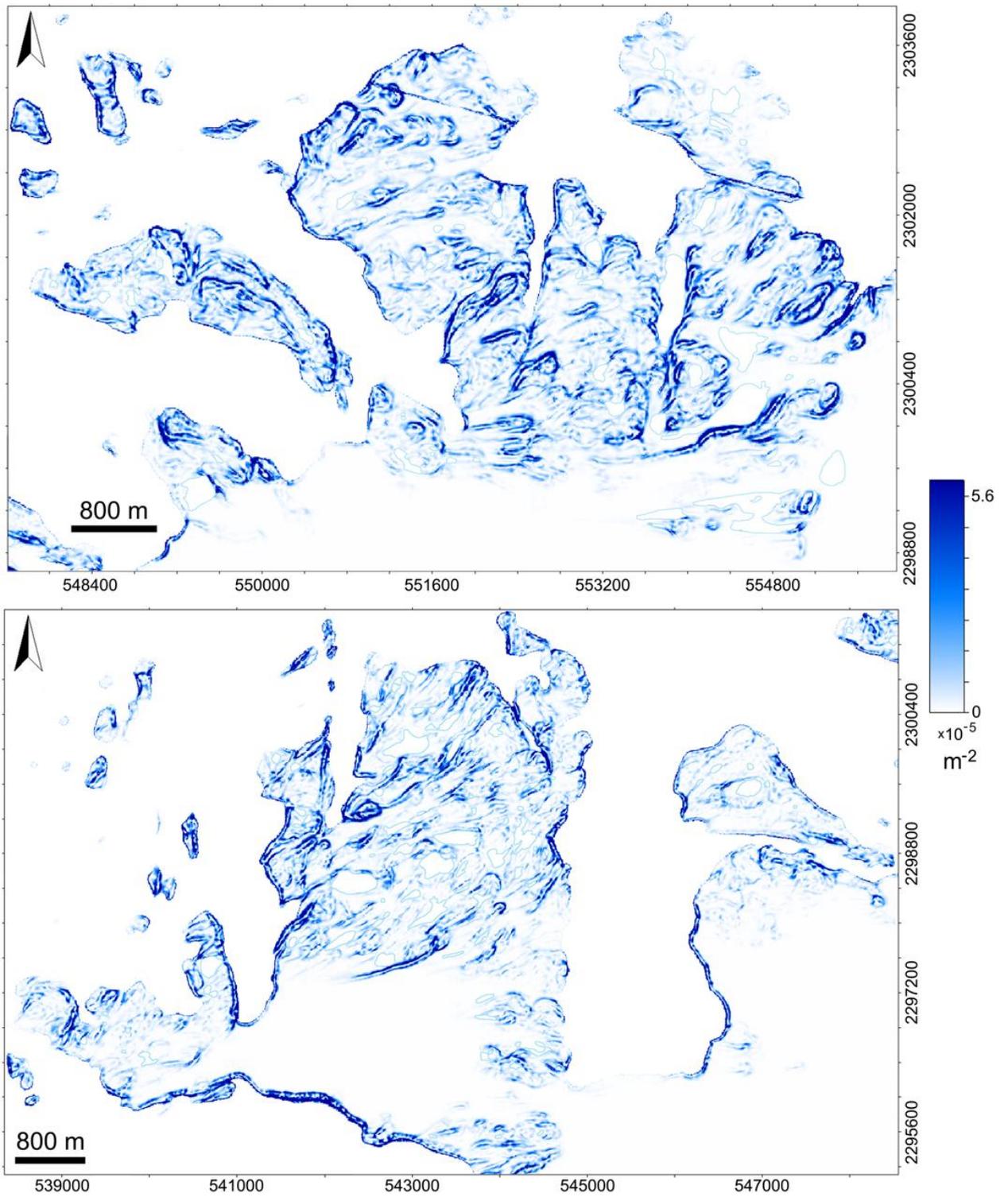

Figure 18. Larsemann Hills, ring curvature.
Upper: Broknes Peninsula. Lower: Stornes Peninsula.

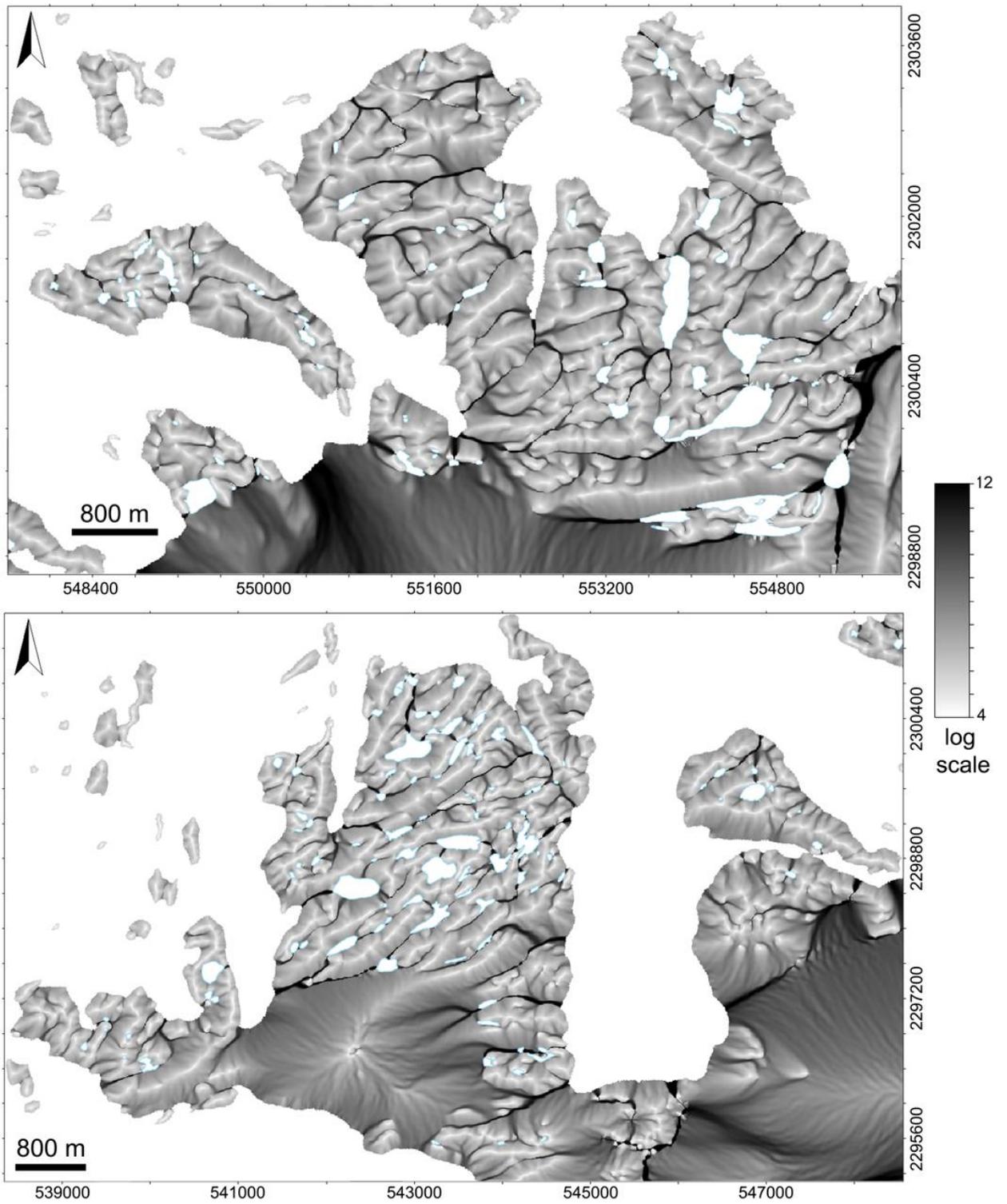

Figure 19. Larsemann Hills, catchment area.
Upper: Broknes Peninsula. Lower: Stornes Peninsula.

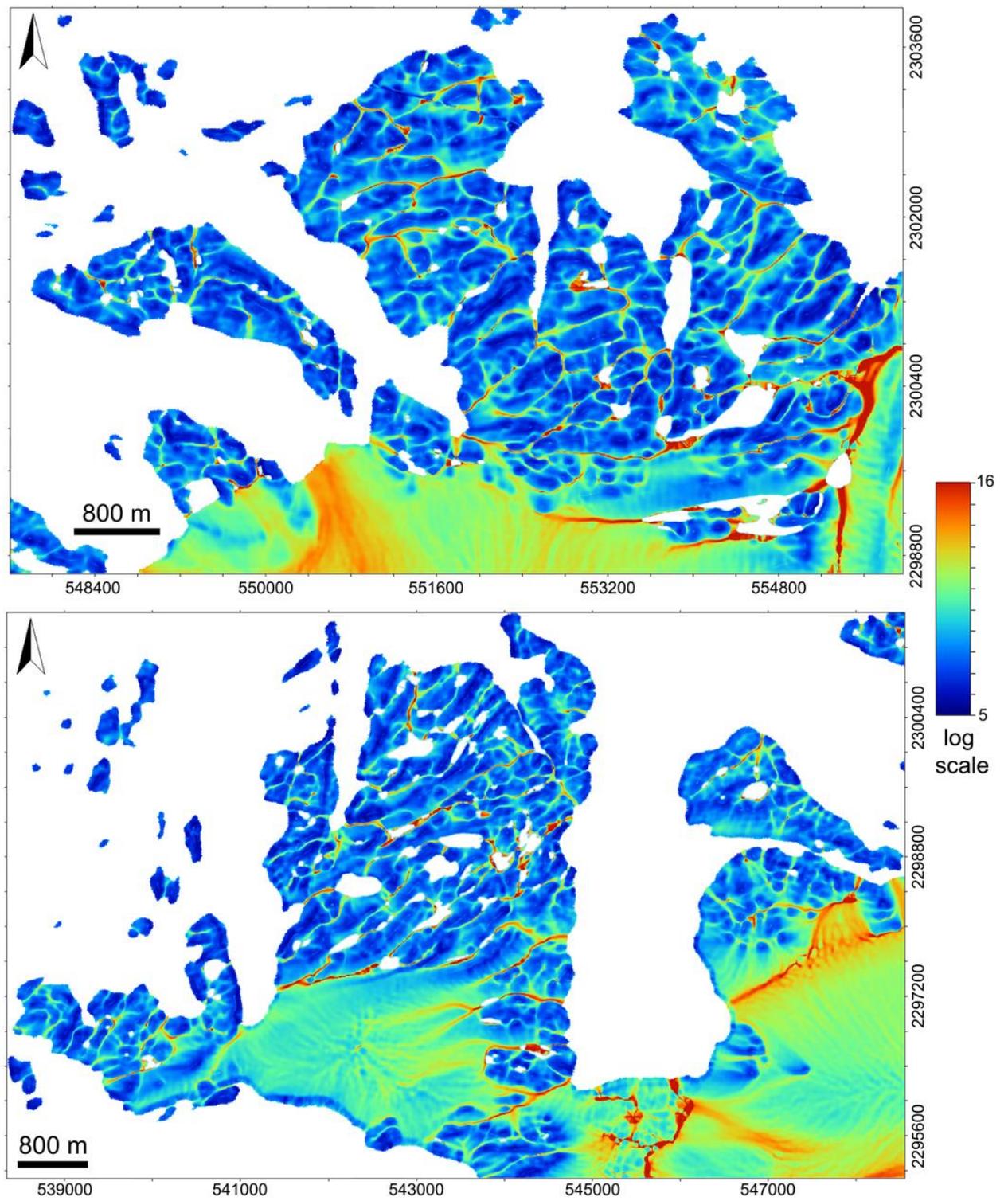

Figure 20. Larsemann Hills, topographic index.
Upper: Broknes Peninsula. Lower: Stornes Peninsula.

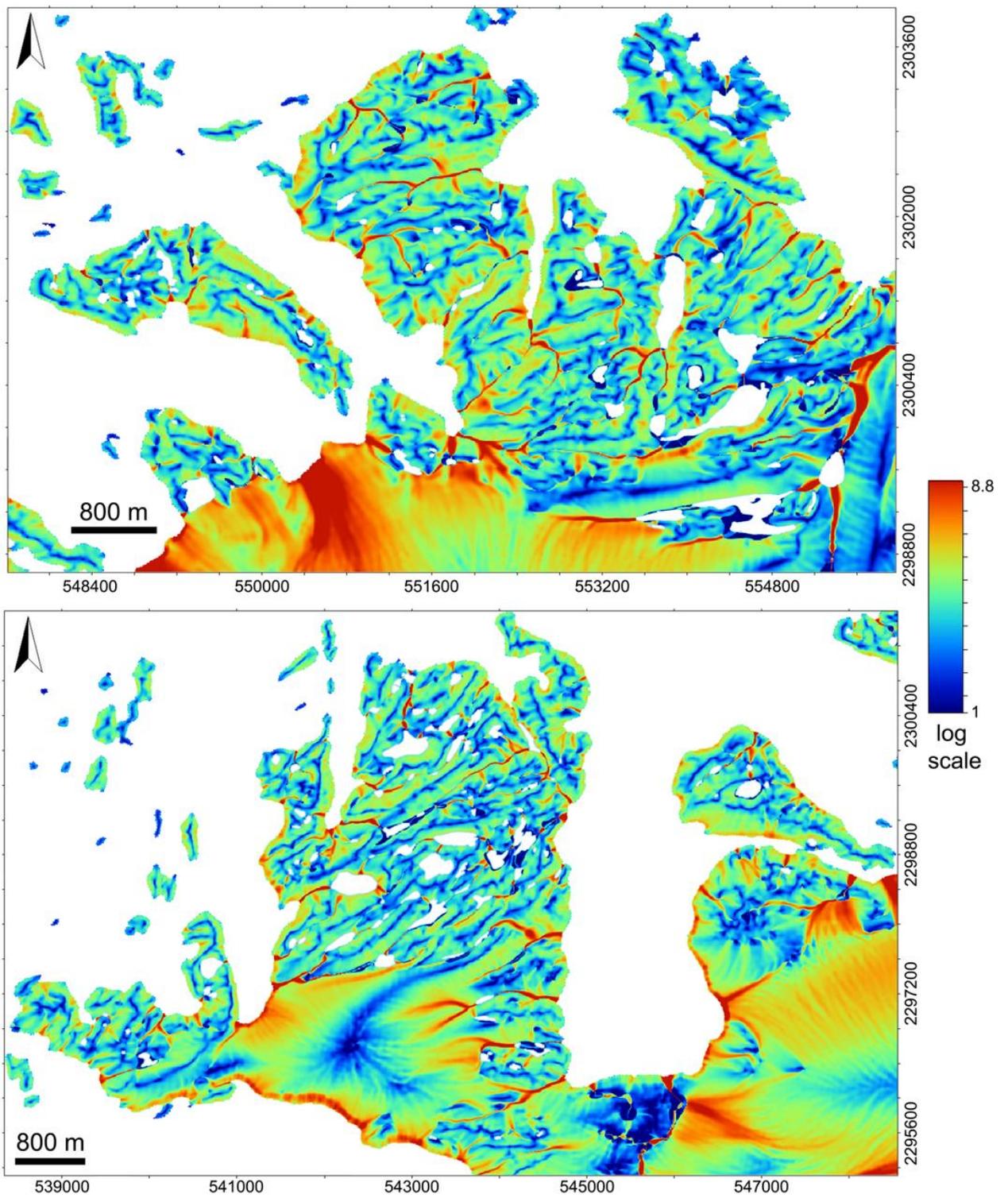

Figure 21. Larsemann Hills, stream power index.
Upper: Broknes Peninsula. Lower: Stornes Peninsula.

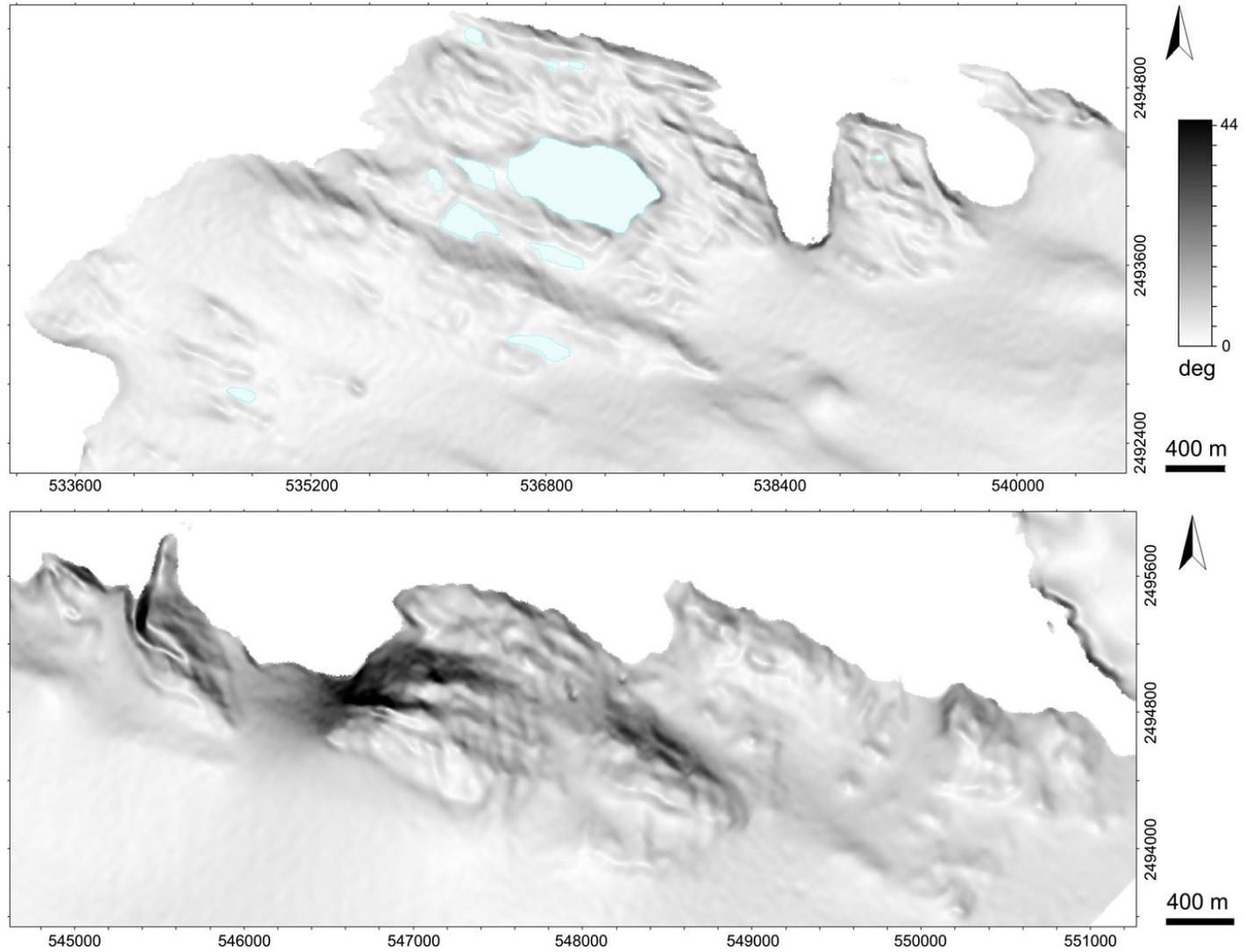

Figure 22. Thala Hills, slope.
Upper: Molodezhny Oasis. Lower: Vecherny Oasis.

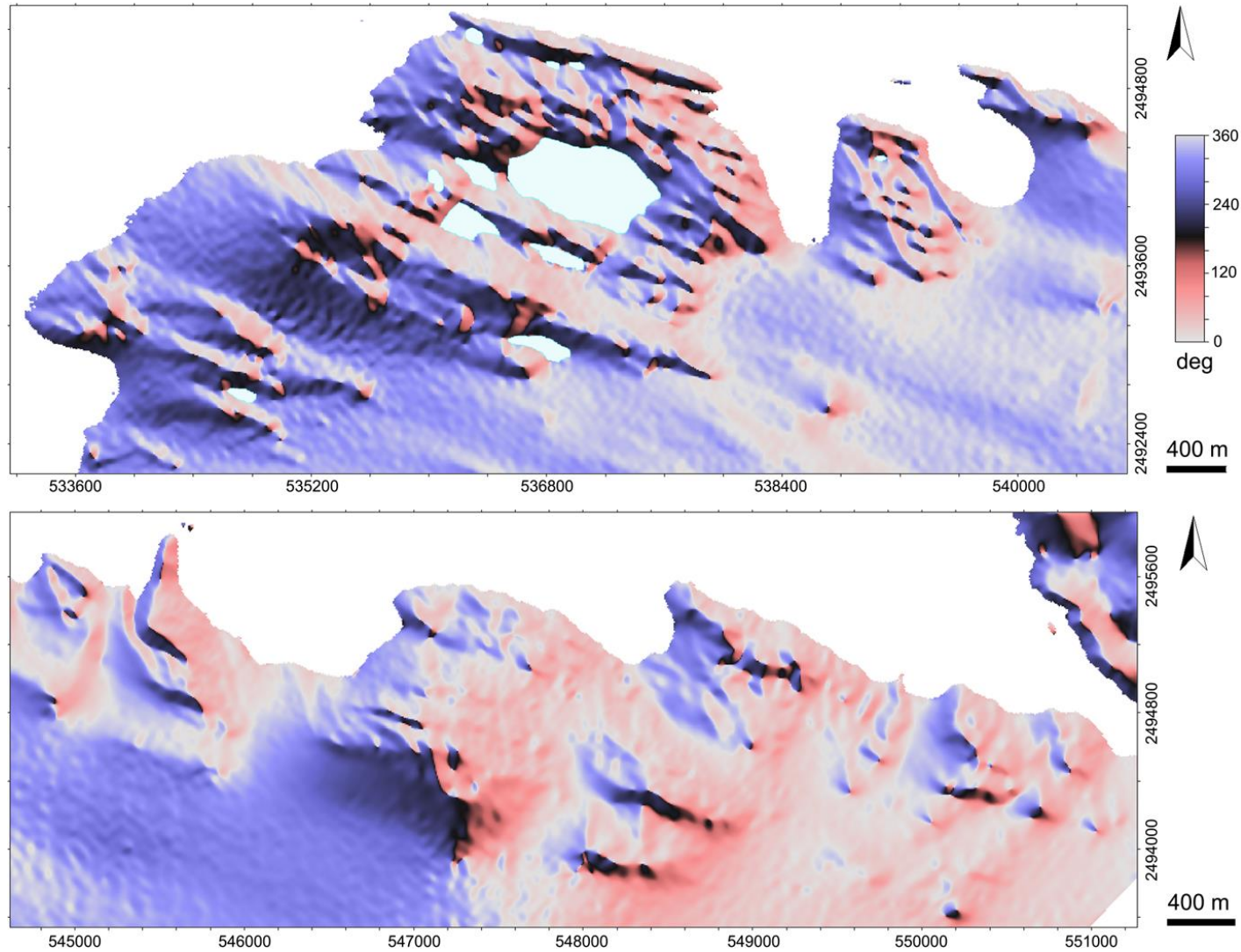

Figure 23. Thala Hills, aspect.
Upper: Molodezhny Oasis. Lower: Vecherny Oasis.

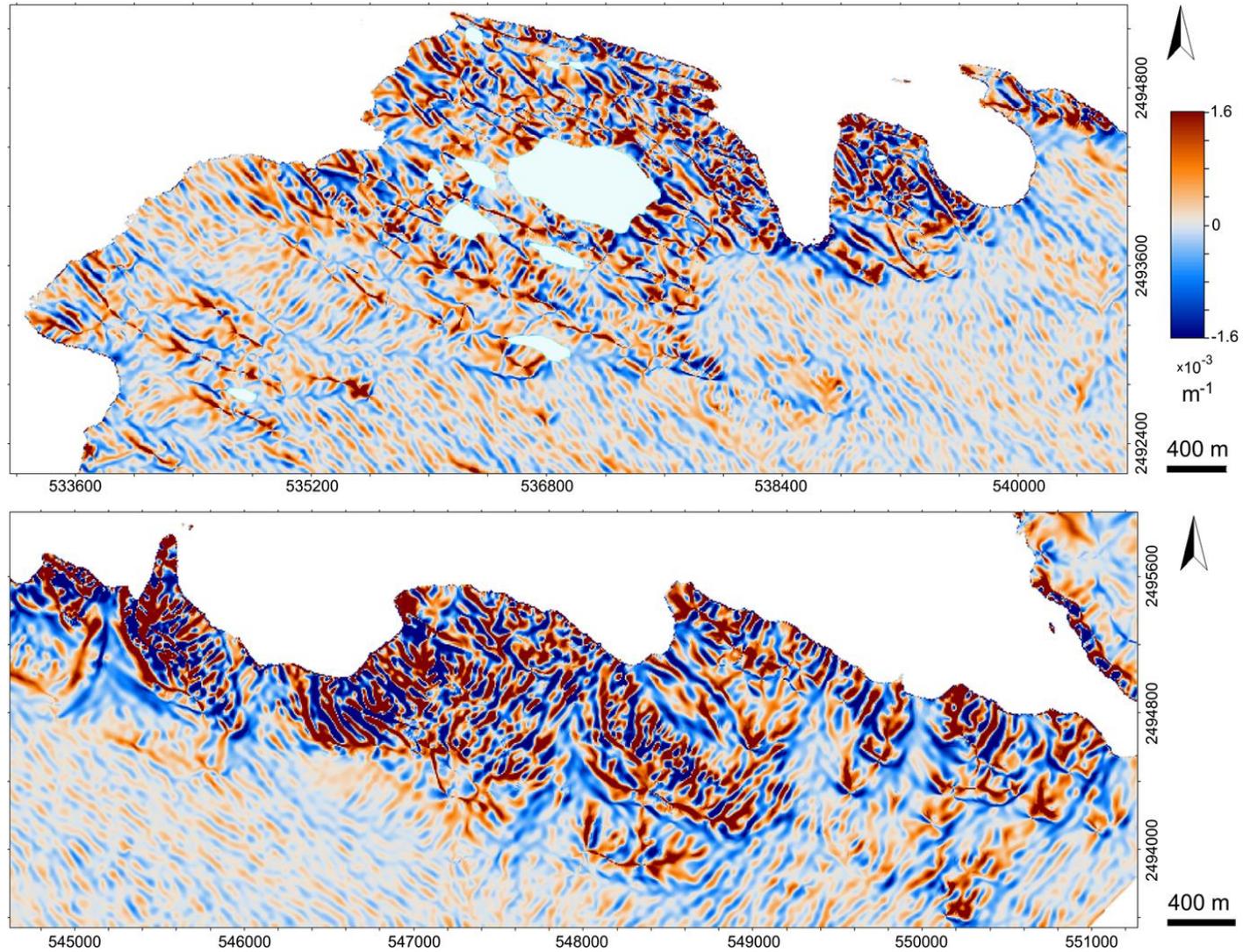

Figure 24. Thala Hills, horizontal curvature.
Upper: Molodezhny Oasis. Lower: Vecherny Oasis.

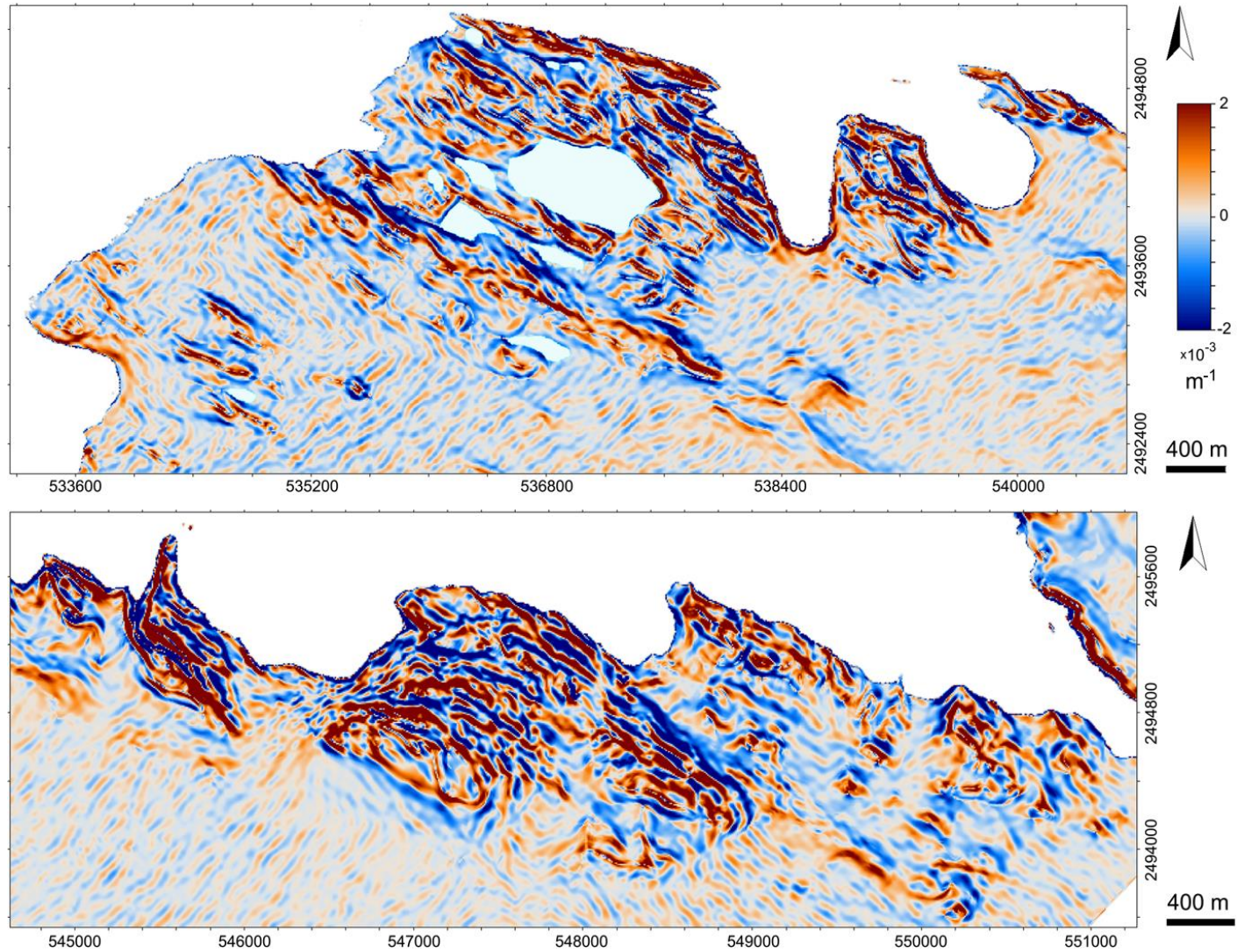

Figure 25. Thala Hills, vertical curvature.
Upper: Molodezhny Oasis. Lower: Vecherny Oasis.

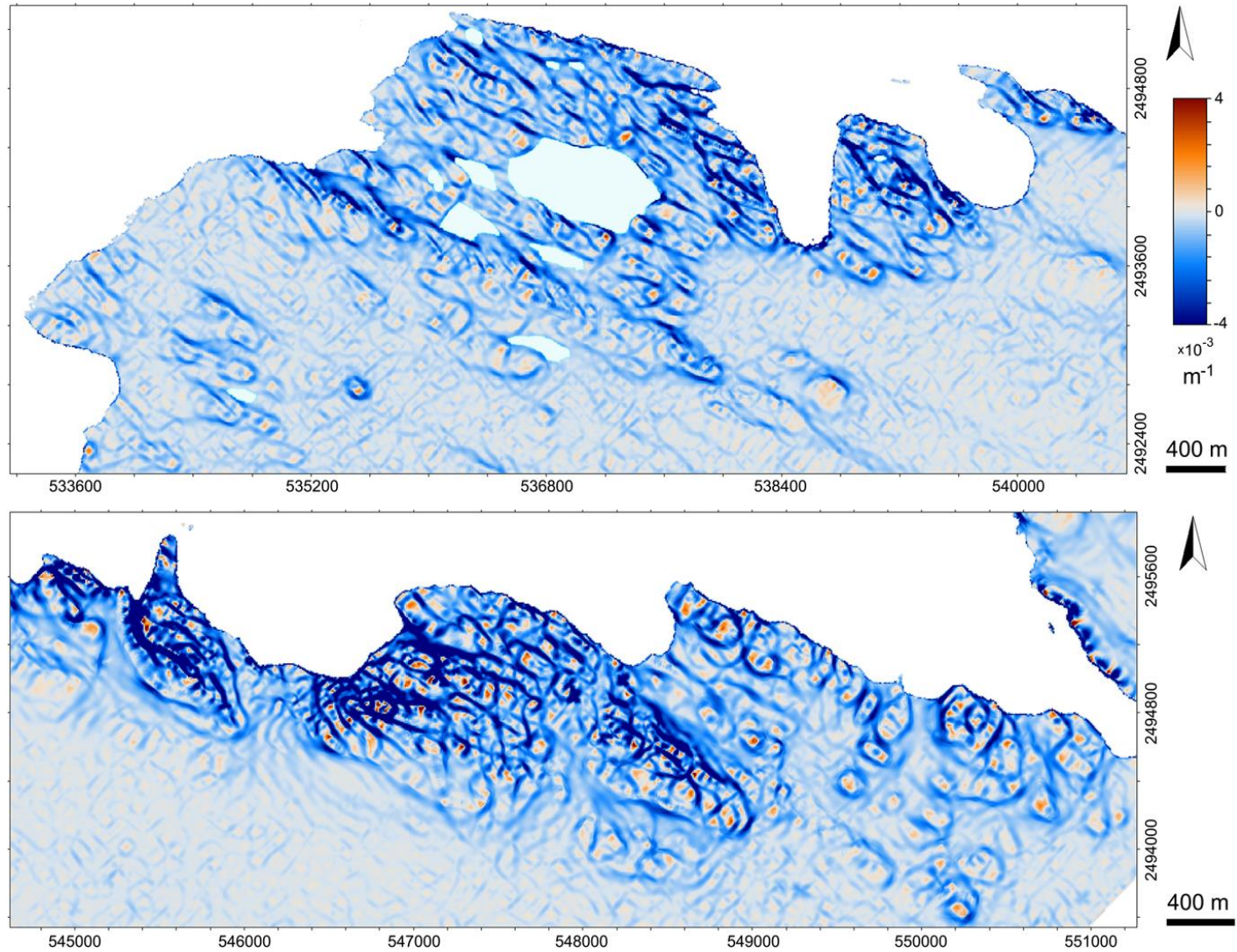

Figure 26. Thala Hills, minimal curvature.
Upper: Molodezhny Oasis. Lower: Vecherny Oasis.

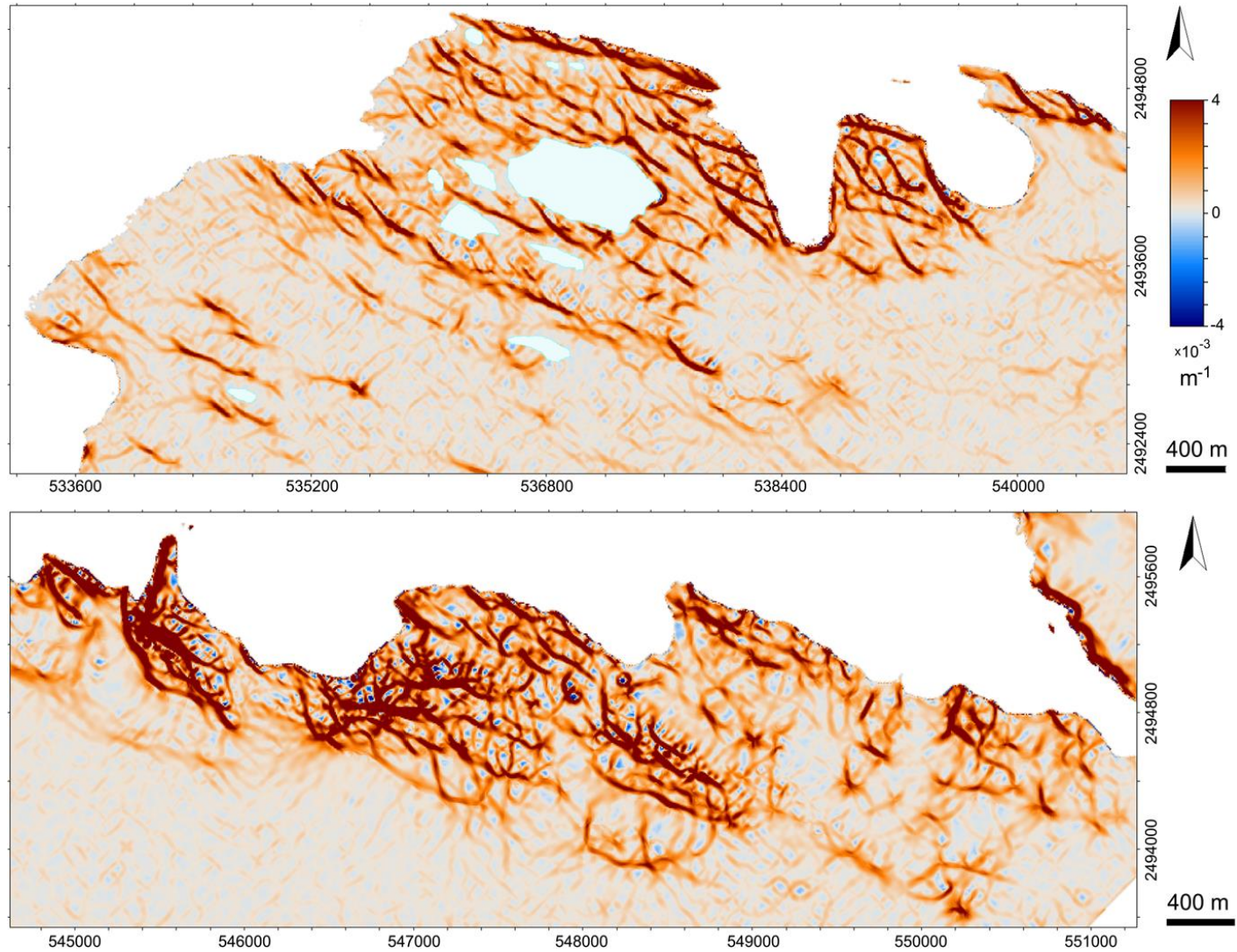

Figure 27. Thala Hills, maximal curvature.
Upper: Molodezhny Oasis. Lower: Vecherny Oasis.

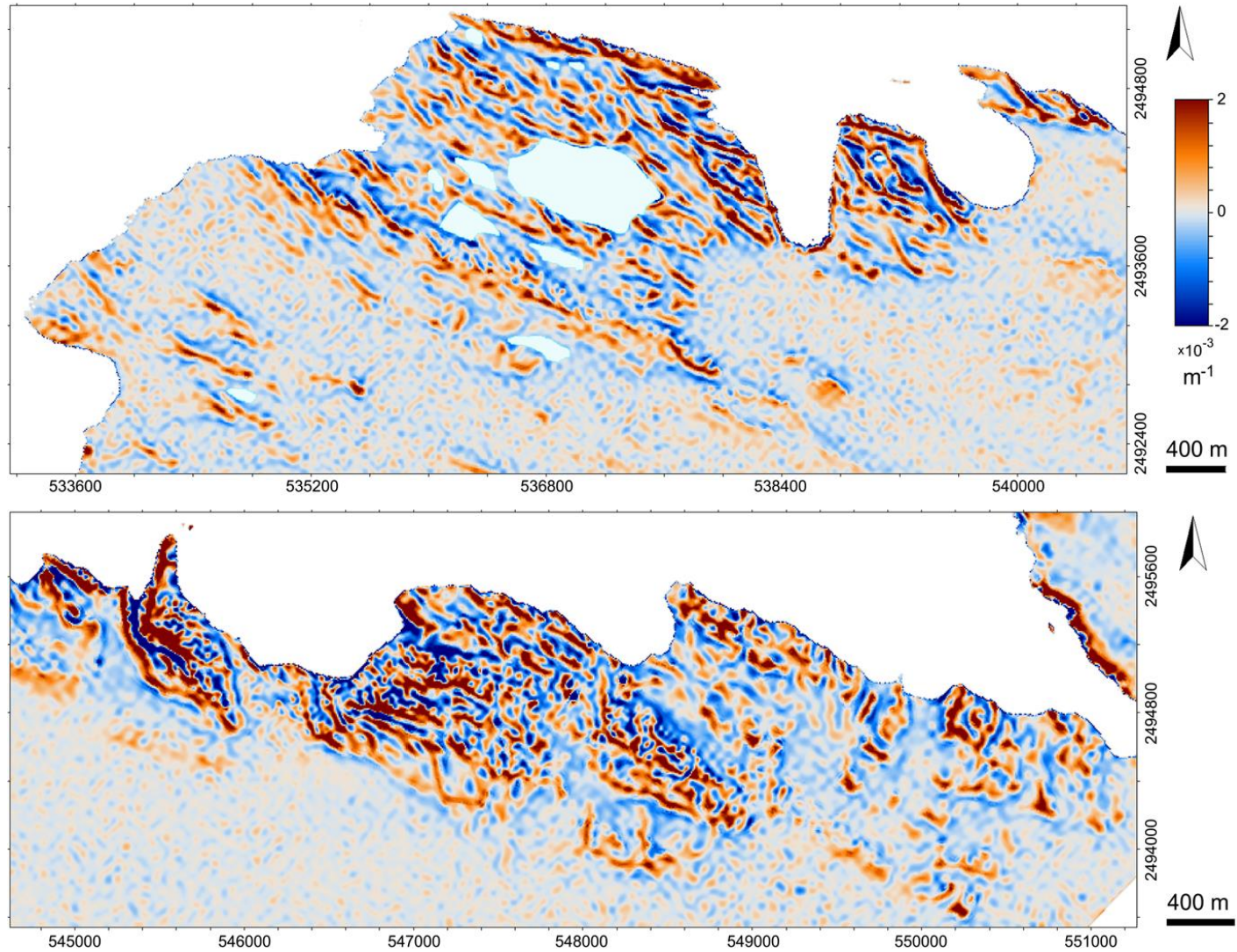

Figure 28. Thala Hills, mean curvature.
Upper: Molodezhny Oasis. Lower: Vecherny Oasis.

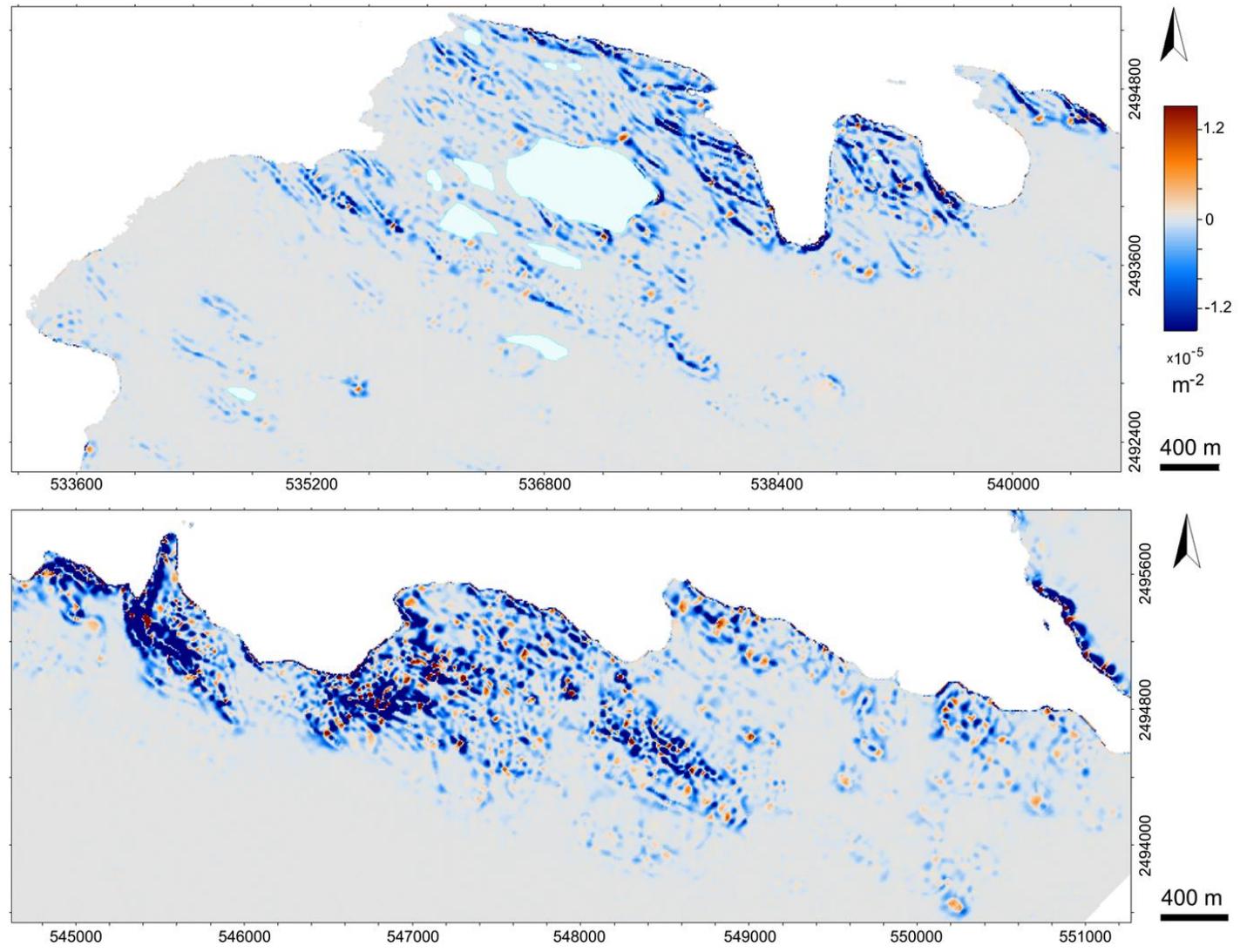

Figure 29. Thala Hills, Gaussian curvature.
Upper: Molodezhny Oasis. Lower: Vecherny Oasis.

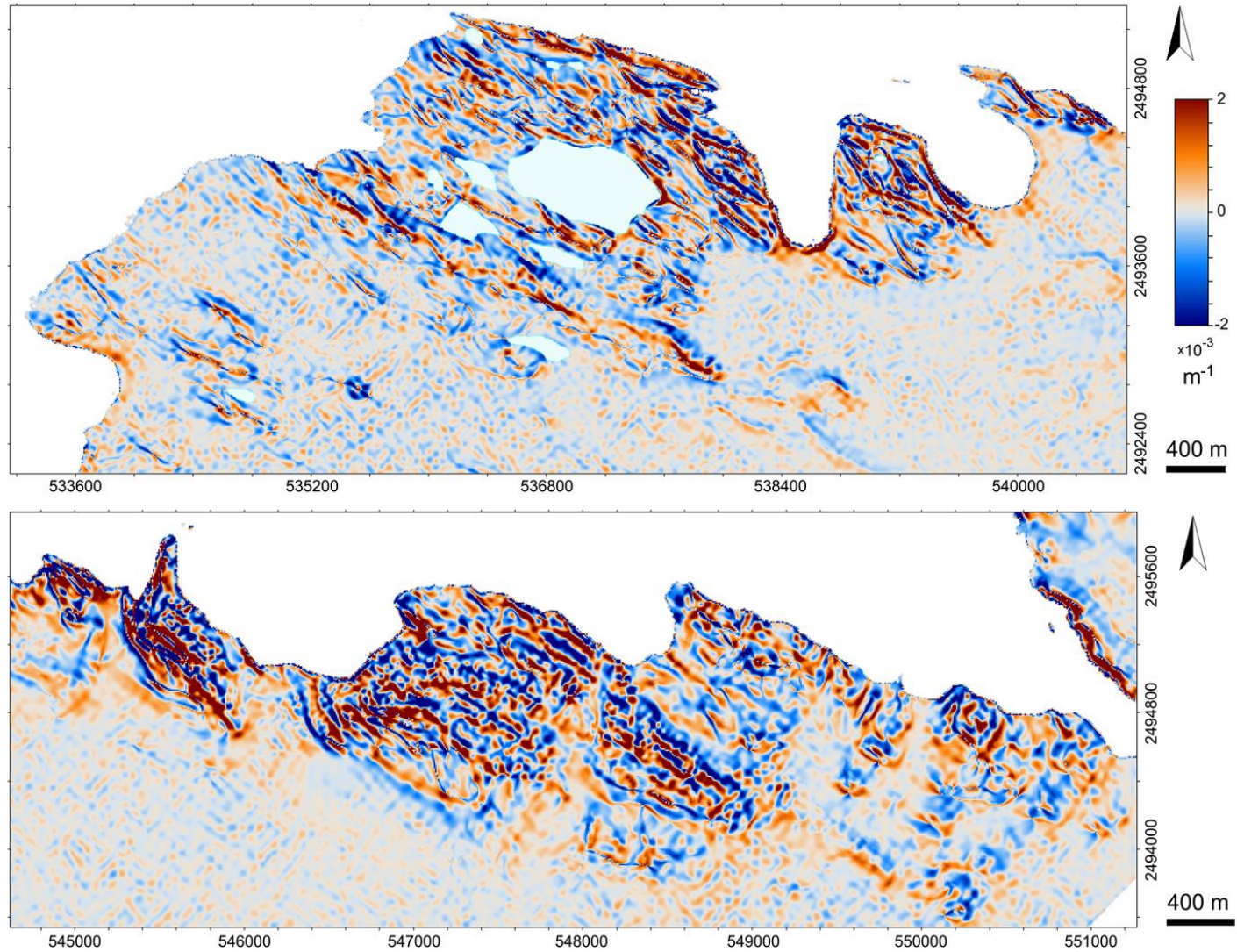

Figure 30. Thala Hills, difference curvature.
Upper: Molodezhny Oasis. Lower: Vecherny Oasis.

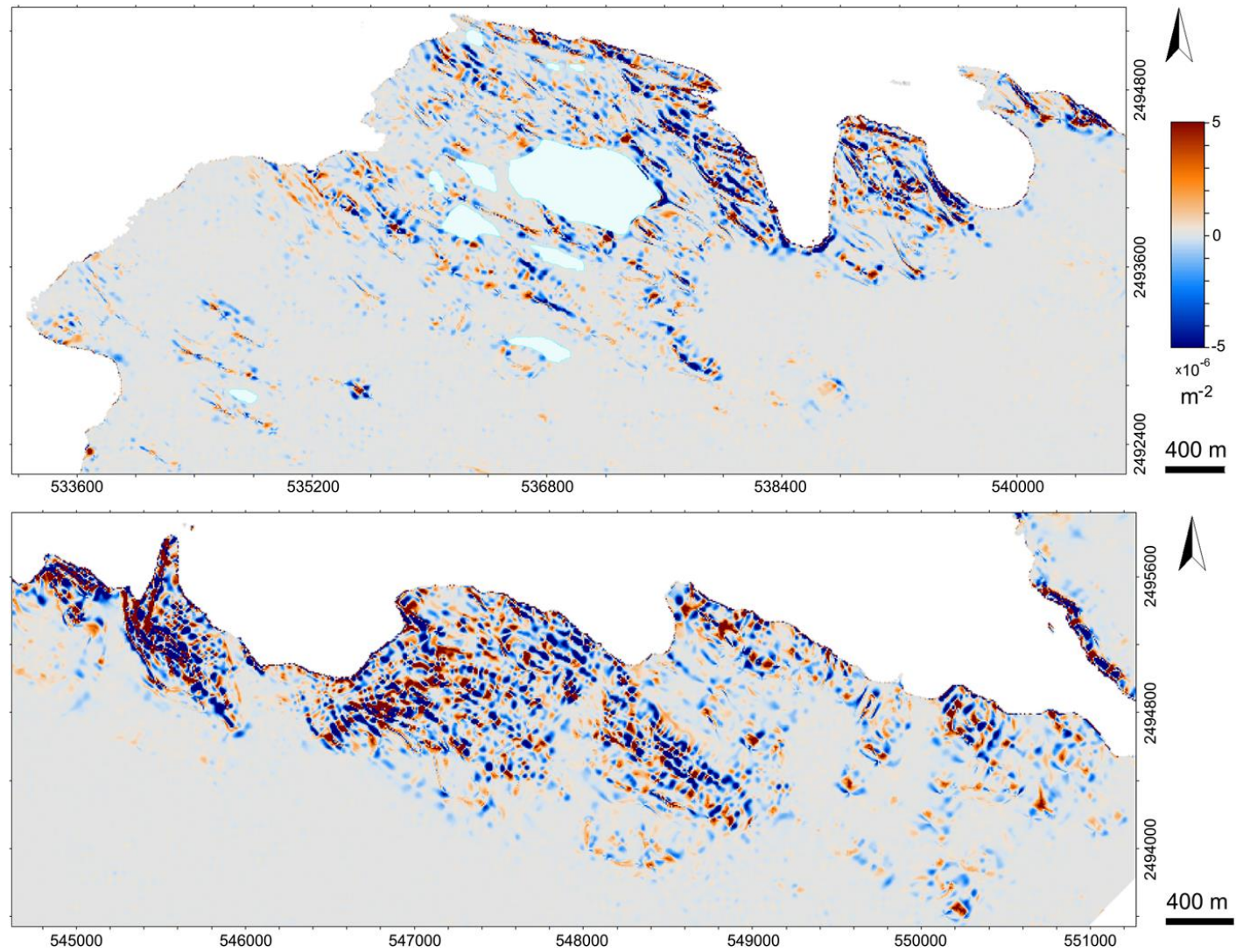

Figure 31. Thala Hills, accumulation curvature.
Upper: Molodezhny Oasis. Lower: Vecherny Oasis.

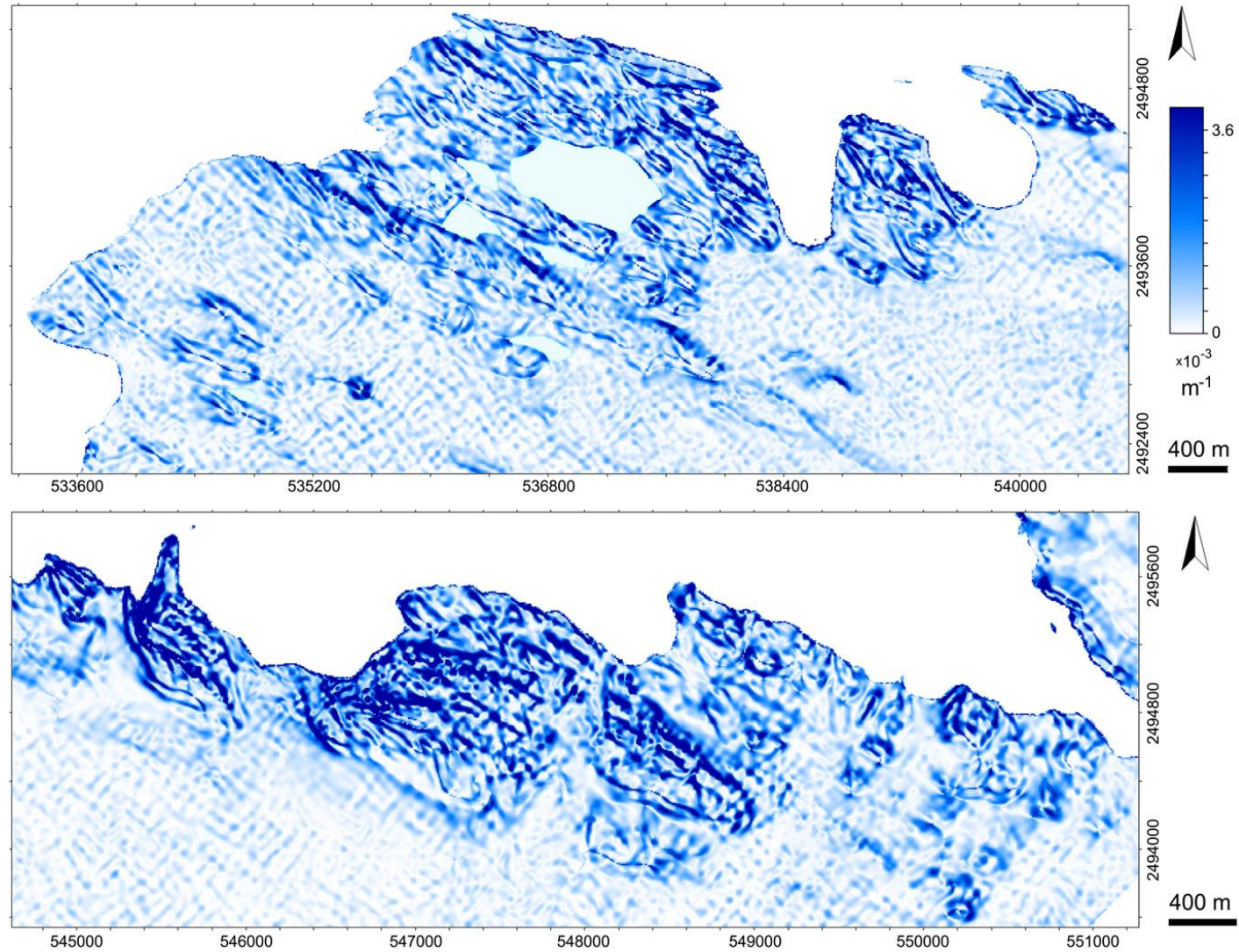

Figure 32. Thala Hills, horizontal excess curvature.
Upper: Molodezhny Oasis. Lower: Vecherny Oasis.

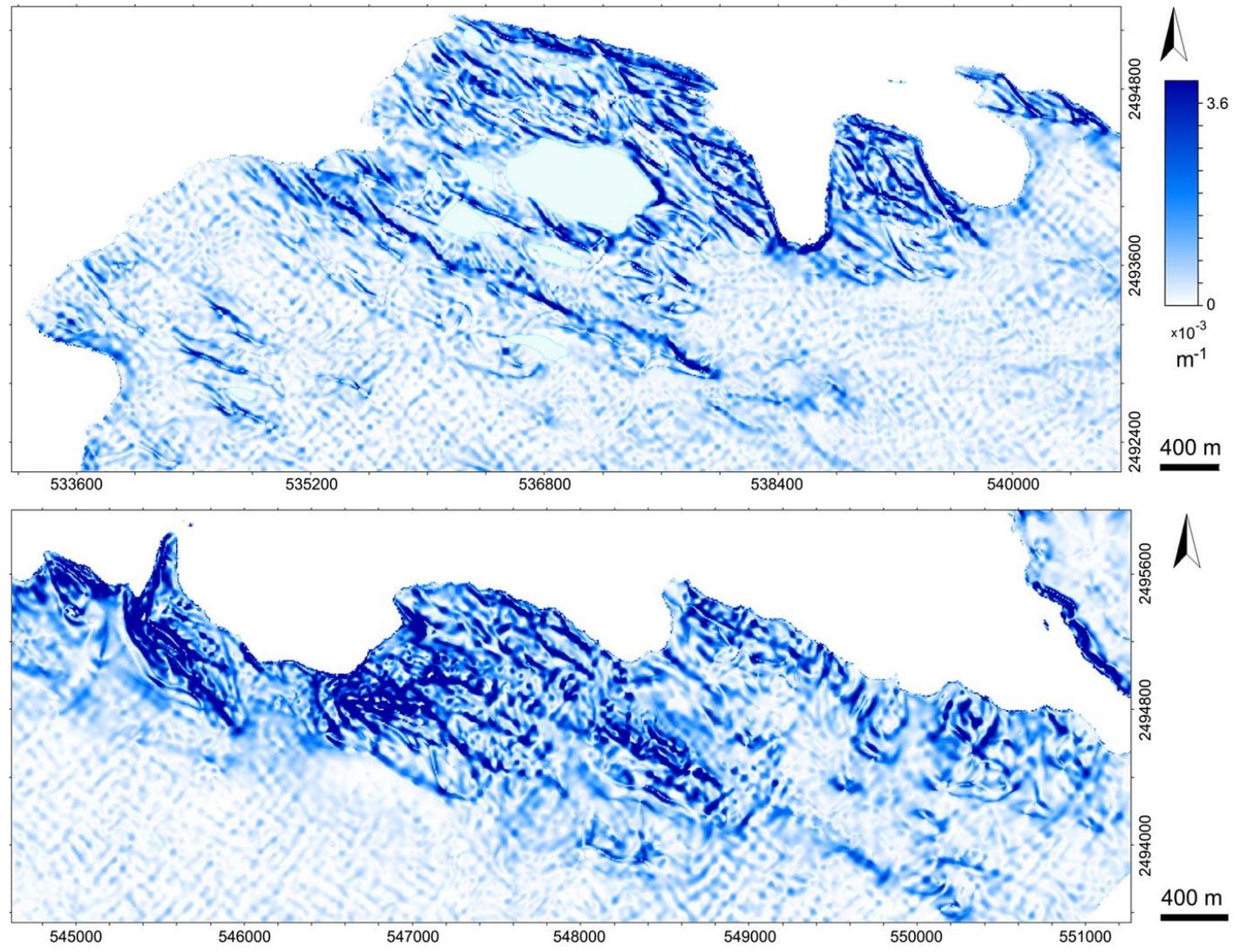

Figure 33. Thala Hills, vertical excess curvature.
Upper: Molodezhny Oasis. Lower: Vecherny Oasis.

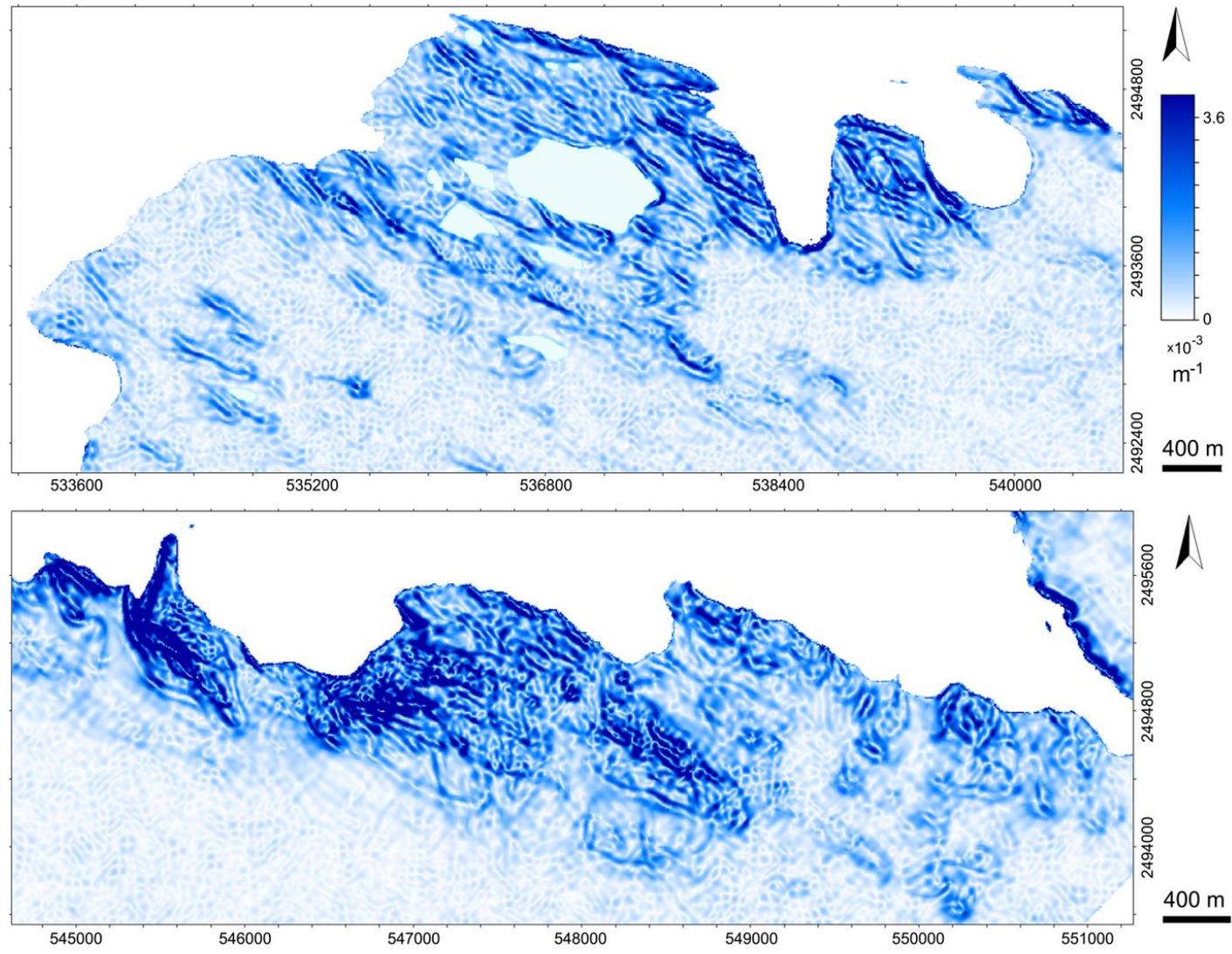

Figure 34. Thala Hills, unsphericity curvature.
Upper: Molodezhny Oasis. Lower: Vecherny Oasis.

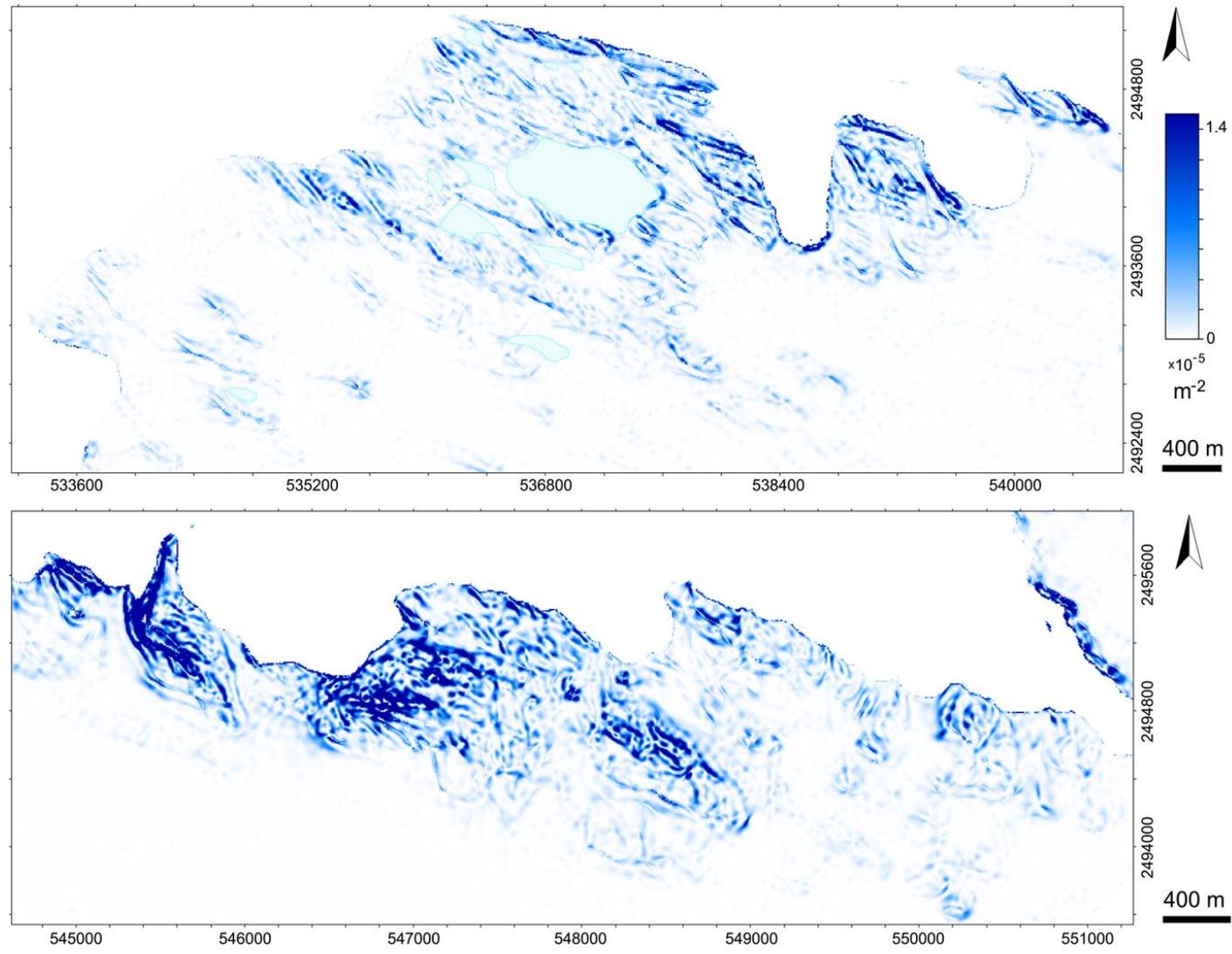

Figure 35. Thala Hills, ring curvature.
Upper: Molodezhny Oasis. Lower: Vecherny Oasis.

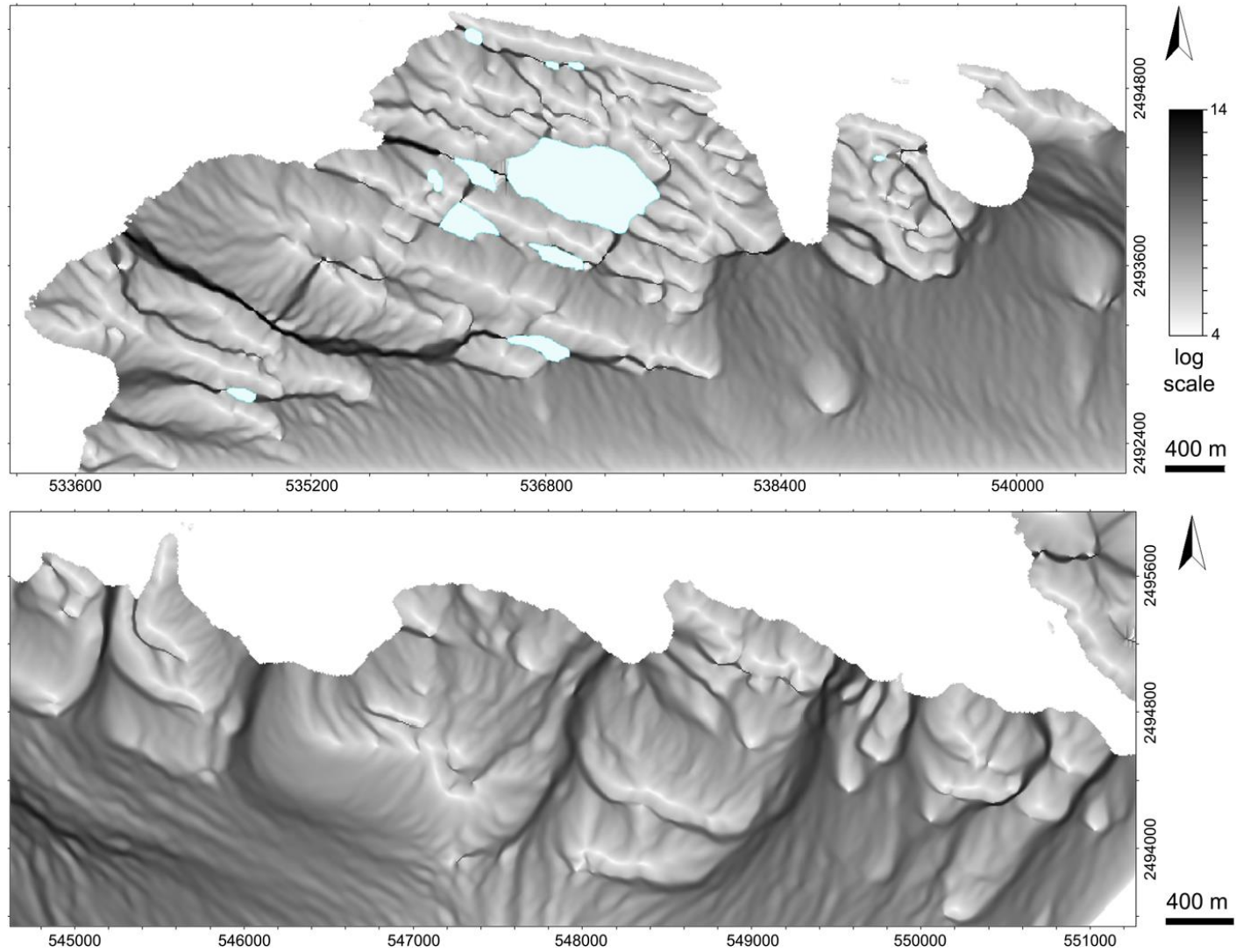

Figure 36. Thala Hills, catchment area.
Upper: Molodezhny Oasis. Lower: Vecherny Oasis.

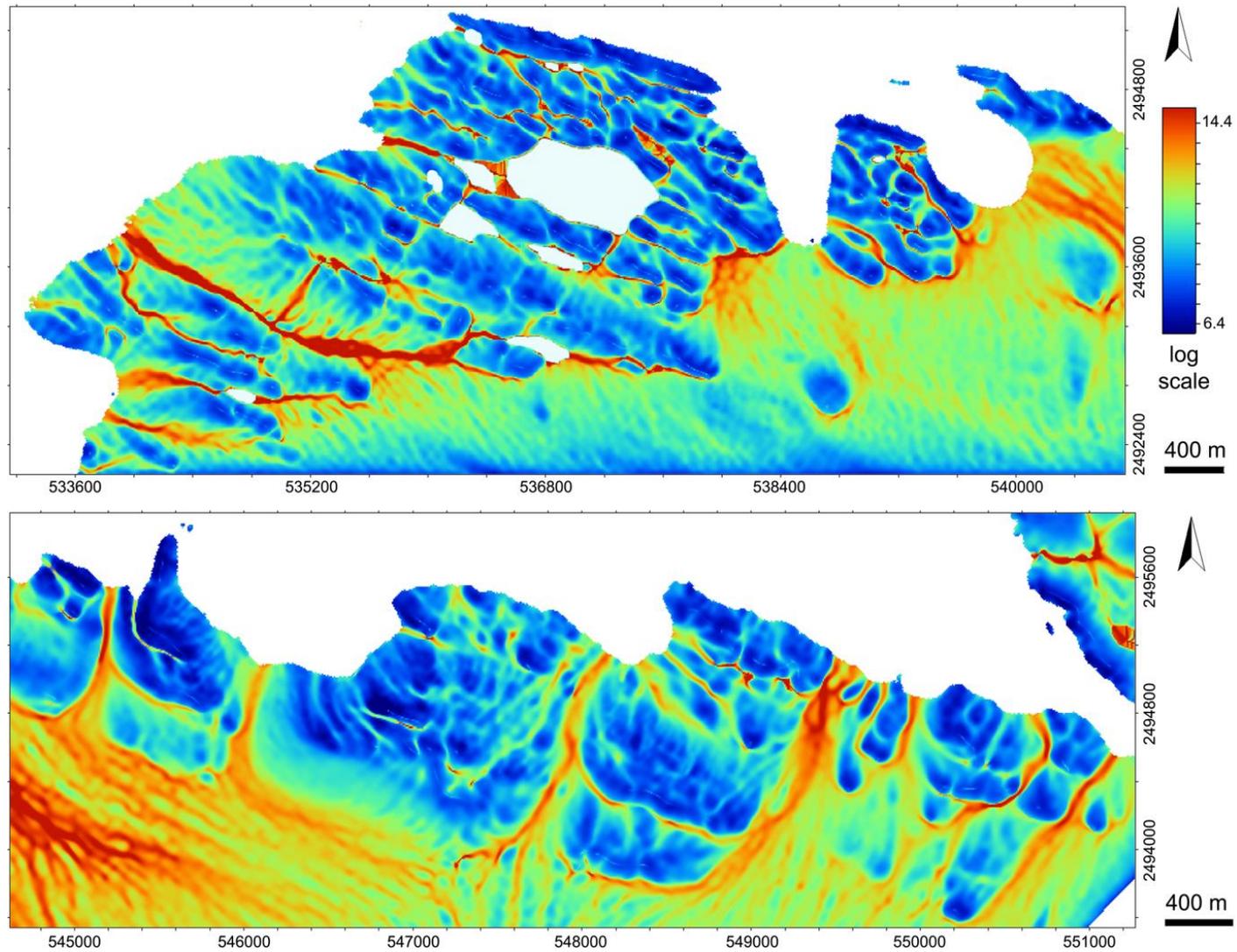

Figure 37. Thala Hills, topographic index.
Upper: Molodezhny Oasis. Lower: Vecherny Oasis.

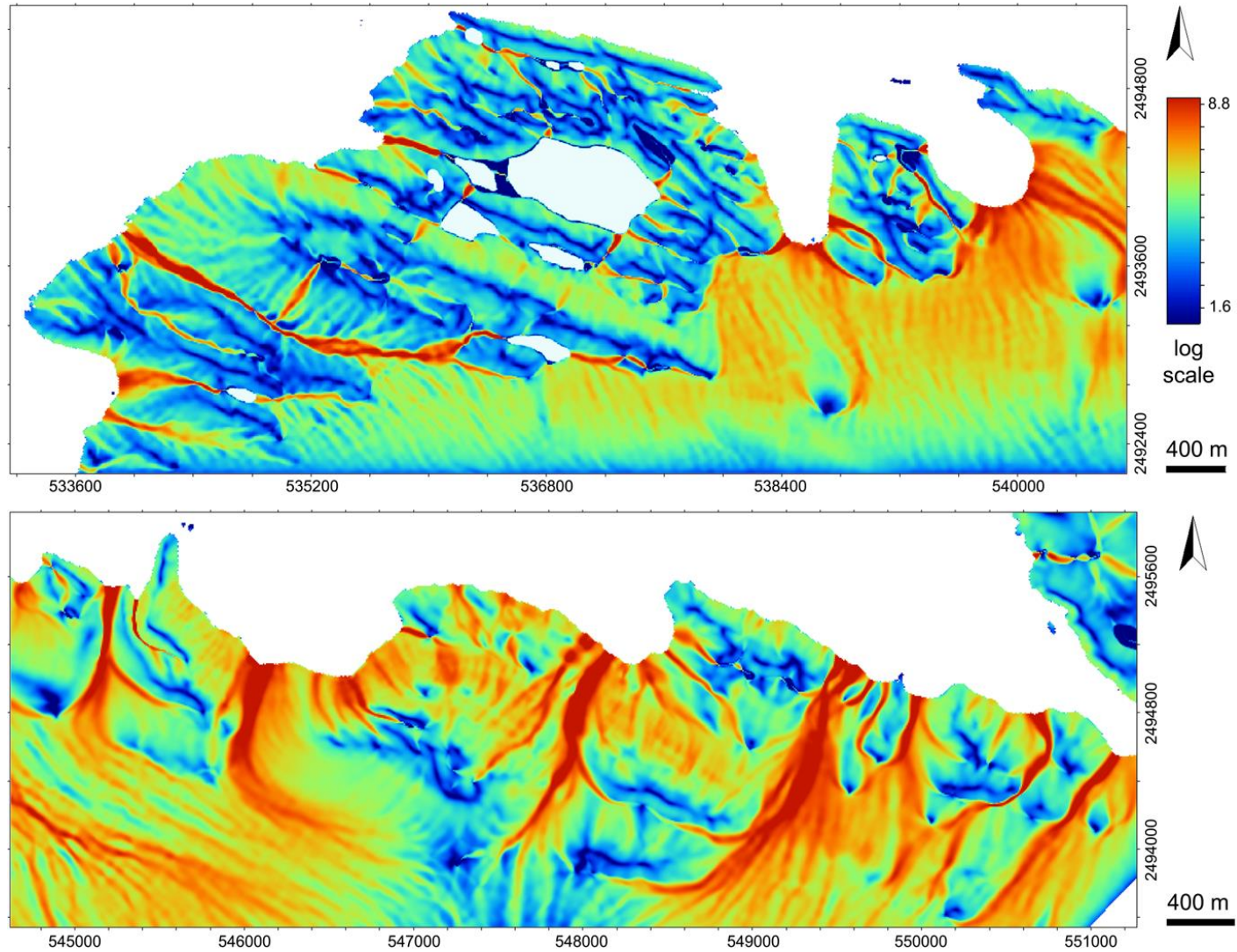

Figure 38. Thala Hills, stream power index.
Upper: Molodezhny Oasis. Lower: Vecherny Oasis.

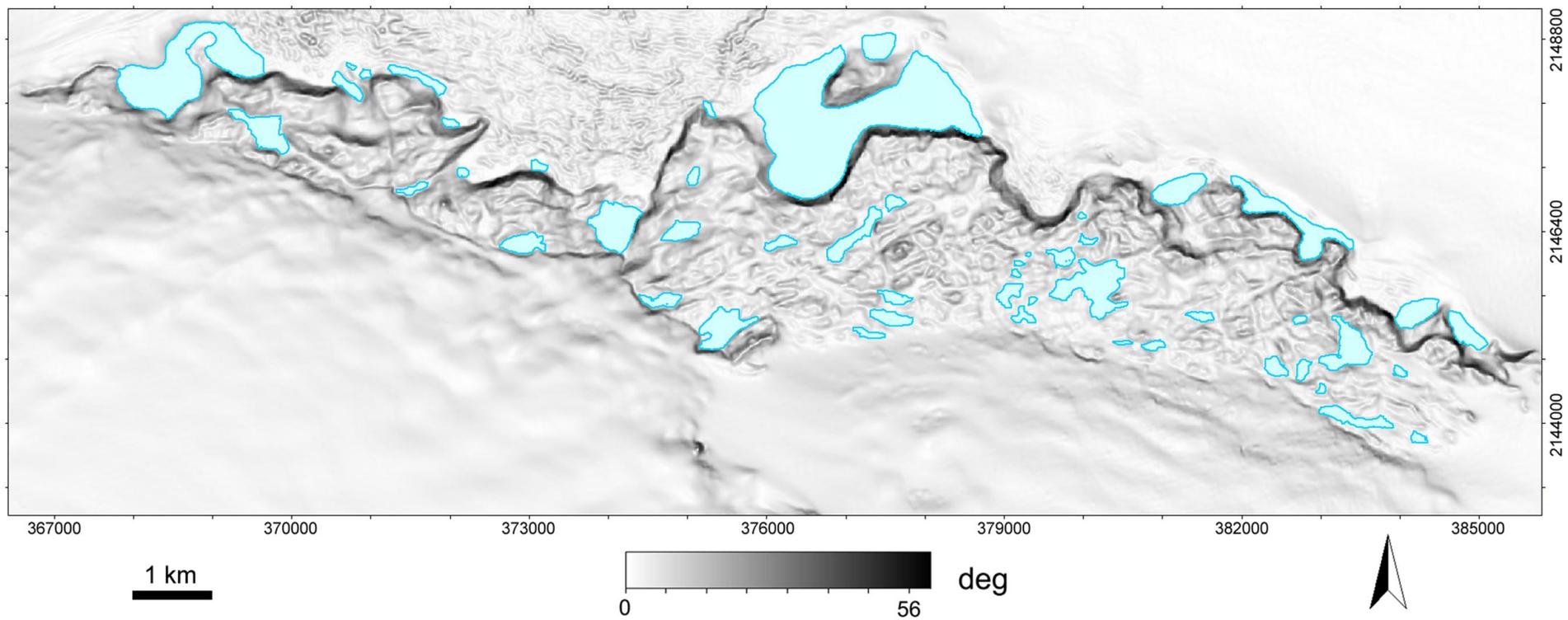

Figure 39. Schirmacher Oasis, slope.

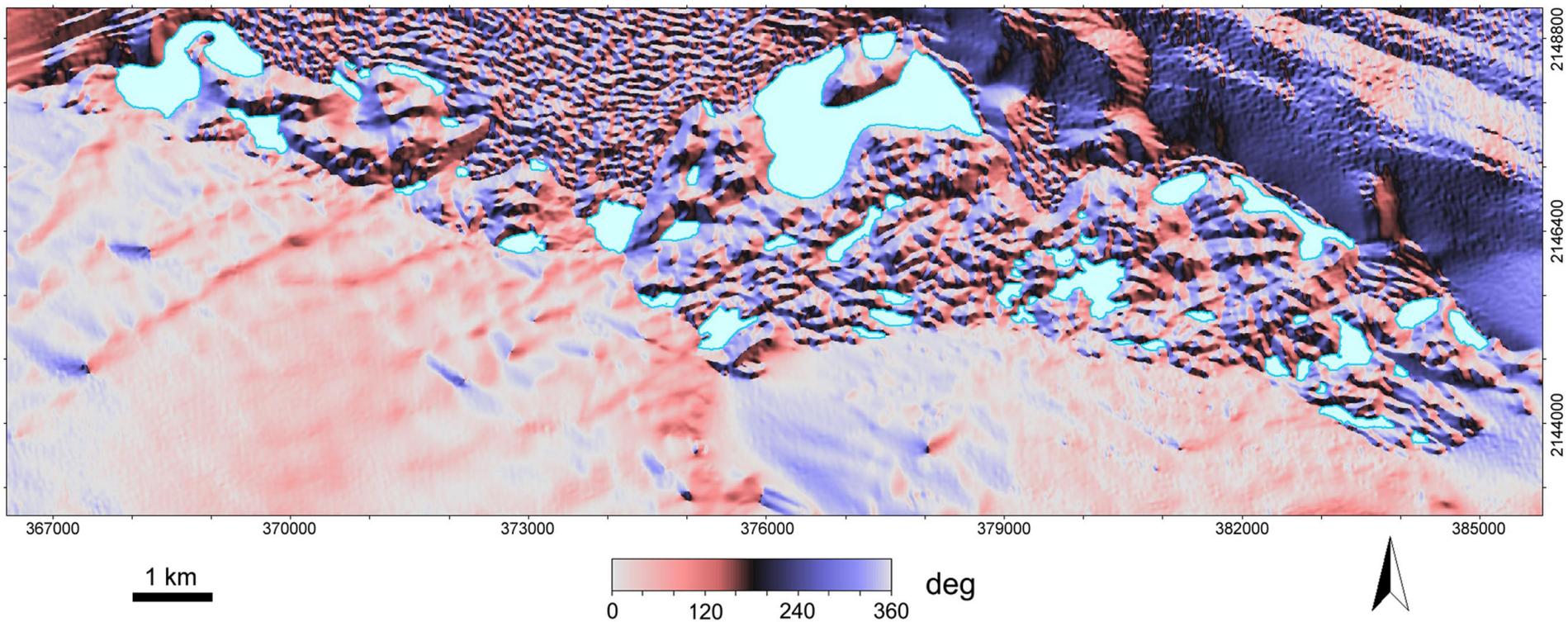

Figure 40. Schirmacher Oasis, aspect.

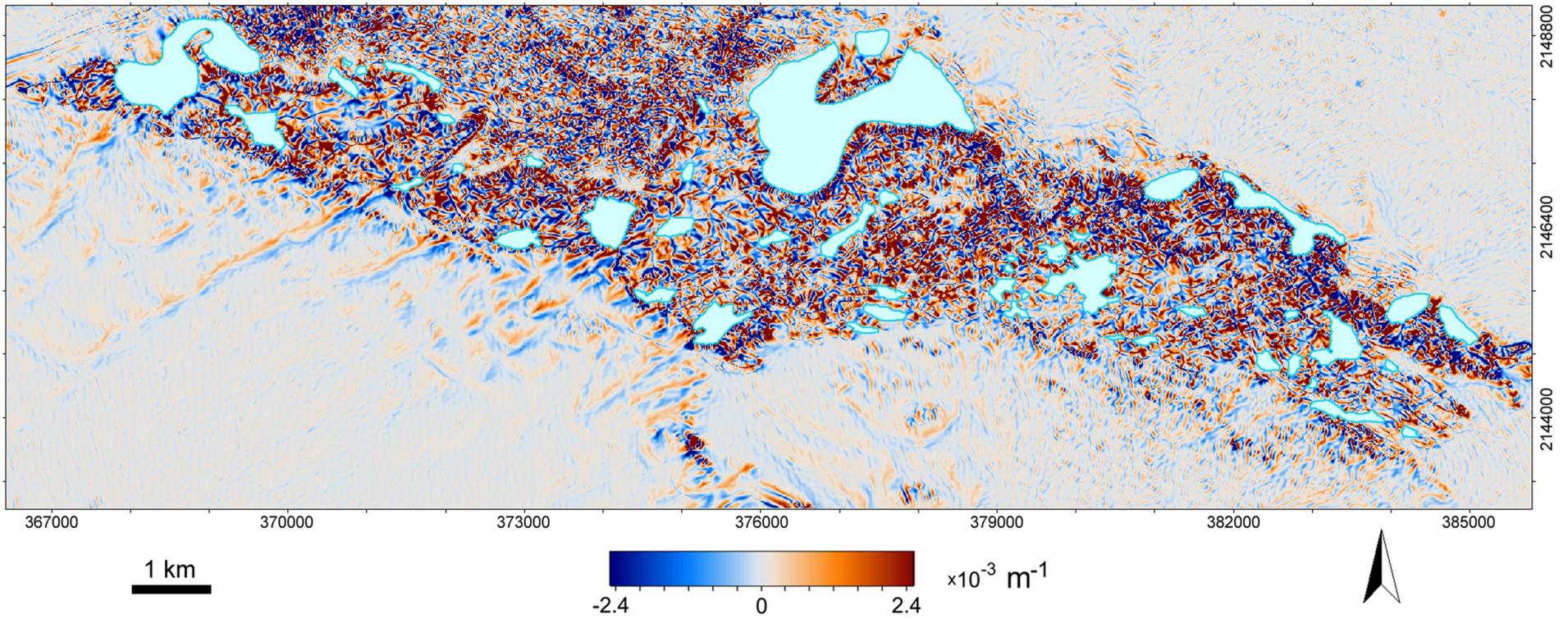

Figure 41. Schirmacher Oasis, horizontal curvature.

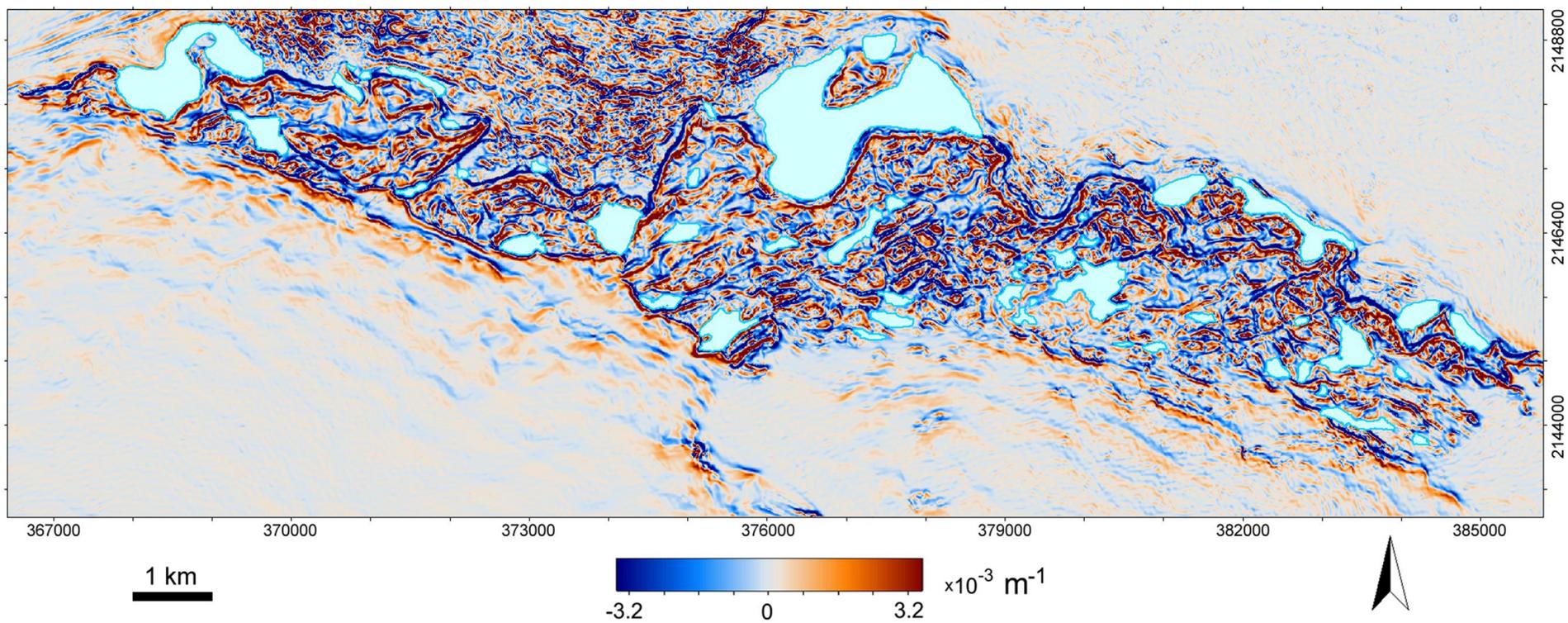

Figure 42. Schirmacher Oasis, vertical curvature.

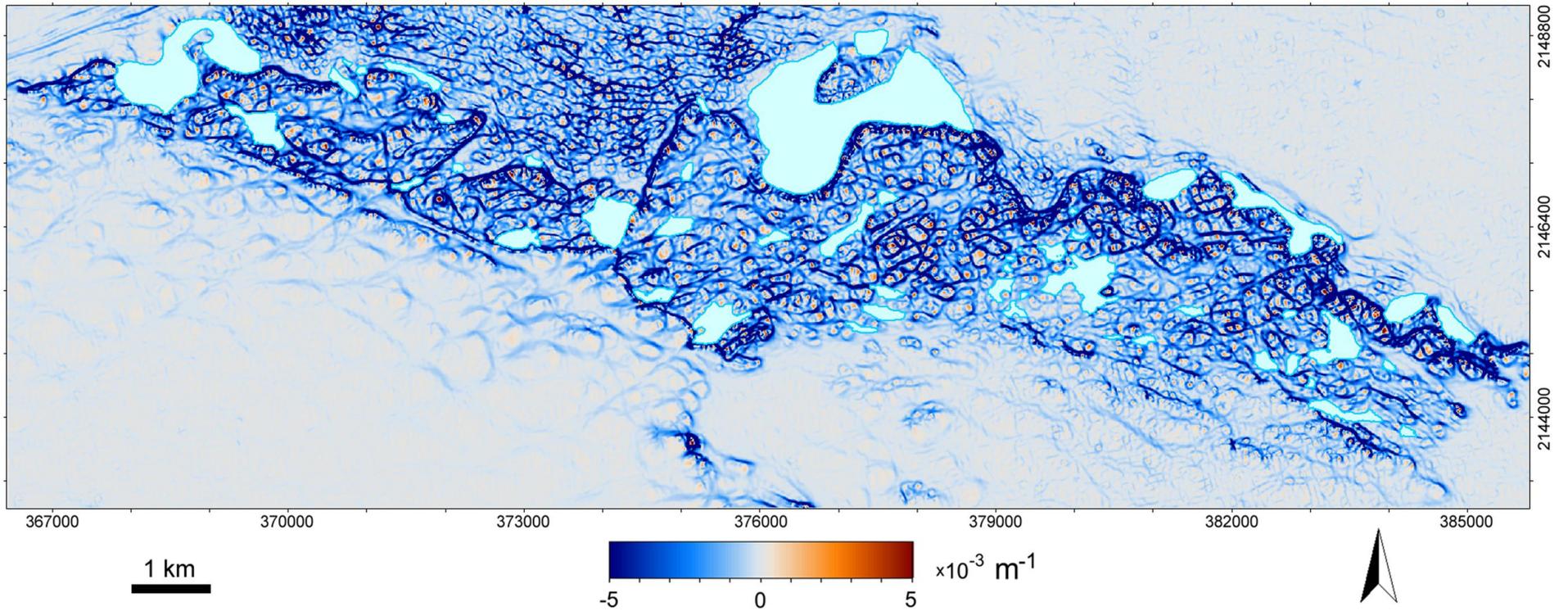

Figure 43. Schirmacher Oasis, minimal curvature.

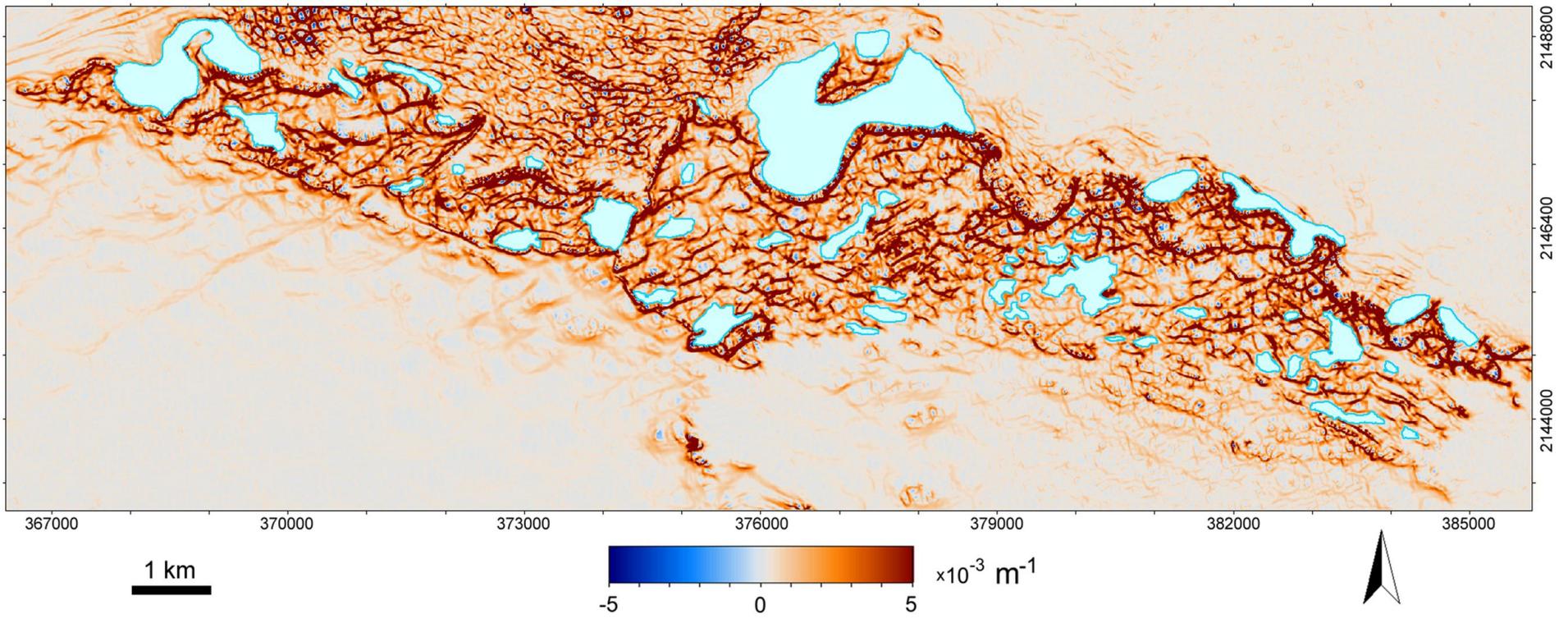

Figure 44. Schirmacher Oasis, maximal curvature.

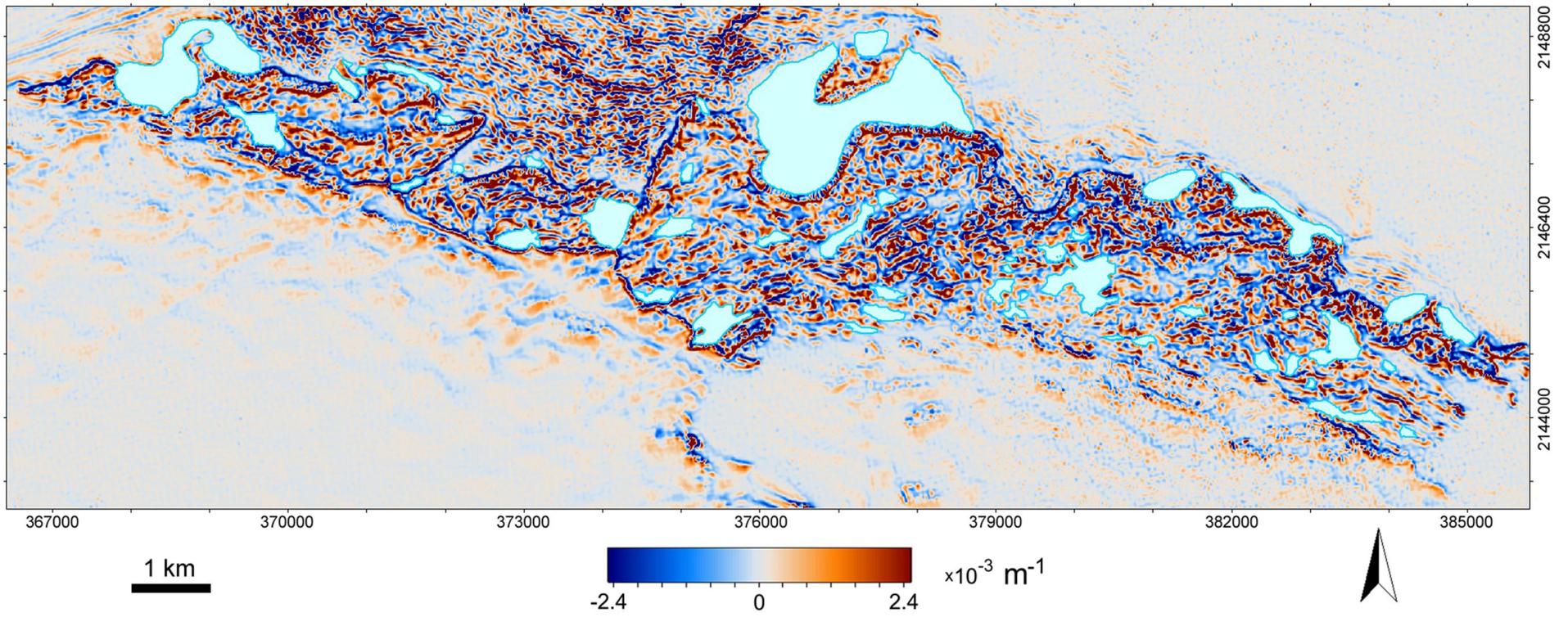

Figure 45. Schirmacher Oasis, mean curvature.

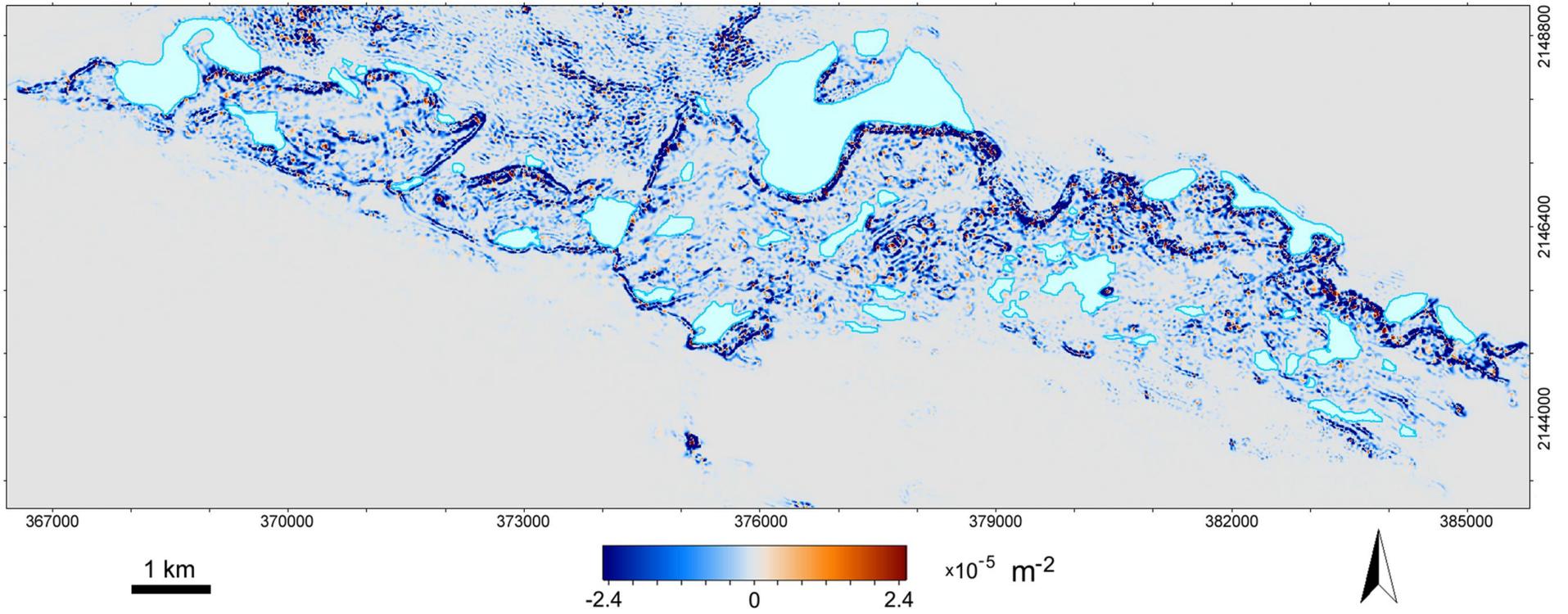

Figure 46. Schirmacher Oasis, Gaussian curvature.

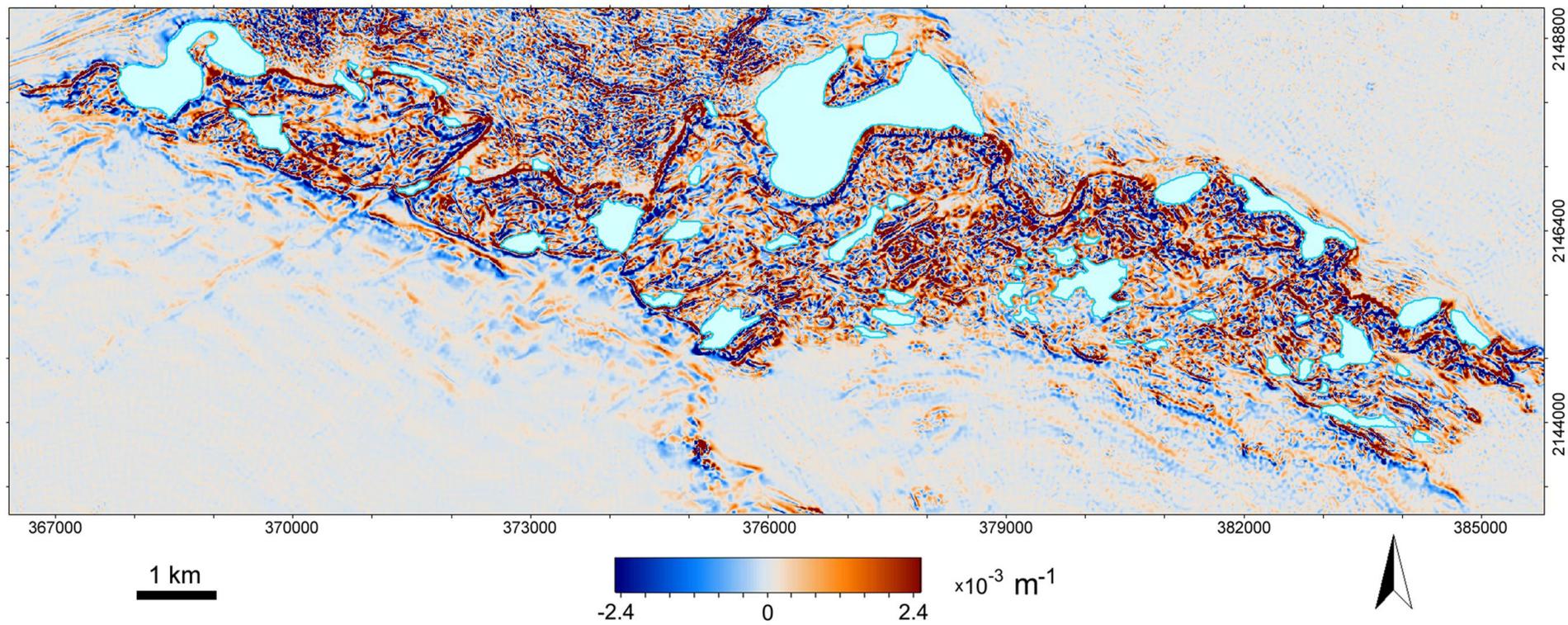

Figure 47. Schirmacher Oasis, difference curvature.

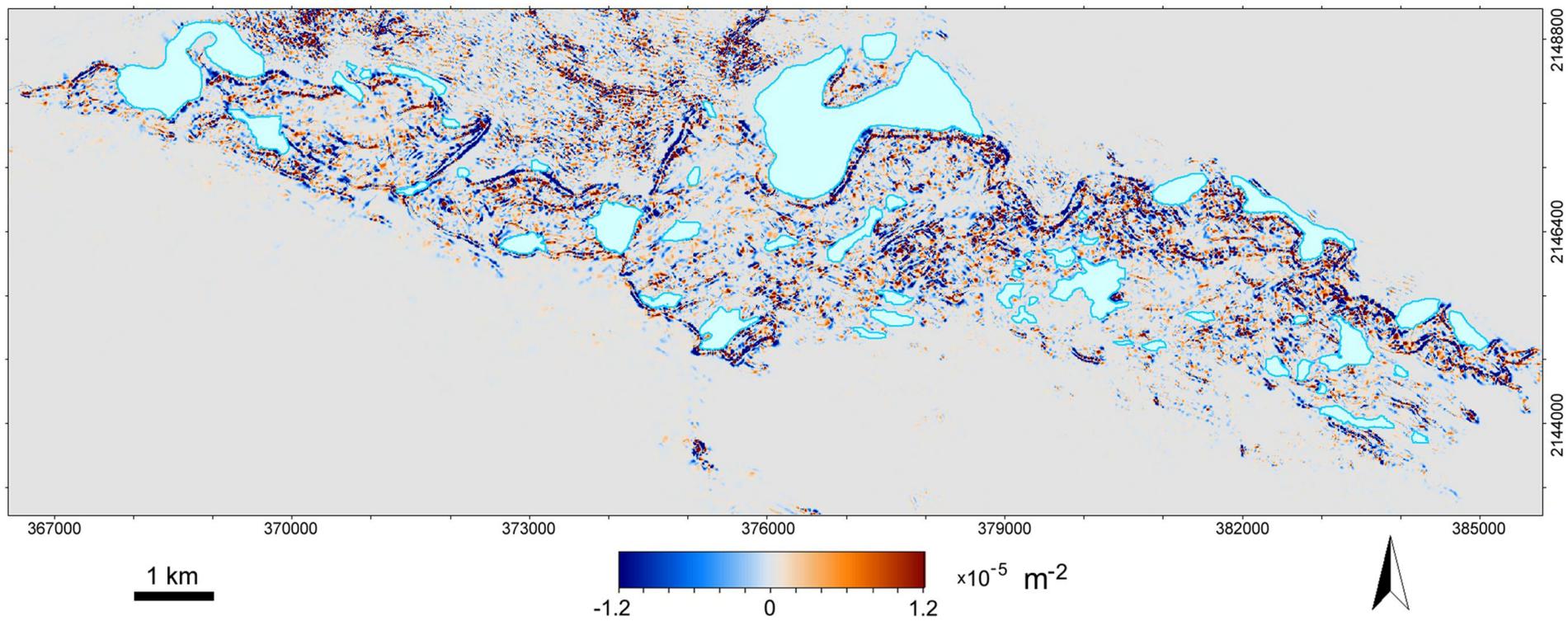

Figure 48. Schirmacher Oasis, accumulation curvature.

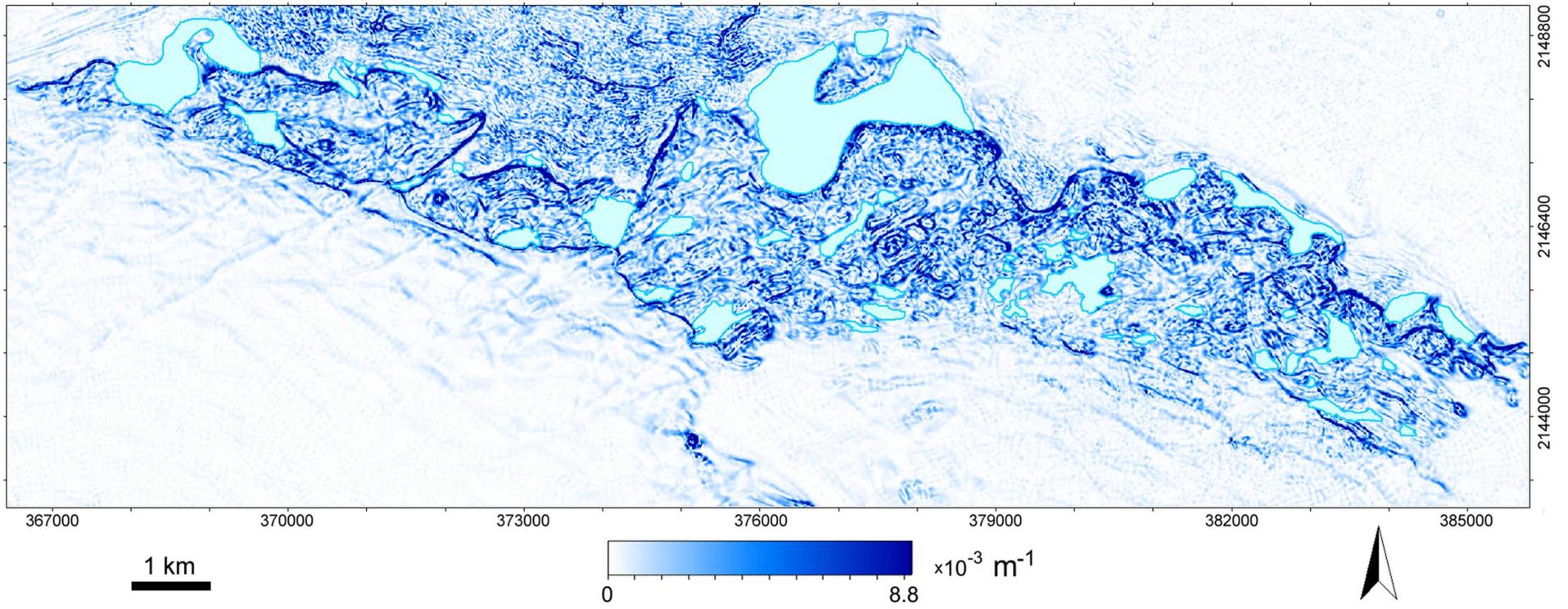

Figure 49. Schirmacher Oasis, horizontal excess curvature.

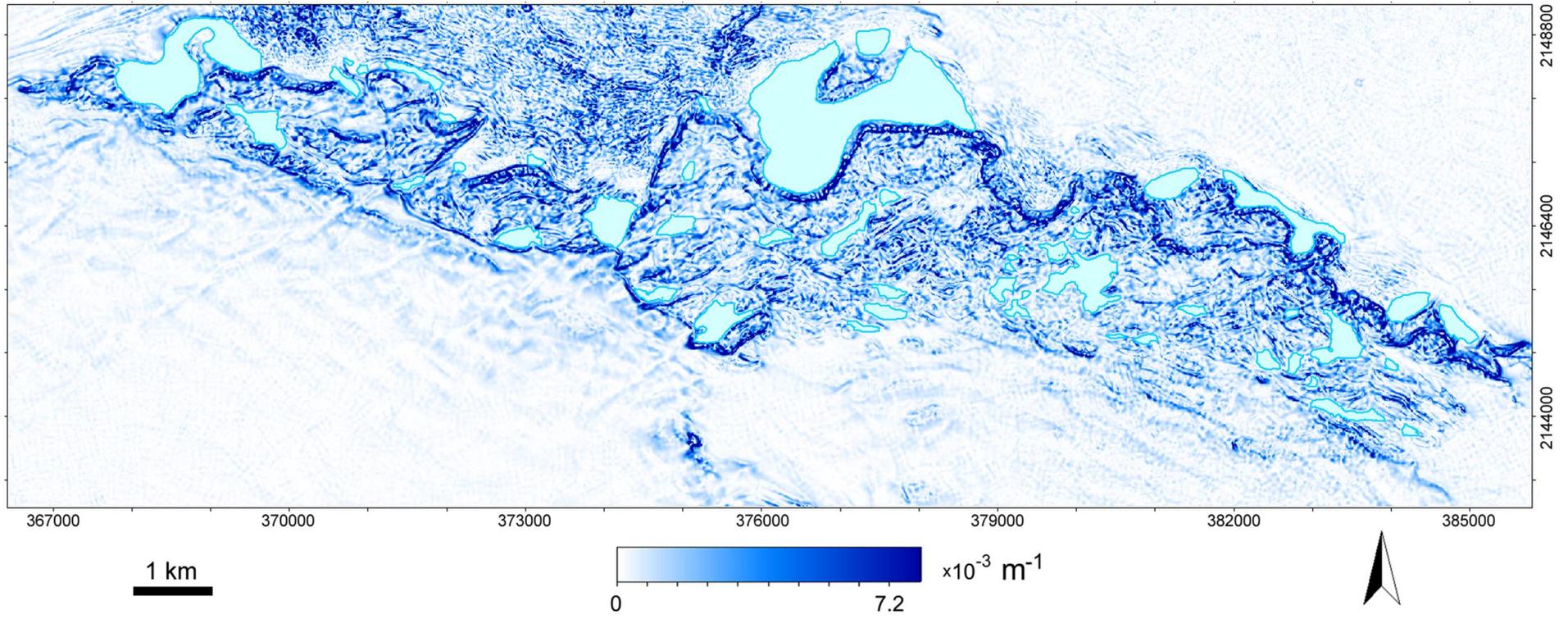

Figure 50. Schirmacher Oasis, vertical excess curvature.

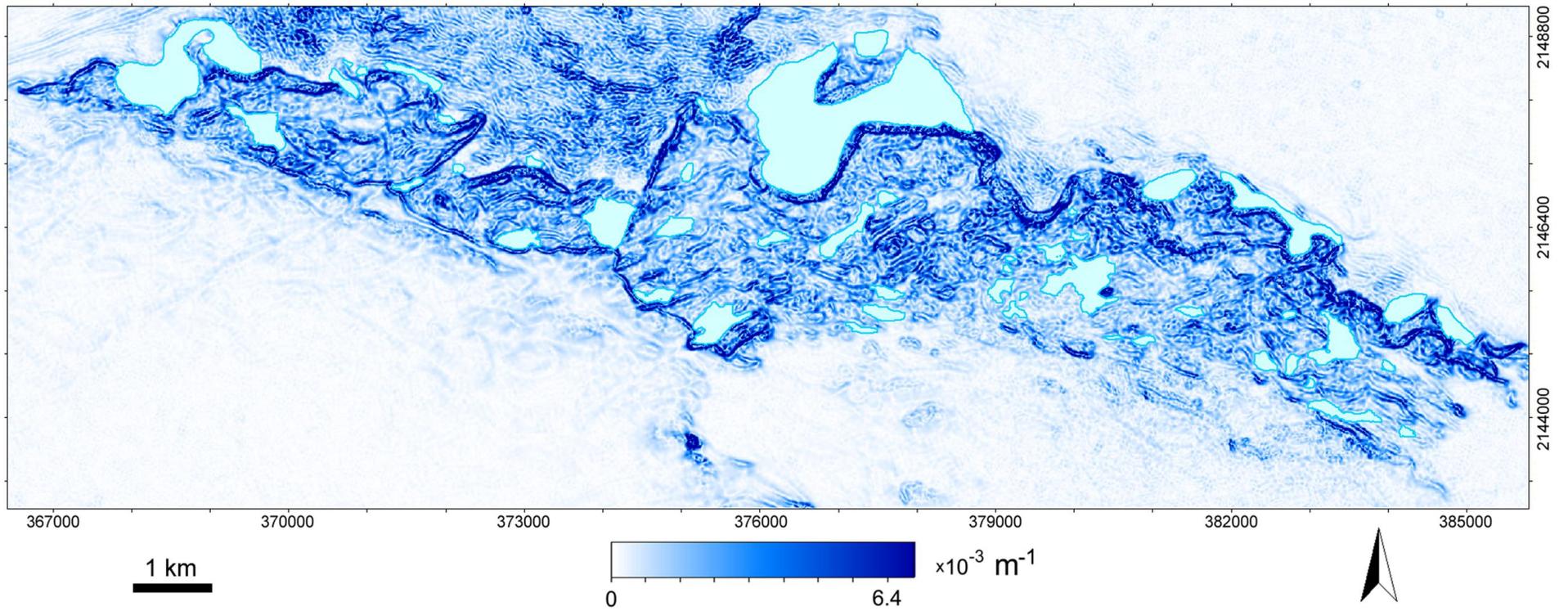

Figure 51. Schirmacher Oasis, unsphericity curvature.

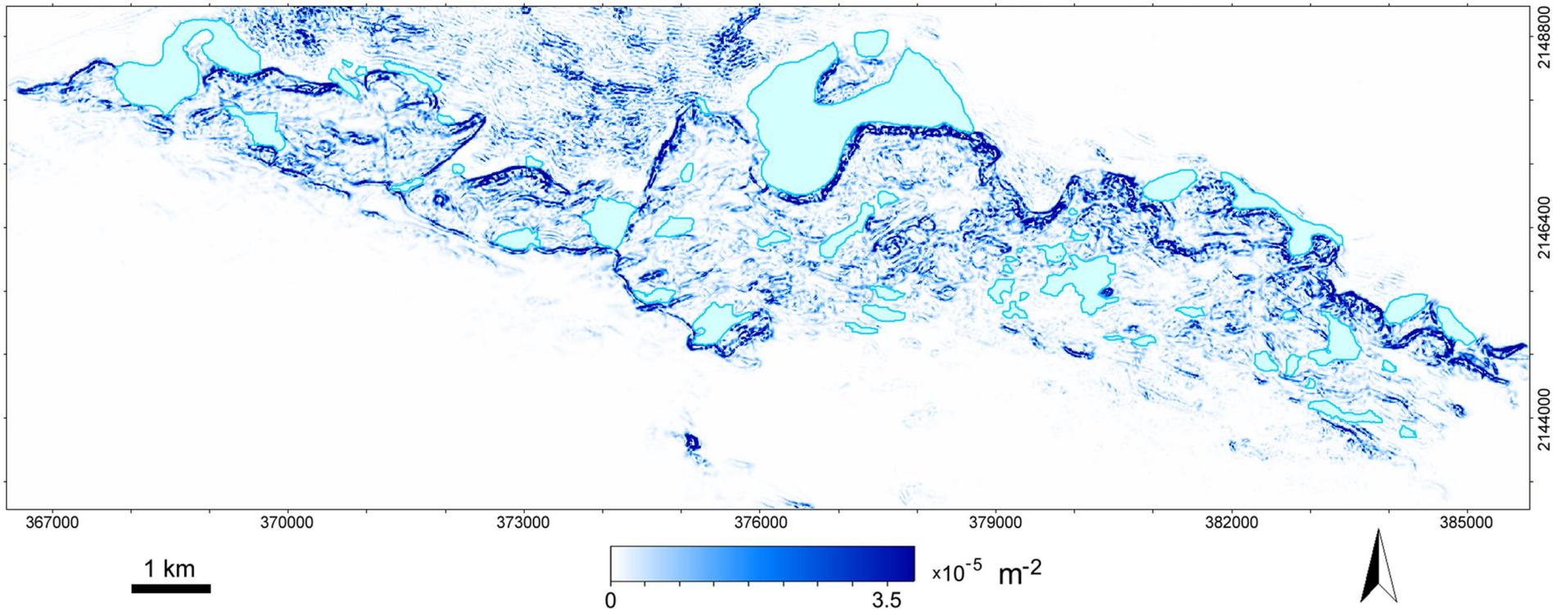

Figure 52. Schirmacher Oasis, ring curvature.

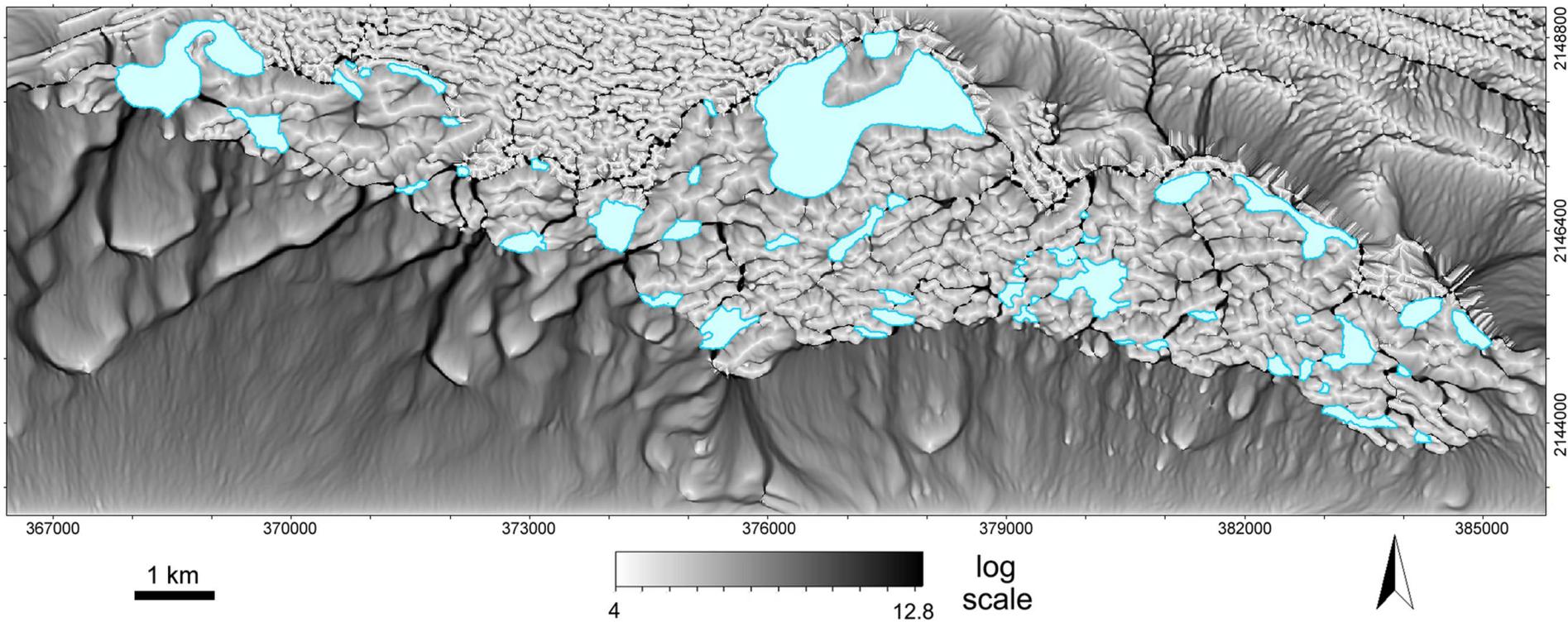

Figure 53. Schirmacher Oasis, catchment area.

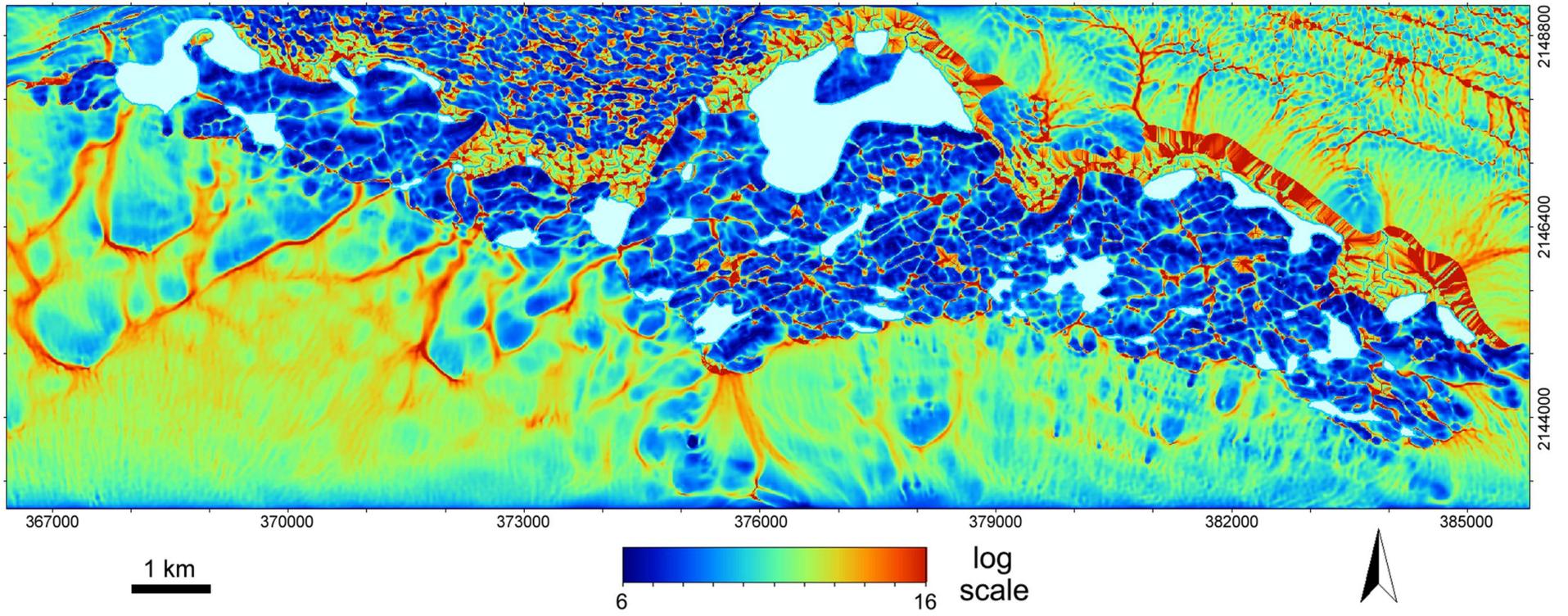

Figure 54. Schirmacher Oasis, topographic index.

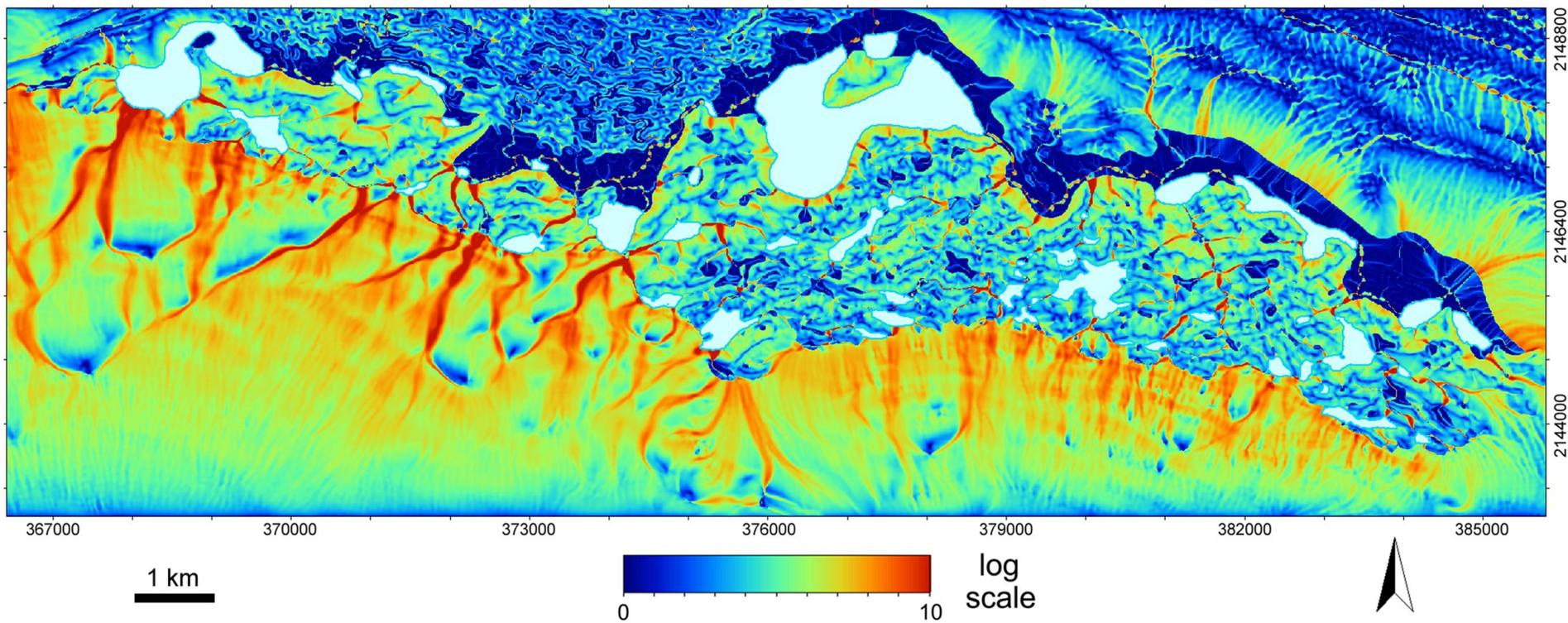

Figure 55. Schirmacher Oasis, stream power index.

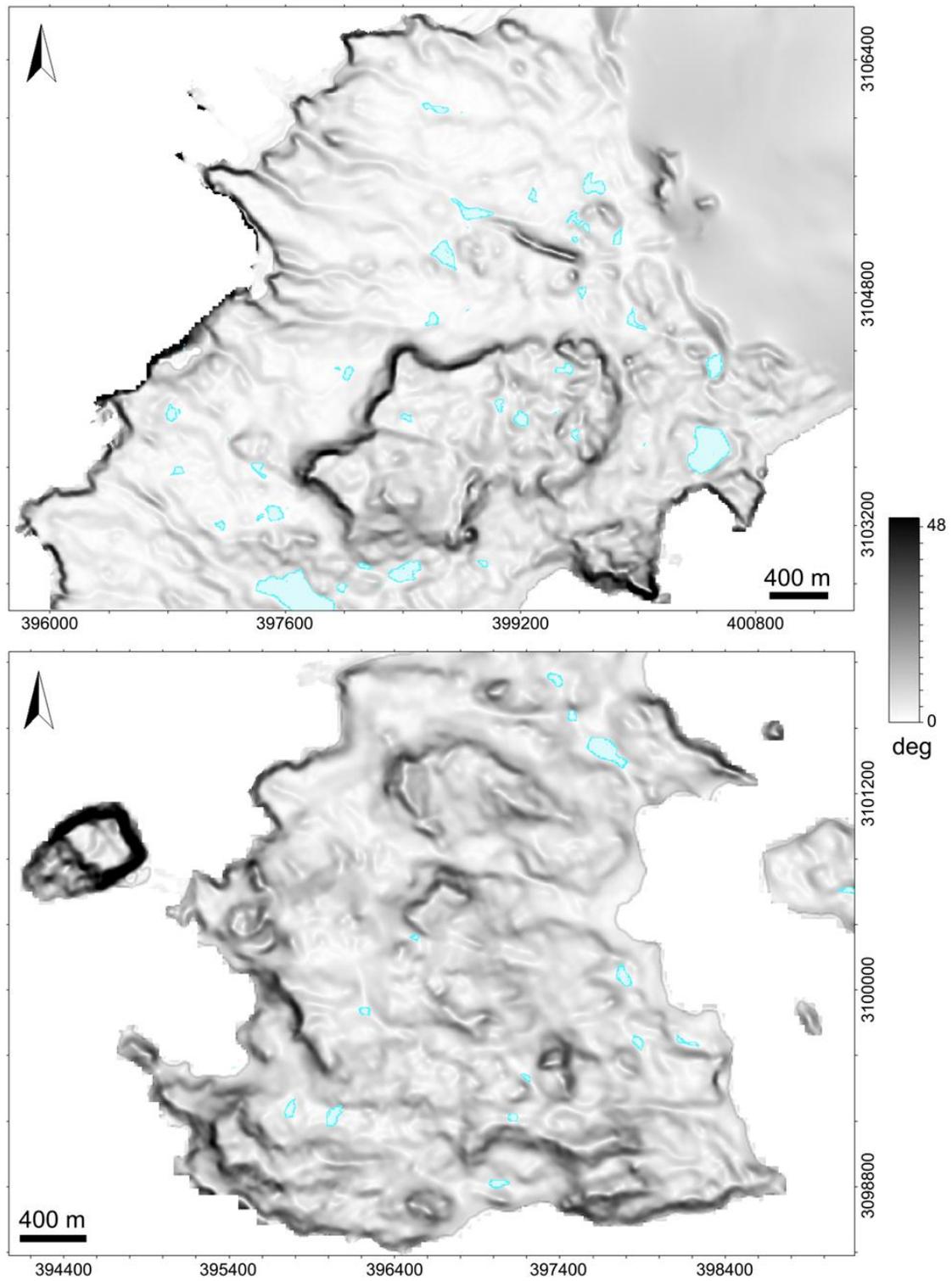

Figure 56. Fildes Peninsula, slope.
Upper: northern part of the peninsula. Lower: southern part of the peninsula

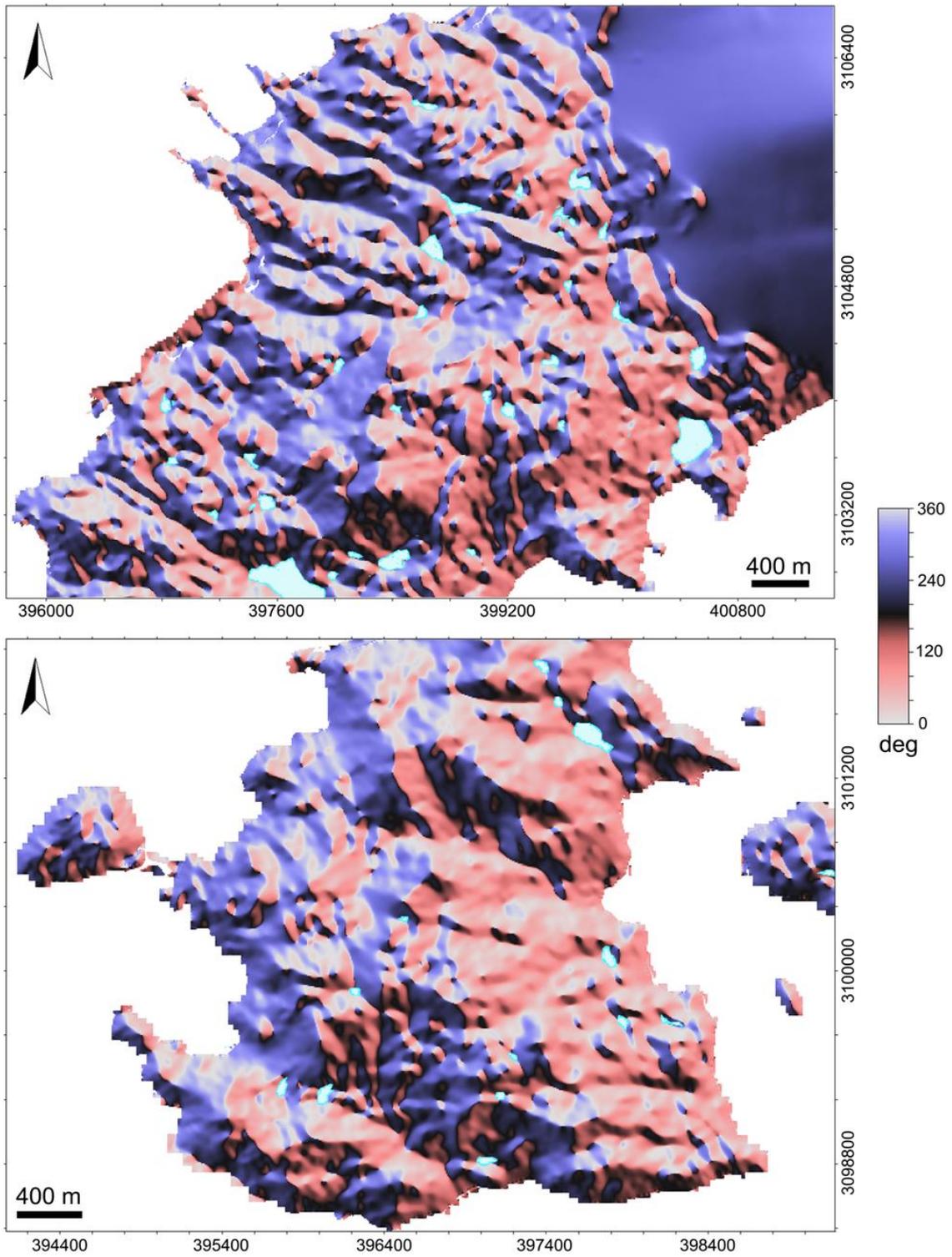

Figure 57. Fildes Peninsula, aspect.
Upper: northern part of the peninsula. Lower: southern part of the peninsula

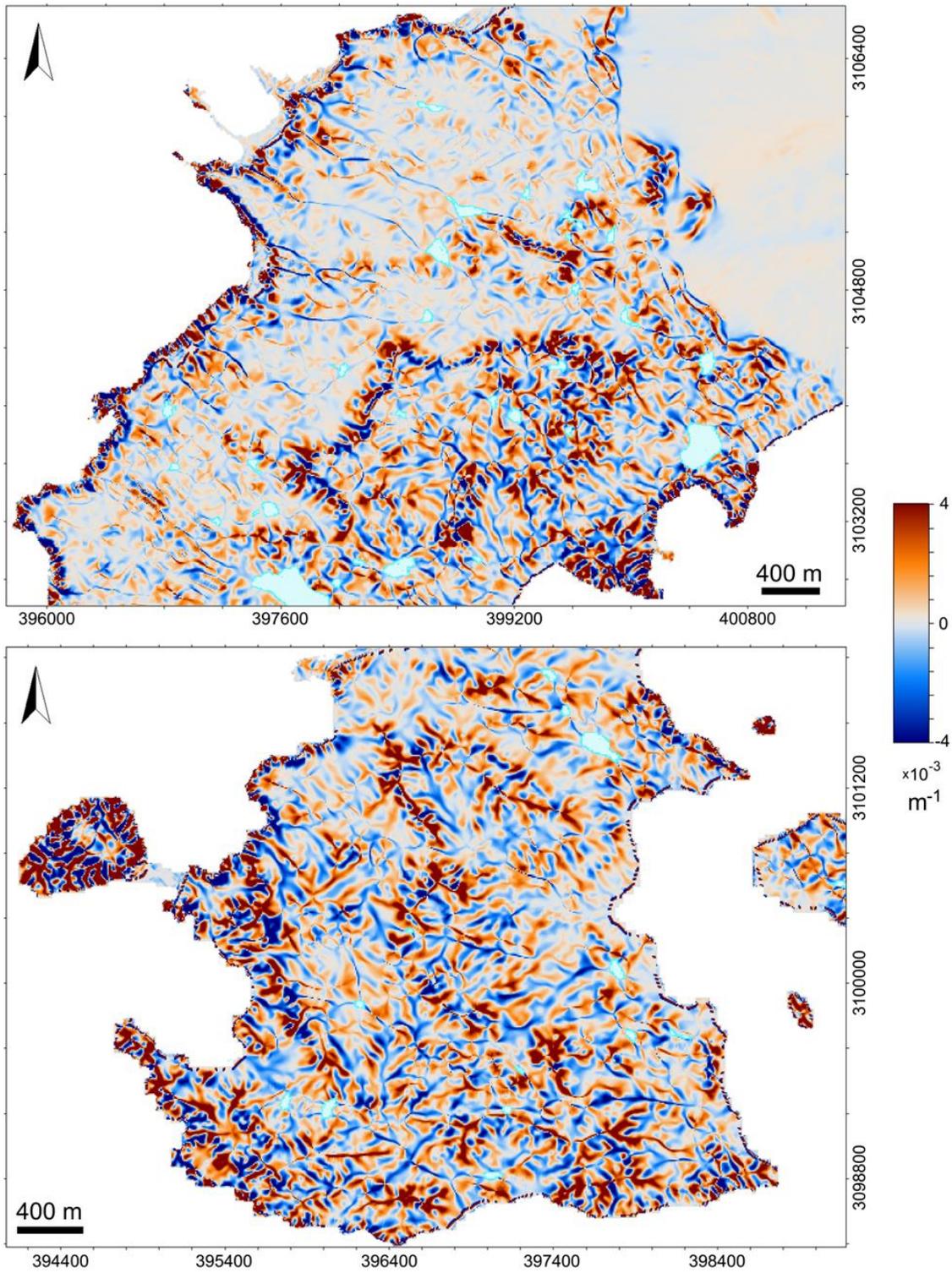

Figure 58. Fildes Peninsula, horizontal curvature.
Upper: northern part of the peninsula. Lower: southern part of the peninsula

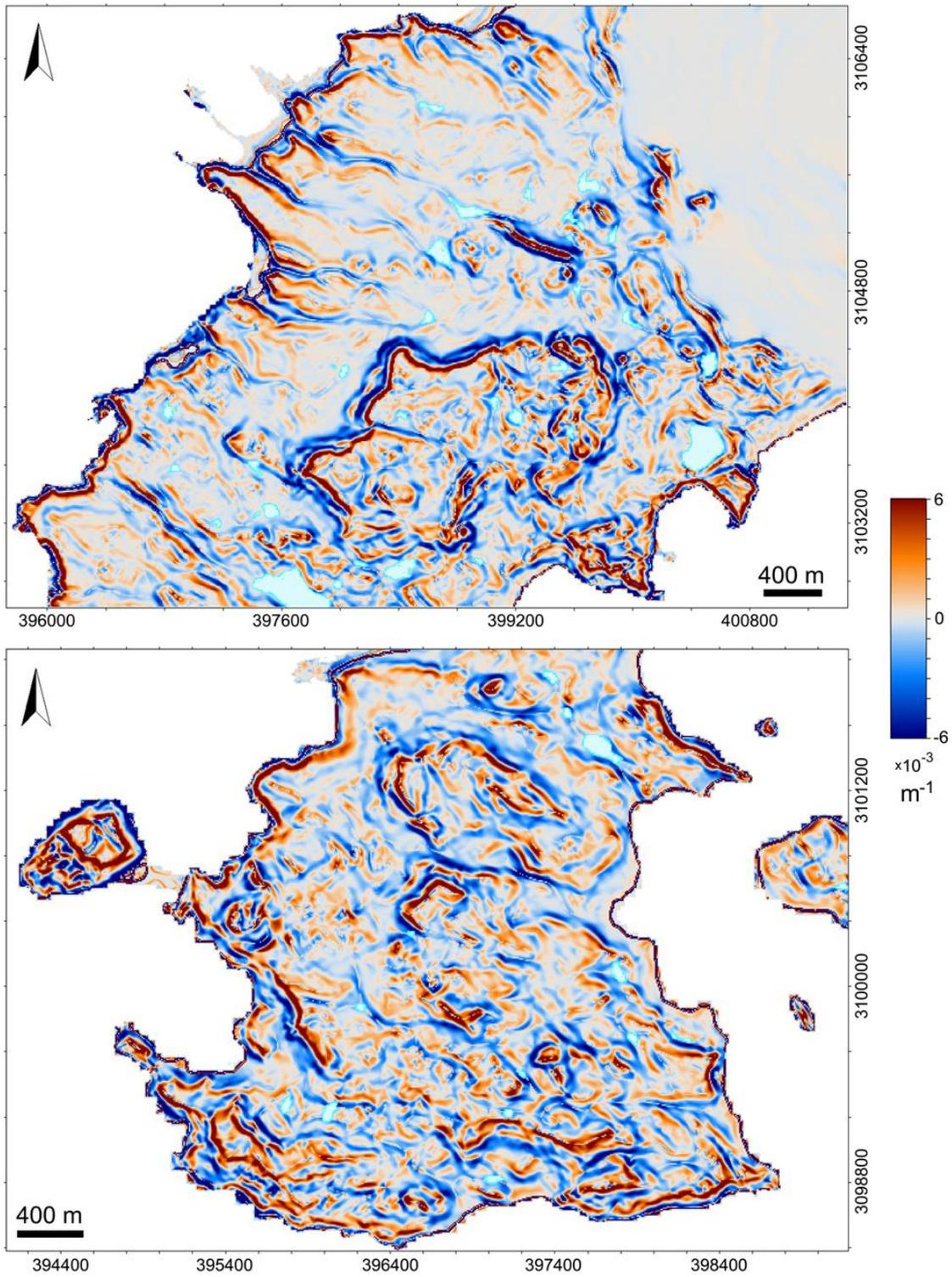

Figure 59. Fildes Peninsula, vertical curvature.
Upper: northern part of the peninsula. Lower: southern part of the peninsula

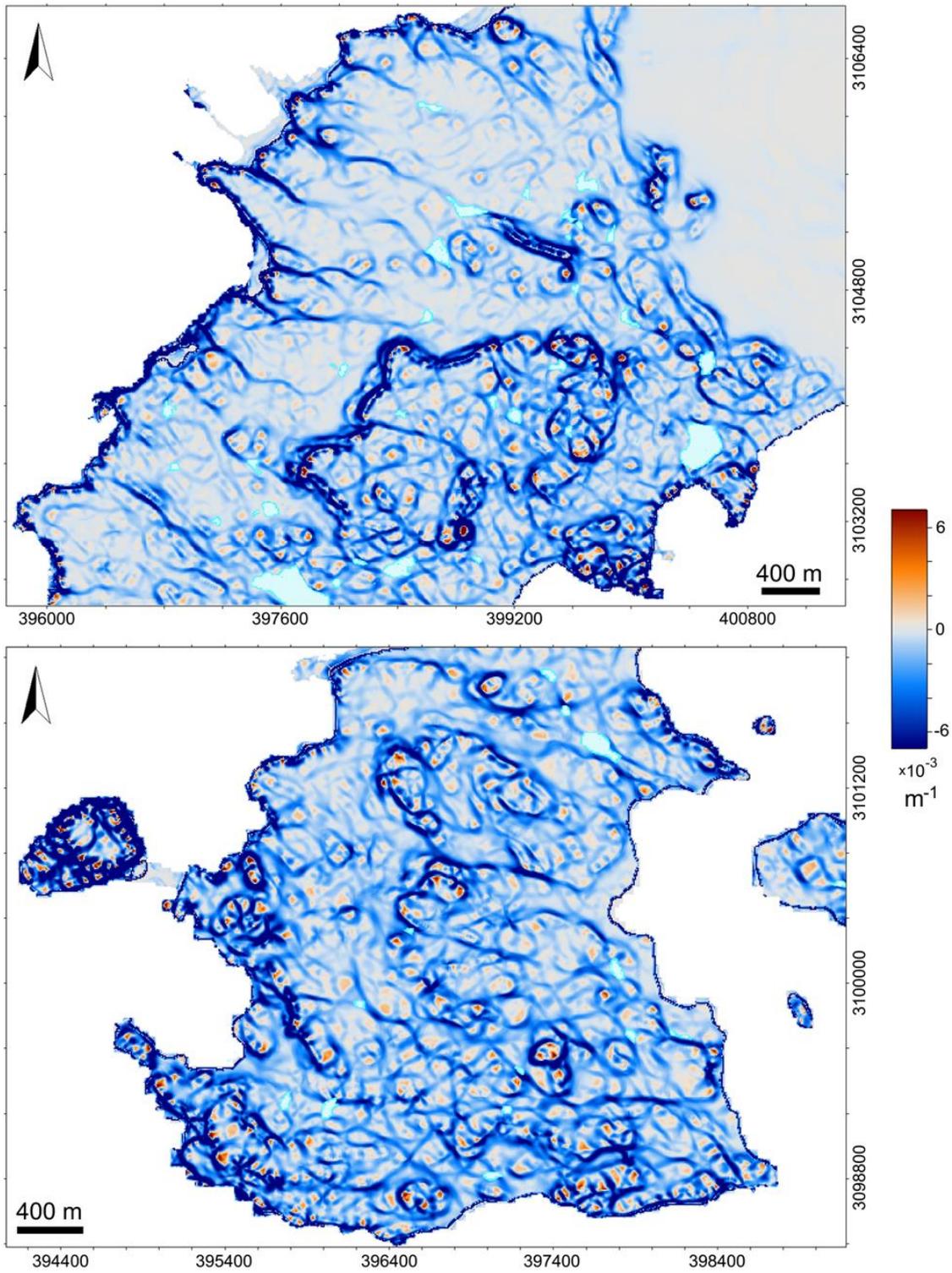

Figure 60. Fildes Peninsula, minimal curvature.
Upper: northern part of the peninsula. Lower: southern part of the peninsula

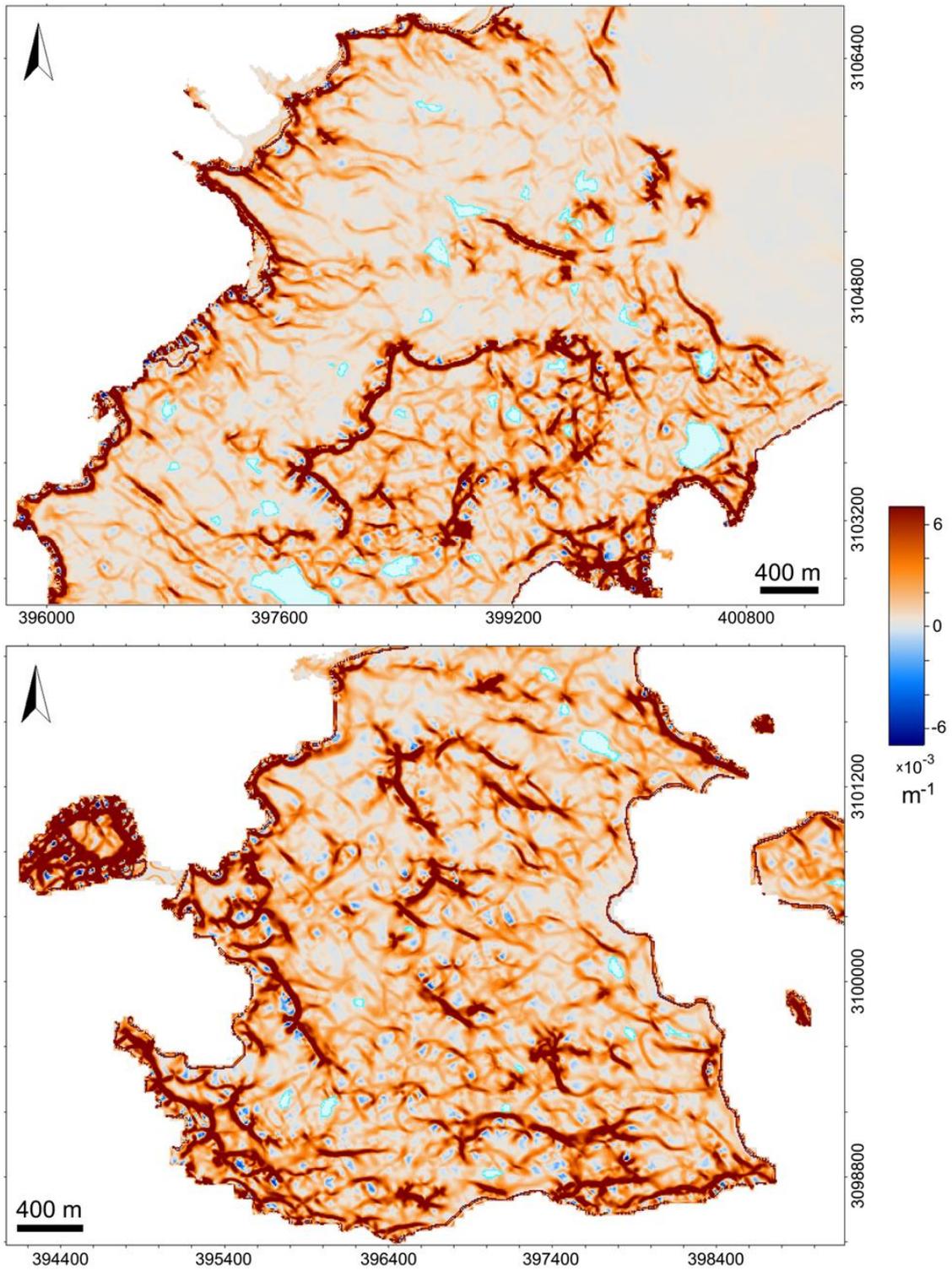

Figure 61. Fildes Peninsula, maximal curvature.
Upper: northern part of the peninsula. Lower: southern part of the peninsula

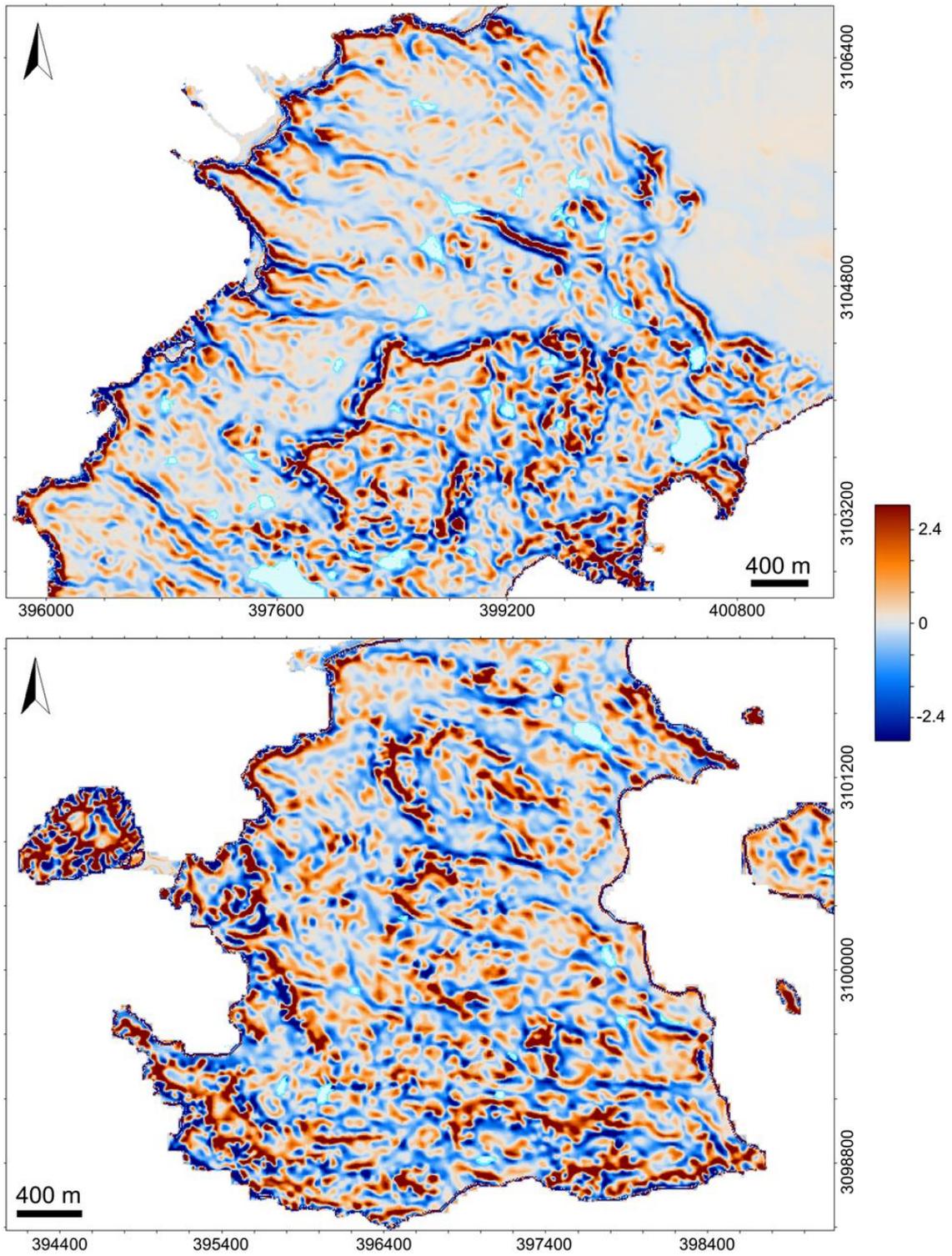

Figure 62. Fildes Peninsula, mean curvature.
Upper: northern part of the peninsula. Lower: southern part of the peninsula

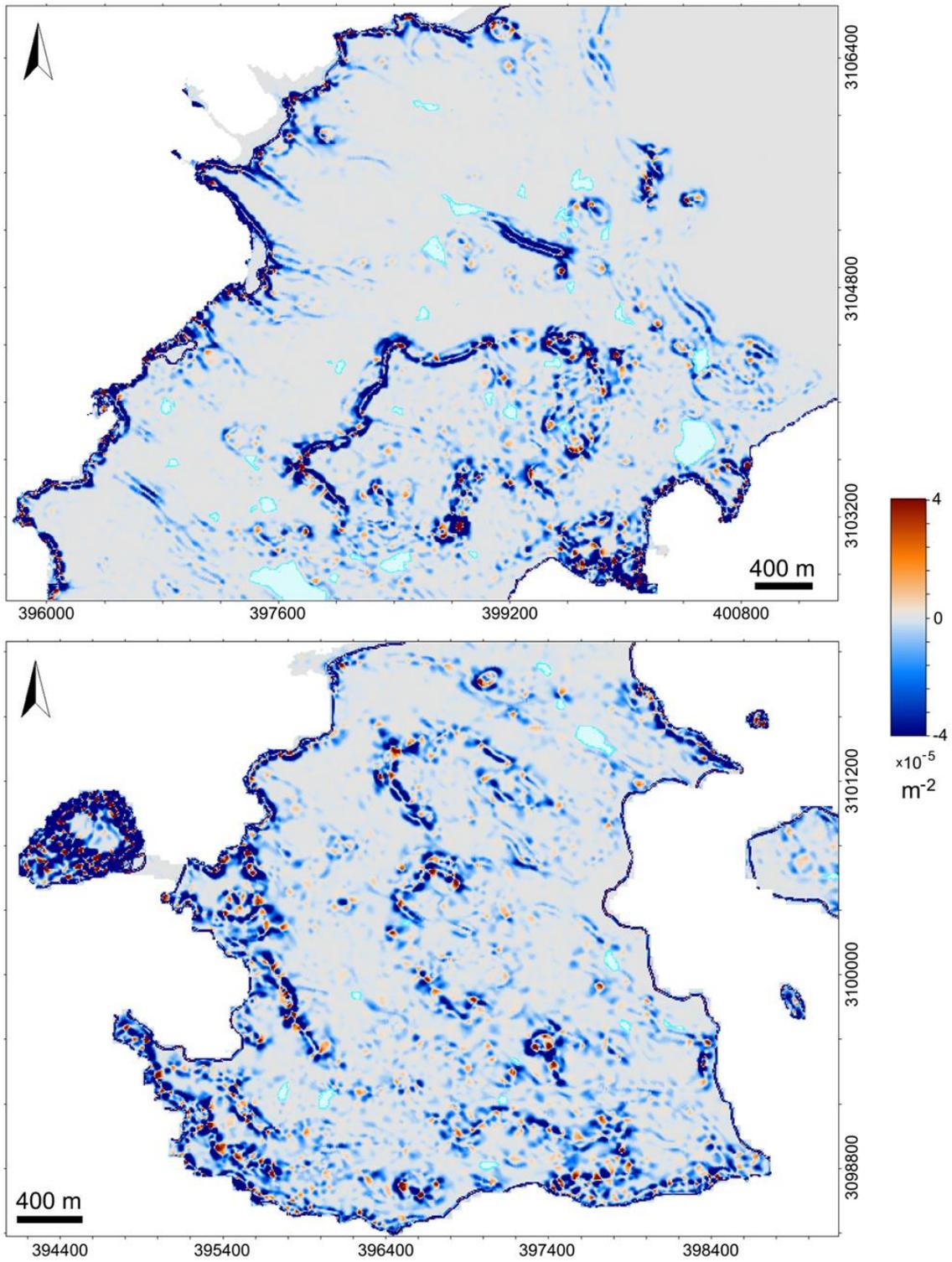

Figure 63. Fildes Peninsula, Gaussian curvature.
Upper: northern part of the peninsula. Lower: southern part of the peninsula

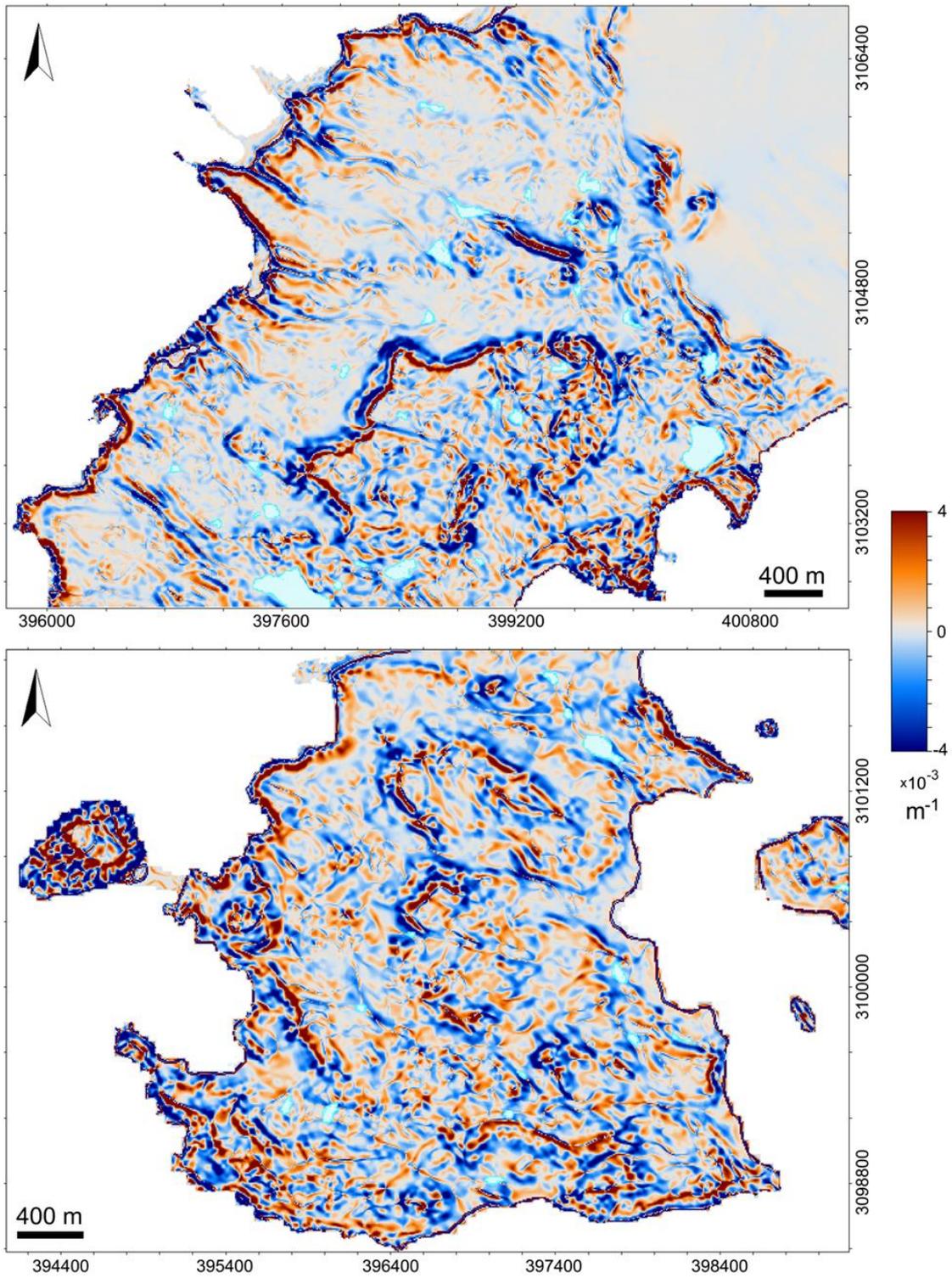

Figure 64. Fildes Peninsula, difference curvature.
Upper: northern part of the peninsula. Lower: southern part of the peninsula

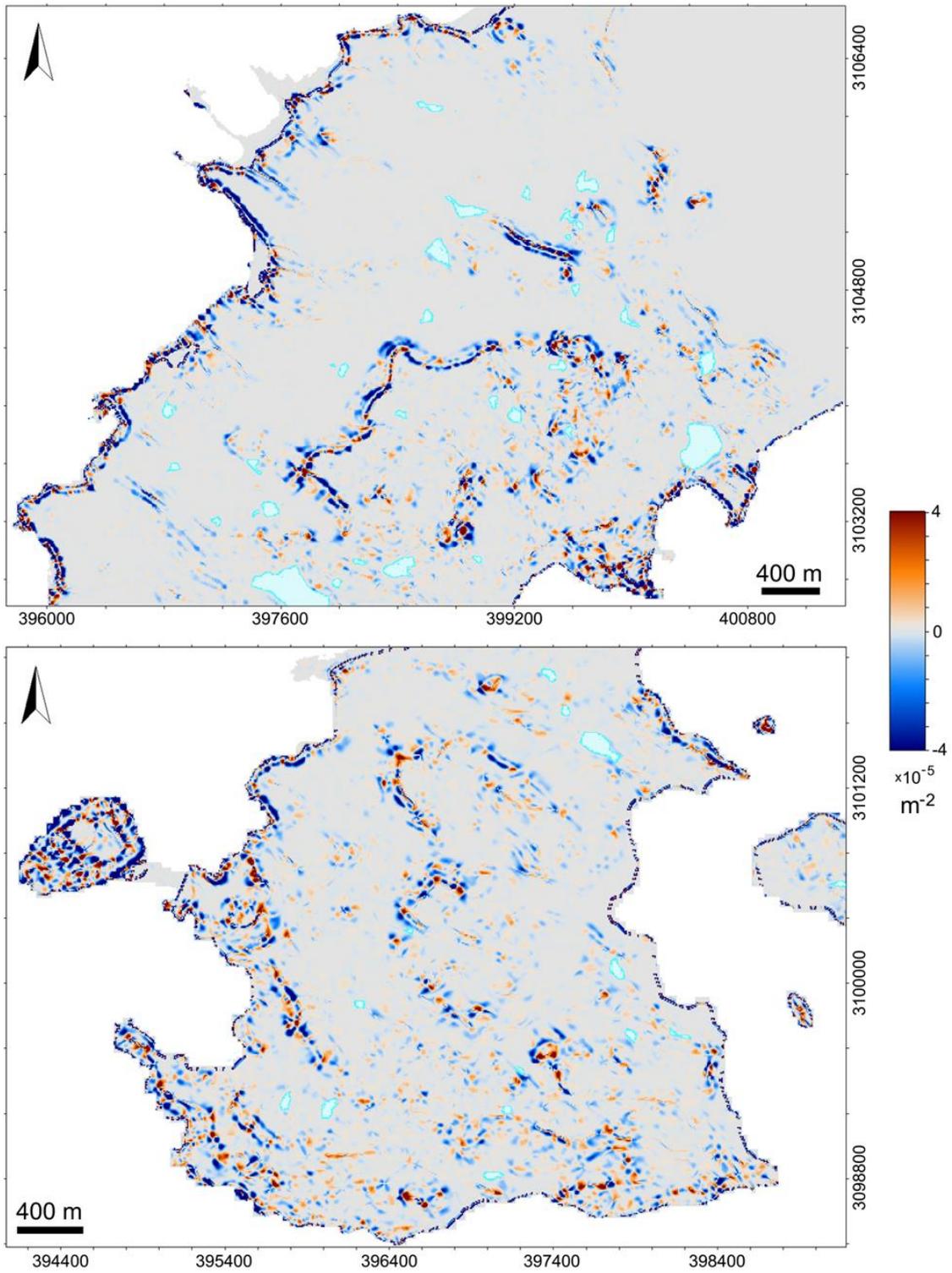

Figure 65. Fildes Peninsula, accumulation curvature.
Upper: northern part of the peninsula. Lower: southern part of the peninsula

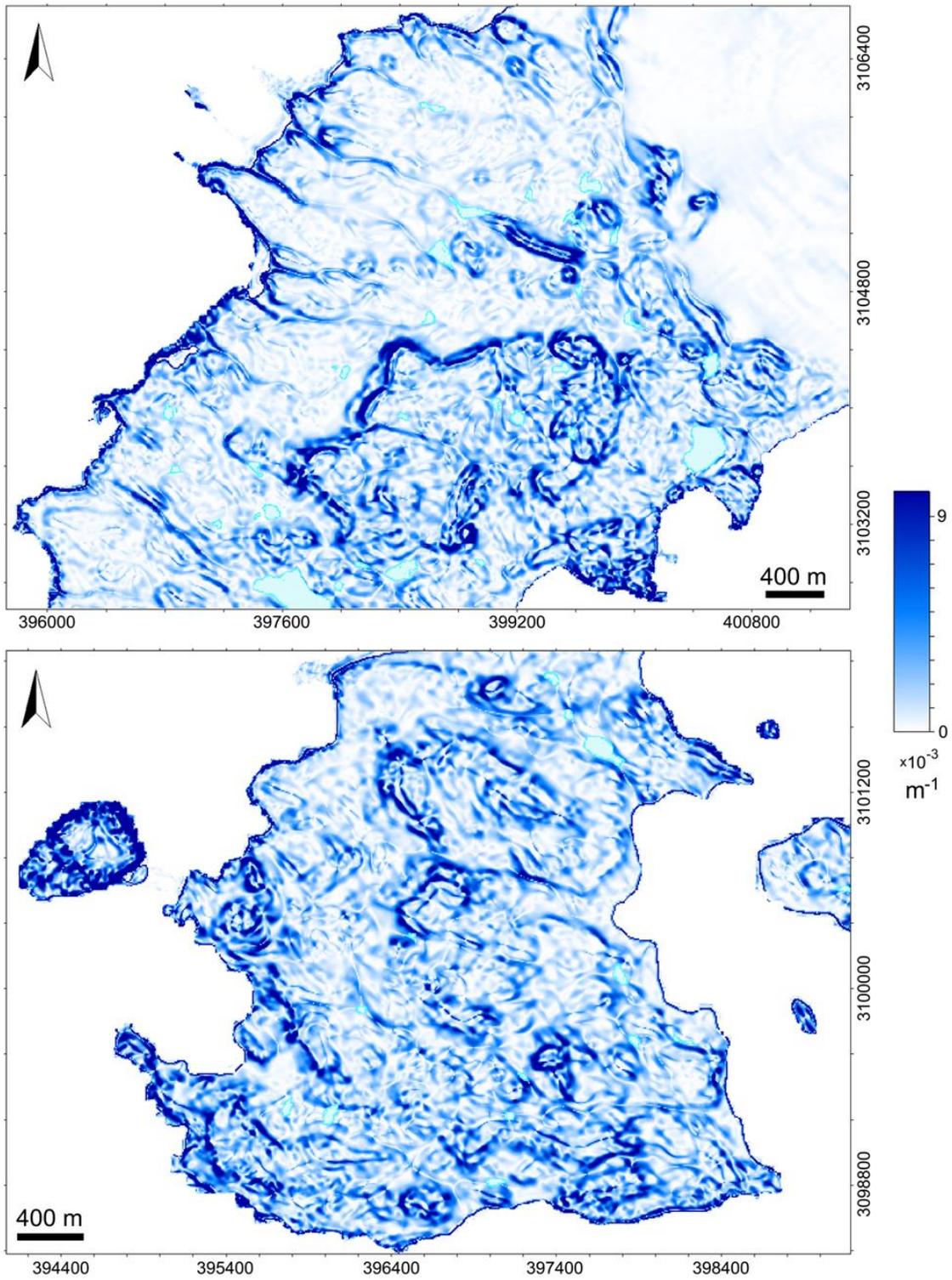

Figure 66. Fildes Peninsula, horizontal excess curvature.
Upper: northern part of the peninsula. Lower: southern part of the peninsula

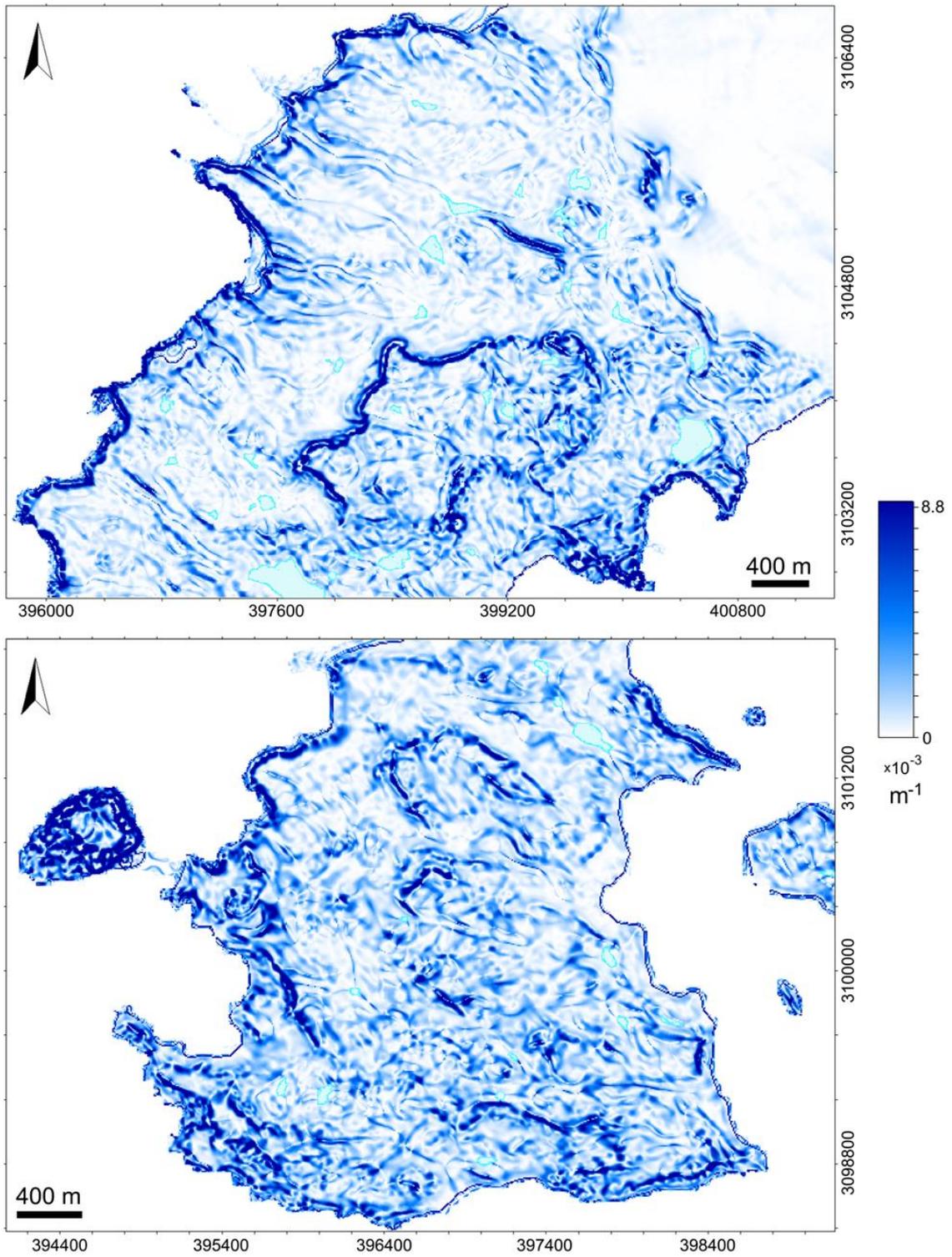

Figure 67. Fildes Peninsula, vertical excess curvature.
Upper: northern part of the peninsula. Lower: southern part of the peninsula

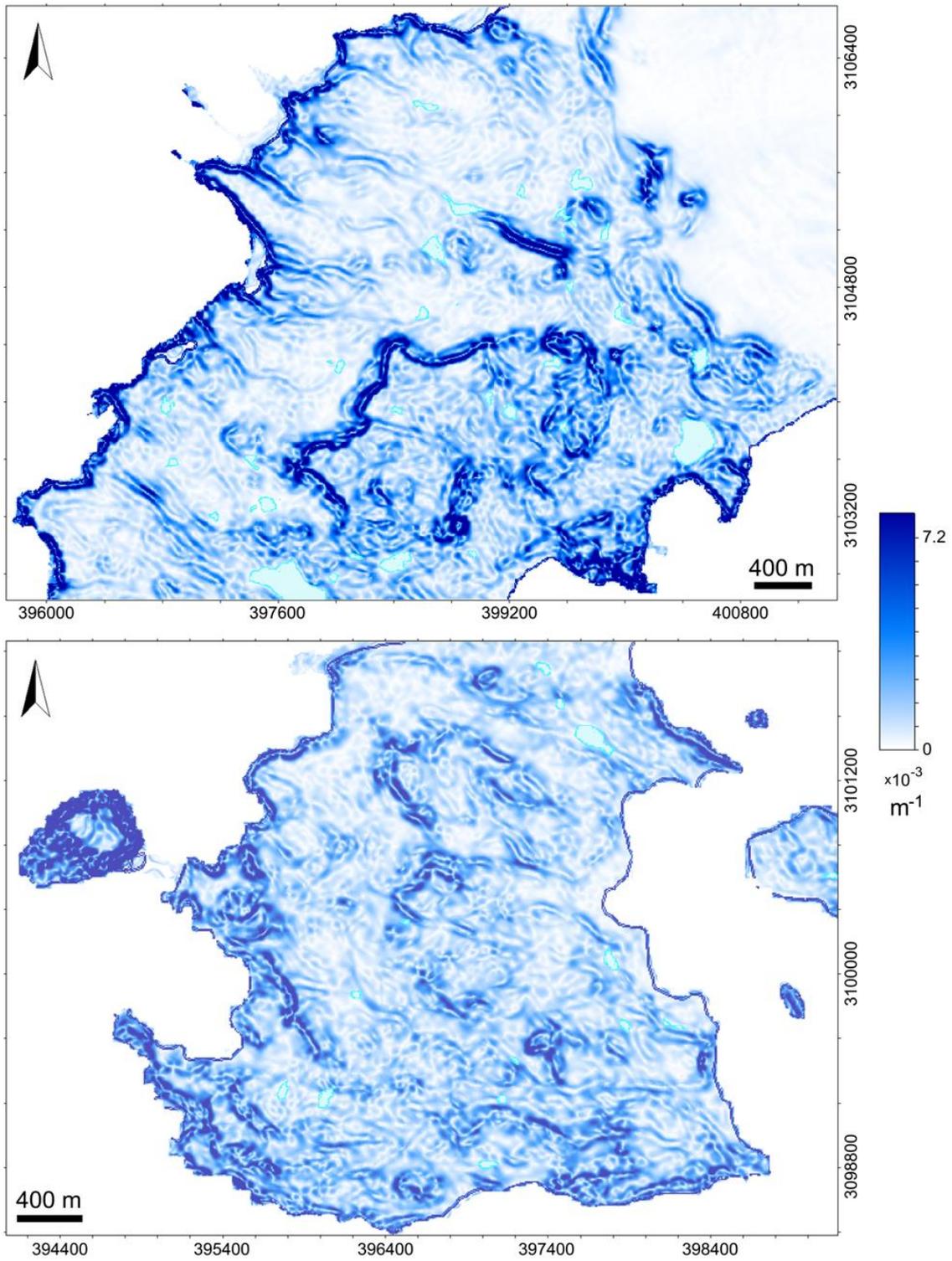

Figure 68. Fildes Peninsula, unsphericity curvature.
Upper: northern part of the peninsula. Lower: southern part of the peninsula

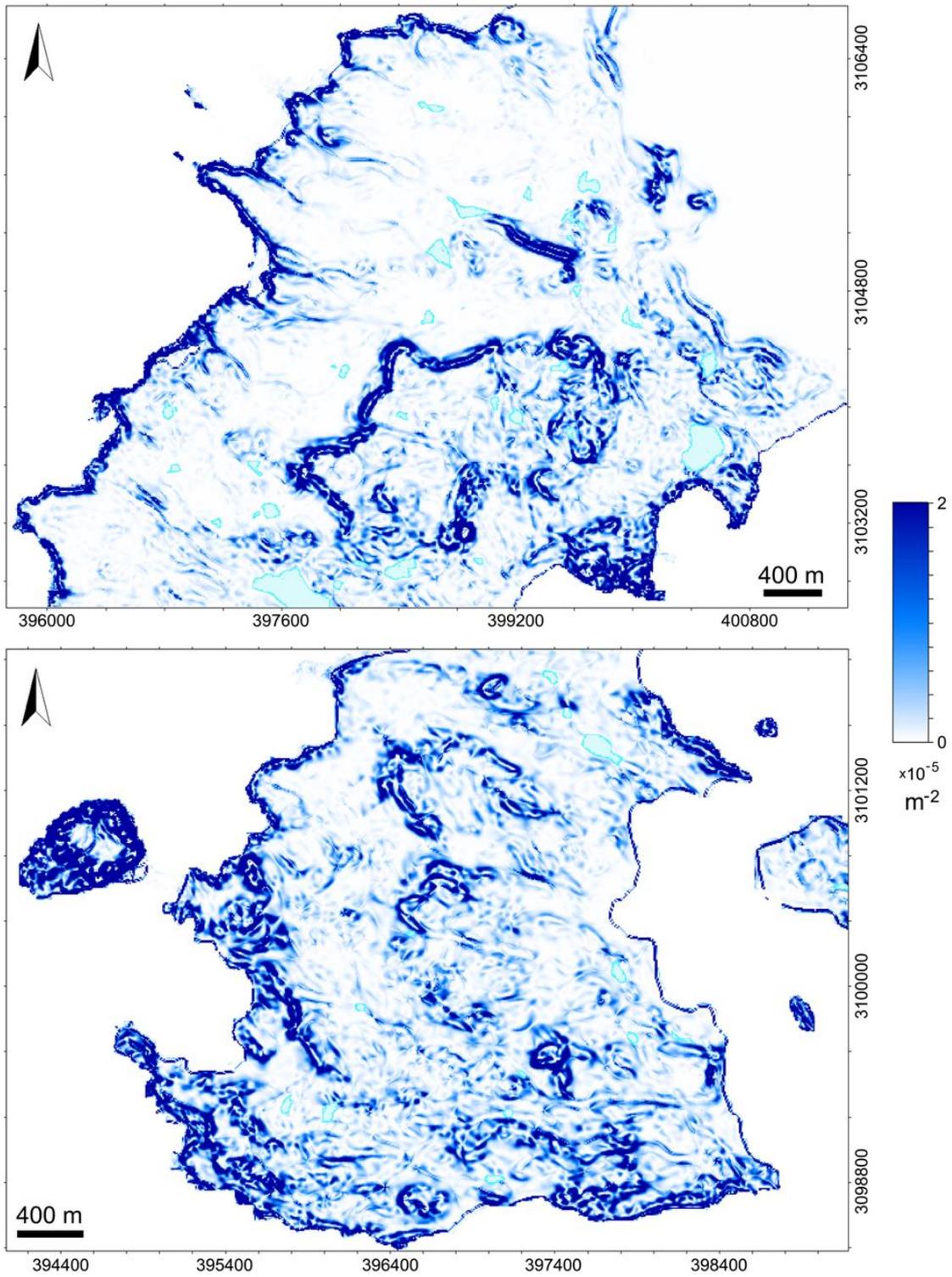

Figure 69. Fildes Peninsula, ring curvature.
Upper: northern part of the peninsula. Lower: southern part of the peninsula

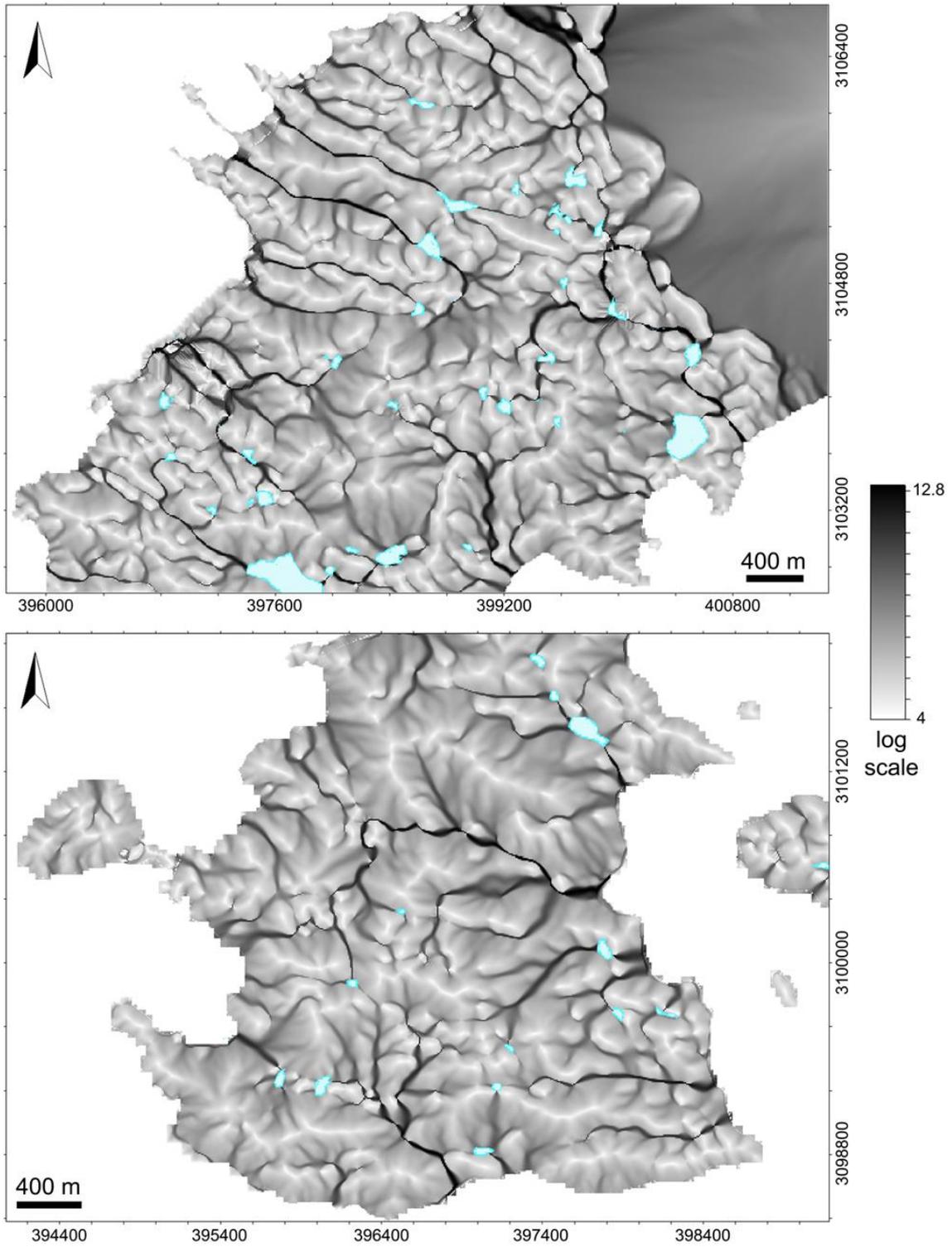

Figure 70. Fildes Peninsula, catchment area.
Upper: northern part of the peninsula. Lower: southern part of the peninsula

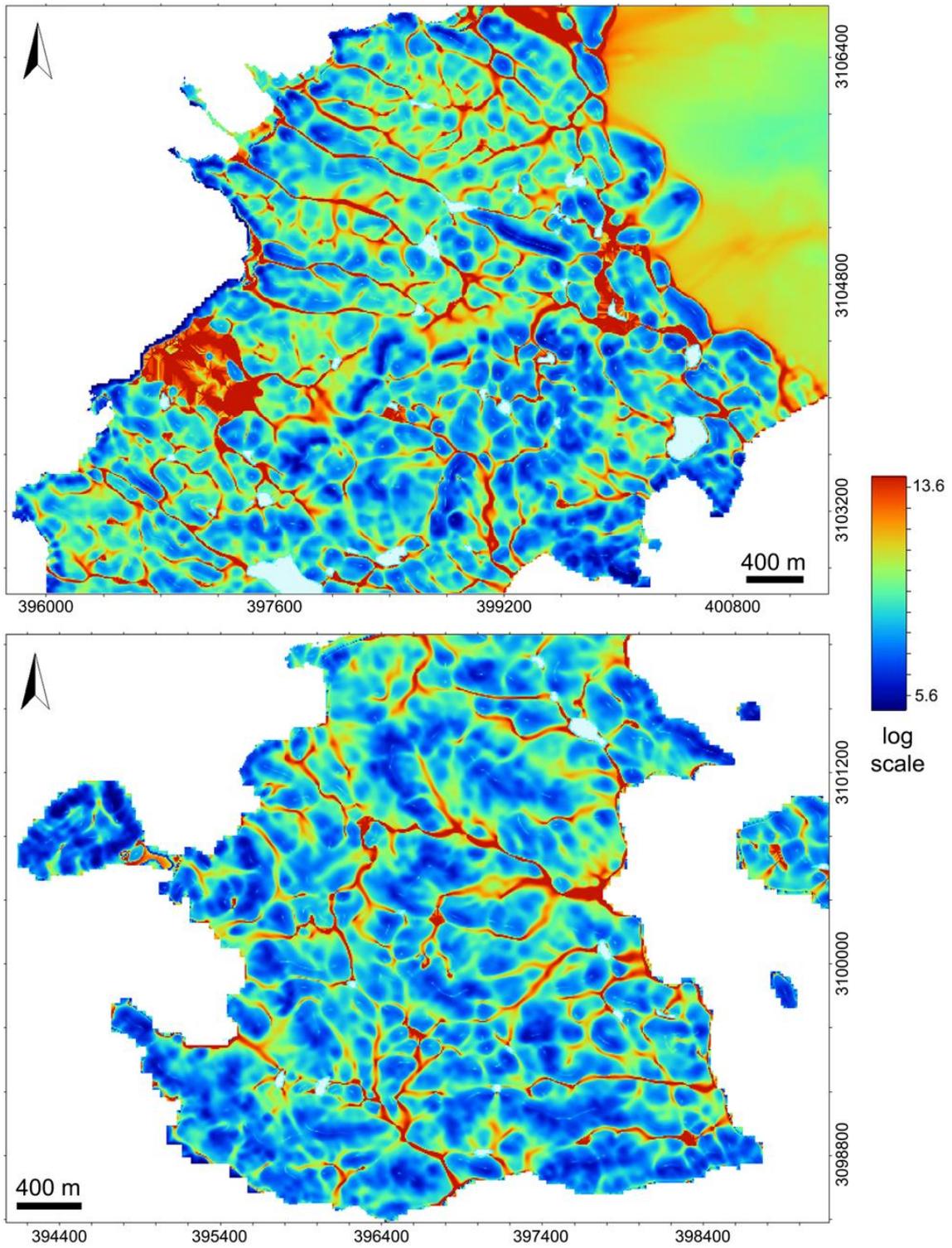

Figure 71. Fildes Peninsula, topographic index.
Upper: northern part of the peninsula. Lower: southern part of the peninsula

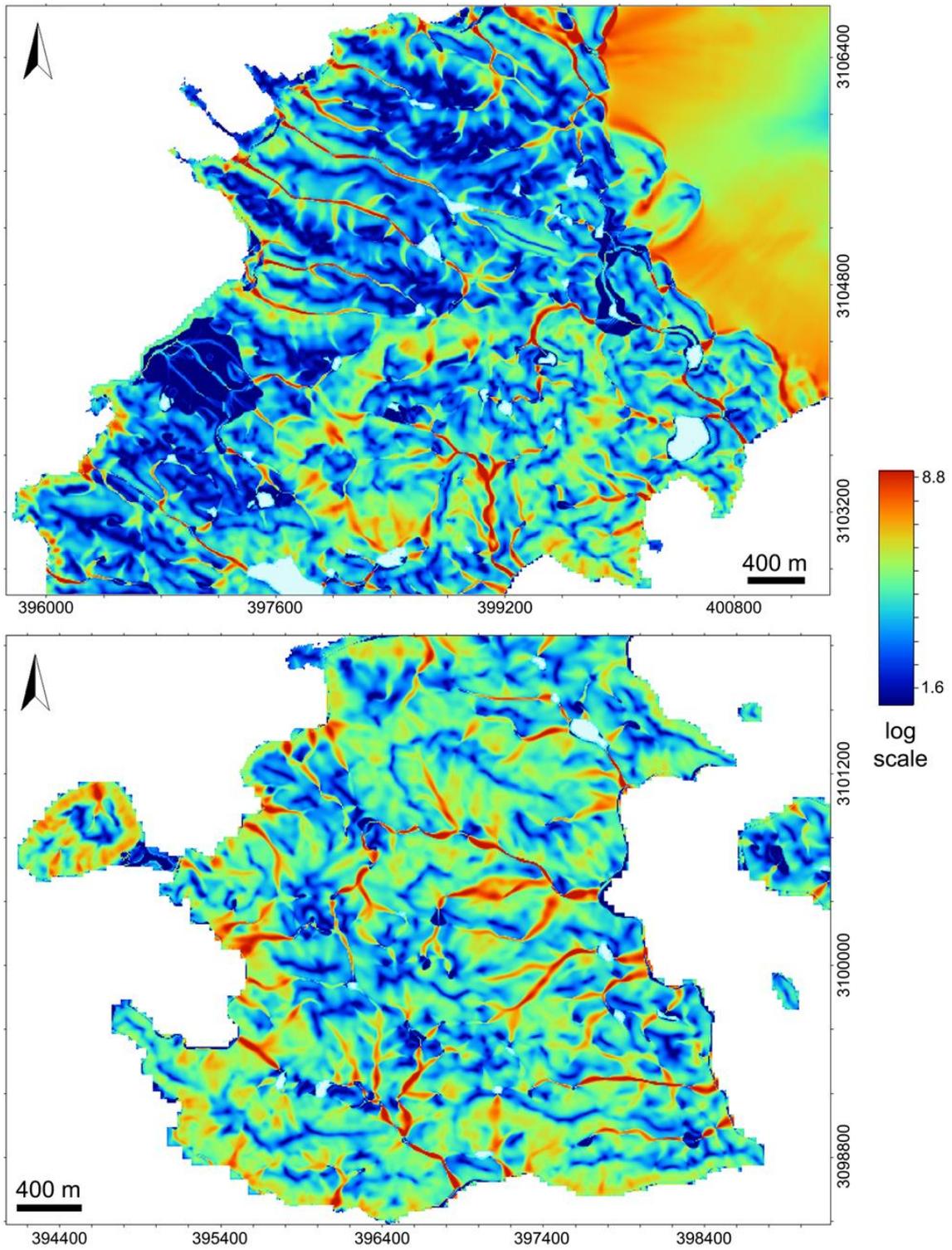

Figure 72. Fildes Peninsula, stream power index.
Upper: northern part of the peninsula. Lower: southern part of the peninsula

## 4. Results

Figures 5–72 represent fragments of the derived geomorphometric maps of the oases.

The $G$ map (Figures 5, 22, 39, and 56) clearly shows the spatial distribution of slope steepness. For example, within the Larsemann Hills, the steepest slopes are typical for the Mount Wall (southern Broknes), Rusty Ridge, and Castle Bluff (western Broknes), the unnamed mountain south of the Canterbury Hill, and the coastal slopes of the northwestern Broknes and southern Stornes (where they coincide with the edge cliff of the passive Stornes glacial dome). The slopes of the hills in the east of the Fisher Island are also steep enough. In general, it should be noted that in spite of low elevations of the Larsemann Hills oasis (maximum elevation is 158 m above sea level at the Blundell Peak of the Stornes Peninsula) and the use of the term 'hills' to describe the local topography, in many parts of the oasis the terrain looks like steep rocky mountains (indeed, some hills are called peaks).

The $A$ Map (Figures 6, 23, 40, and 57) shows the spatial distribution of slope aspect. For example, within the Larsemann Hills, a large part of the slopes of the structurally predetermined elongated hills of Broknes and Stornes Peninsulas are oriented to the northwest and southeast, corresponding to the general fault (lineament) structure of the region. The hill slopes of the Grovnes, Brattnevet, Stinear Peninsulas, and Fisher Island, which seem to wedge between the Broknes and Stornes Peninsulas, are oriented predominantly to the southwest and northeast. This may serve as an indirect indication of the difference in the geological origin of these two groups of territories.

The $k_h$ map (Figures 7, 24, 41, and 58) shows the spatial distribution of convergence and divergence zones of surface flows ($k_h < 0$ and $k_h > 0$, respectively). In the geomorphological sense, these are spurs of valleys and ridges (blue and brown colors, respectively). In combination, the convergence and divergence zones represent the flow structure of the terrain, which is obviously structurally predetermined there.

The $k_v$ map (Figures 8, 25, 42, and 59) shows the spatial distribution of zones of relative deceleration and acceleration of surface flows ($k_v < 0$ and $k_v > 0$, respectively). Geomorphologically, this map shows areas of sharp change of elevation (e.g., scarps). The fault (lineament) structure of the territory is also well readable on this map.

It is reasonable to consider maps of $k_{min}$ and $k_{max}$ (Figures 9 and 10, 26 and 27, 43 and 44, 60 and 61) together. Their analysis reveals two types of elongated linear forms: concave (e.g., valleys) represented by dark blue lines (Figures 9, 26, 43, and 60, $k_{min} < 0$) and convex (e.g., ridges) represented by dark brown lines (Figures 10, 27, 44, and 61, $k_{max} > 0$). In terms of revealing geological lineaments (faults, folds, etc.) of the oases territories, the $k_{min}$ and $k_{max}$ maps are the most informative material, since all the lineaments are visible on these maps with the naked eye and their identification does not require additional processing of these map images.

The maps of the other curvatures (Figures 11–18, 28–35, 45–52, and 62–69), at first glance, do not provide significantly new information compared to the maps of the four discussed curvatures − $k_h$, $k_v$, $k_{min}$, and $k_{max}$, which are most popular in geosciences. It should be noted here that the 12 calculated types of curvatures are part of a mathematically complete system of surface curvatures (Shary, 1995). From the viewpoint of the geomorphometric theory, their joint application and analysis have a fundamental character. Let us further discuss briefly the derived maps of other curvatures.

$H$ represents two flow accumulation mechanisms, convergence and relative deceleration, with equal weights. Therefore, it is not surprising that the $H$ maps (Figures 11, 28, 45, and 62) visually represents a kind of average picture of the $k_h$ and $k_v$ maps (Figures 7 and 8, 24 and 25, 41 and 42, 58 and 59). $K$ retains its values after bending a surface if it occurred without tension, compression, or rupture. The $K$ maps (Figures 12, 29, 46, and 63) may be useful in geological studies of the oases topography evolution.

$E$ shows to what extent the relative deceleration of flows (measured by $k_v$) is higher than flow convergence (measured by $k_h$) at a given point of the topographic surface. The $E$ maps (Figures 13, 30, 47, and 64) can be used for comparative analysis of the prevalence of substance accumulation mechanisms in sedimentology. The $K_a$ maps (Figures 14, 31, 48, and 65) may be used for the same purpose, as this morphometric variable assesses the degree of flux accumulation.

$k_{he}$ shows to what extent the bending of a normal section tangential to a contour line is larger than the minimal bending at a given point of the topographic surface (Figures 15, 32, 49, and 66). $k_{ve}$ shows to what extent the bending of a normal section having a common tangent line with a slope line is larger than the minimal bending at a given point of the topographic surface (Figures 16, 33, 50, and 67). $M$ shows how much the shape of the land surface element deviates from the spherical shape (Figures 17, 34, 51, and 68). $K_r$ describes flow line twisting (Figures 18, 35, 52, and 69).

$CA$ is a measure of the contributing area. As a result, on the $CA$ maps (Figures 1.19, 1.36, 1.53, and 1.70), ridges and valleys are clearly distinguished as light and dark lines (low and high $CA$ values), correspondingly. On the $CA$ maps, one can see flow structures of glaciers, which significantly differ from the oases structure.

$TI$ is a measure of the potential degree of water accumulation in the landscape. $TI$ reaches high values in areas with high values of $CA$ at low values of $G$ (e.g., a terrain with a large upslope contributing area and flat local topography) (Figures 20, 37, 54, and 71). $SI$ is a measure of the potential flow erosion and related landscape processes. $SI$ reaches high values in areas with high values of both $CA$ and $G$ (e.g., a highly sloped terrain with a large upslope contributing area) (Figures 21, 38, 55, and 72).

## 5. Discussion

The calculated digital geomorphometric models and maps of the oases can be useful, first of all, for revealing and mapping of topographically manifested faults and folds (Florinsky, 2016, chaps. 13 and 14), which are represented as lineaments on the curvature maps. As noted in Section 4, the $k_{min}$ and $k_{max}$ maps are the most useful for these purposes in the terrain conditions of regions under study (Figures 9 and 10, 26 and 27, 43 and 44, 60 and 61). The maps of other curvature types can be used to clarify position, geometry, and type of faults and folds revealed on the $k_{min}$ and $k_{max}$ maps.

The oases areas are generally 'open' geologically, i.e., there is no soil and vegetation cover, Quaternary deposits are found in some valleys and depressions, and bedrock outcrops are observed everywhere (snowfields are the only problem for field studies). Therefore, field verification of the revealed lineaments cannot in most cases cause significant difficulties.

The combined use of $k_h$ and $k_v$ models (Figures 7 and 8, 24 and 25, 41 and 42, 58 and 59) allows one to identify zones of relative accumulation of surface flows, which coincide with the sites of fault intersection and are characterized by increased fragmentation and permeability of rocks (Florinsky, 2016, chap. 15). In these zones, there is an interaction and exchange between two types of substance flows: (1) lateral, gravity-driven substance flows moved along the land surface and in the near-surface layer (water, dissolved and suspended substances), and (2) vertical, upward substance flows (fluids, groundwater of different mineralization and temperature). Maps of accumulation zones / sites of fault intersection derived from the $k_h$ and $k_v$ models could be useful for geochemical and mineralogical studies of the oases, especially the Stornes Peninsula of the Larsemann Hills, where borosilicate and phosphate complexes are developed, the origin of which is not yet completely clear.

It is well known that topography largely determines the thermal, wind, and hydrological regimes of slopes, thus controlling the distribution and properties of soils and vegetation covers (Florinsky, 2016, chap. 9). In particular, the information on $G$ and $A$ (Figures 5 and 6, 22 and 23, 39 and 40, 56 and 57), together with the Sun position data, can be used to calculate the terrain insolation models, and, together with the wind rose data, can be utilized to create maps of windward and leeward slopes, to predict the spatial differentiation of snow accumulation, and to identify areas protected from the wind impact.

Such information is important for soil, geobotanical, and ecological studies of the regions discussed, where one of the main meteorological factors determining the microclimate is a strong periodic katabatic wind. Information on the differentiation of slopes by insolation level and on areas protected from wind impact may be useful to clarify and predict the spatial distribution of primitive soils (Mergelov, 2014; Dolgikh et al., 2015; Lupachev et al., 2020a, 2020b) and vegetation cover (i.e., mosses, lichens, algae) (Dolgikh et al., 2015; Gupta, 2015) in the oases. To refine this prediction, the *TI* map (Figures 20, 37, 54, and 71) can be useful, which shows the morphometric prerequisites of migration pathways and areas of moisture accumulation in the landscape during the austral summer.

For the oases, it is urgent to monitor and predict outbursts of lakes, a number of which are cascade systems (Boronina, 2022). On the Broknes Peninsula of the Larsemann Hills, such outbursts may particularly pose a threat to the transport connectivity of the Progress Station with the Zenith Airfield and Vostok Station. For the eastern and southeastern part of the Broknes Peninsula, the *TI* map (Figure 20) clearly shows potential floodwater discharge pathways (narrow red areas) from the Lake LH73 into the Lake Progress, then into the Lake Sibthorpe, and then through a canyon to the Dålkoy Bay, as well as from the Boulder Lake to the zone of the 2017 catastrophic subsidence in the Dålk Glacier (Florinsky and Bliakharskii, 2019), and then towards the Dålkoy Bay. We believe that the *TI* model can be applied to identification and analysis of topographic prerequisites of floodwater discharge pathways along the oasis lake cascades.

All curvature types are functions of the first and second partial derivatives of elevation (see formulas in Table 1). As a result, curvatures are very sensitive to the smallest changes in elevation values (Florinsky, 2016, chap. 5) including high-frequency noise and artifacts. As we noted above, the used fragment of the REMA DEM is characterized by sufficient smoothness and almost complete absence of visible high-frequency noise (the calculated geomorphometric maps are not noisy and are well readable). At the same time, all of the calculated curvature maps of the Larsemann Hills (Figures 7–18) clearly show an artifact that we did not specially remove: a narrow lineament running at an oblique angle in the north of the western Broknes Peninsula and cutting off the Mirror Peninsula from the rest of the eastern Broknes Peninsula. This is a trace of incorrect stitching of two adjacent tiles of the REMA DEM.

The sensitivity of the curvatures to subtle elevation differences is also manifested in different representation 'style' of the ice-free areas and the glaciers on the curvature maps. The oases areas are colored with contrasting colors (or shades of the same color) due to high differences in curvature values, while the glaciers are colored with faint colors (or shades of the same color) due to extremely low differences in curvature values (Figures 7–18, 28–35, 41–52, and 58–69).

## 6. Conclusions

We carried out geomorphometric modeling and mapping of territories of several Antarctic oases including the Larsemann Hills, Thala Hills, Schirmacher Oasis, and Fildes Peninsula. For each territory, digital models of 17 morphometric variables were derived from the REMA DEM including the complete system of curvatures. Geomorphometric modeling and mapping of Antarctic oases have not been carried out before.

The study is conducted in the framework of the author's multiyear project on mathematical and cartographic (geomorphometric) digital modeling of the topography of oases and other ice-free territories of Antarctica, the final goal of which is to create a digital, large-scale geomorphometric atlas of such territories.

The calculated morphometric maps can be useful for structural geological and process-oriented hydrological studies of the oases and other ice-free Antarctic territories.

## Acknowledgements

The author is grateful to A.V. Klepikov and V.L. Martyanov (Russian Antarctic Expedition, Arctic and Antarctic Research Institute, St. Petersburg) for the possibility to participate in the seasonal works of the 68th Russian Antarctic Expedition (RAE). Constant comprehensive support and assistance were provided by A.N. Nikolaev (68th RAE), D.A. Mamadaliev (67th RAE), and D.V. Shepelev (68th RAE). Special thanks to T.N. Skrypitsyna (Moscow State University of Geodesy and Cartography), T.Yu. Petrova (Institute of Mathematical Problems of Biology, Russian Academy of Sciences, Pushchino), M.V. Gribok (Geographical Faculty, Lomonosov Moscow State University), and A.Yu. Pshenichny (68th RAE) for moral support.

## References

ATCM, 2009a. Management plan for Antarctic Specially Protected Area No. 125 Fildes Peninsula, King George Island (25 de Mayo) (Fossil Hill, Holz Stream (Madera Stream), Glacier Dome Bellingshausen (Collins Glacier), Halfthree Point, Suffield Point, Fossil Point, Gradzinski Cove and Skua Cove). In: *Final Report of the Thirty-Second Antarctic Treaty Consultative Meeting. Measure 6. Annex*. Secretariat of the Antarctic Treaty, Buenos Aires, Argentina, 21 p.

ATCM, 2009b. Management plan for Antarctic Specially Protected Area No. 150 Ardley Island, Maxwell Bay, King George Island (25 de Mayo). In: *Final Report of the Thirty-Second Antarctic Treaty Consultative Meeting. Measure 6. Annex*. Secretariat of the Antarctic Treaty, Buenos Aires, Argentina, 13 p.

ATCM, 2014. Management plan for Antarctic Specially Protected Area No. 174 Stornes, Larsemann Hills, Princess Elizabeth Land. In: *Antarctic Treaty Consultative Meeting (ATCM) XXXVII Final Report, Measure 12 Annex*. Secretariat of the Antarctic Treaty, Buenos Aires, Argentina, 13 p.

ATCM, 2021. Larsemann Hills, East Antarctica Antarctic Specially Managed Area No. 6 Management Plan. In: *Antarctic Treaty Consultative Meeting (ATCM) XLIII Final Report. Measure 1*. Secretariat of the Antarctic Treaty, Buenos Aires, Argentina, 35 p.

Australian Antarctic Division, 2005. *Larsemann Hills, Princess Elizabeth Land, Antarctica. Environmental Management Map, scale 1 : 25 000*. Australian Antarctic Division, Kingston, TS.

Bolshiyanov, D.Yu., 2011. Geomorphic structure of the Larsemann Hills. In: Lastochkin, A.N. (Ed.), *Antarctica. Geomorphological Atlas*. Karta, St. Petersburg, Russia, pp. 222–225 (in Russian).

Bormann, P., Fritzsche, D. (Eds.), 1995. *The Schirmacher Oasis, Queen Maude Land, East Antarctica, and Its Surroundings*. Justus Perthes Verlag, Gotha, Germany.

Boronina, A.S., 2022. Large-scale outbursts of lakes in the Antarctic oases: current knowledge. *Ice and Snow* 62: 141–160 (in Russian, with English abstract). doi:10.31857/S2076673422010122.

Burgess, J.S., Spate, A.P., Shevlin, J., 1994. The onset of deglaciation in the Larsemann Hills, Eastern Antarctica. *Antarctic Science* 6: 491–495. doi:10.1017/S095410209400074X.

Carson, C.J., Grew, E.S., 2007. *Geology of the Larsemann Hills, Princess Elizabeth Land, Antarctica. 1 : 25 000 scale map*. Geoscience Australia, Canberra, Australia.

Conrad, O., Bechtel, B., Bock, M., Dietrich, H., Fischer, E., Gerlitz, L., Wehberg, J., Wichmann, V., Boehner, J., 2015. System for Automated Geoscientific Analyses (SAGA) v. 2.1.4. *Geoscientific Model Development* 8: 1991–2007. doi:10.5194/gmd-8-1991-2015.

Dharwadkar, A., Shukla, S.P., Verma, A., Gajbhiye, D., 2018. Geomorphic evolution of Schirmacher Oasis, central Dronning Maud Land, East Antarctica. *Polar Science* 18: 57–62. doi:10.1016/j.polar.2018.06.006.

Dolgikh, A.V., Mergelov, N.S., Abramov, A.A., Lupachev, A.V., Goryachkin, S.V., 2015. Soils of Enderby Land. In: Bockheim, G. (Ed.), *The Soils of Antarctica*. Springer, Cham, Switzerland, pp. 45–63. doi:10.1007/978-3-319-05497-1_4.

Evans, I.S., 1972. General geomorphometry, derivatives of altitude, and descriptive statistics. In:

Chorley, R.J. (Ed.), *Spatial Analysis in Geomorphology*. Methuen, London, UK, pp. 17–90. doi:10.4324/9780429273346-2.

Florinsky, I.V., 2016. *Digital Terrain Analysis in Soil Science and Geology, 2nd ed*. Academic Press, Amsterdam, the Netherlands.

Florinsky, I.V., 2017. An illustrated introduction to general geomorphometry. *Progress in Physical Geography* 41: 723–752. doi:10.1177/0309133317733667.

Florinsky, I.V., 2021. Geomorphometry today. *InterCarto. InterGIS* 27(2): 394–448 (in Russian, with English abstract). doi:10.35595/2414-9179-2021-2-27-394-448.

Florinsky, I.V., 2022. Unmanned aerial survey in the summer season of the 67th Russian Antarctic Expedition. *InterCarto. InterGIS* 28(1): 284–304 (in Russian, with English abstract). doi:10.35595/2414-9179-2022-1-28-284-304.

Florinsky, I.V., Bliakharskii, D.P., 2019. The 2017 catastrophic subsidence in the Dålk Glacier, East Antarctica: unmanned aerial survey and terrain modelling. *Remote Sensing Letters* 10: 333–342. doi:10.1080/2150704X.2018.1552810.

Florinsky, I.V., Skrypitsyna, T.N., 2022. Unmanned aerial survey of the Molodezhny and Vecherny oases area in the season of the 67th Russian Antarctic Expedition: first results. In: Loginov, V.F., Lysenko, S.A., Ryzhikov V.A., and Giginyak Yu.G. (Eds.), *The Natural Environment of Antarctica: Cross-Disciplinary Study Approaches: Proceedings of the IV International Scientific and Practical Conference, Domzheritsy, Belarus, 21–23 Sept. 2022*. Belorus State Technological University, Minsk, pp. 250–253 (in Russian, with English abstract).

Grew, E.S., 1978. Precambrian basement at Molodezhnaya Station, East Antarctica. *Geological Society of America Bulletin* 89: 801–813. doi:10.1130/0016-7606(1978)89<801:PBAMSE>2.0.CO;2.

Gupta, P., 2015. Biodiversity of Larsemann Hills, Antarctica. *Climate Change* 1(3): 174–183.

Guth, P.L., Van Niekerk, A., Grohmann, C.H., Muller, J.-P., Hawker, L., Florinsky, I.V., Gesch, D., Reuter, H.I., Herrera-Cruz, V., Riazanoff, S., López-Vázquez, C., Carabajal, C.C., Albinet, C., Strobl, P., 2021. Digital elevation models: terminology and definitions. *Remote Sensing* 13: 3581. doi:10.3390/rs13183581.

Hengl, T., Reuter, H.I. (Eds.), 2009. *Geomorphometry: Concepts, Software, Applications*. Elsevier, Amsterdam, the Netherlands.

Hodgson, D.A., Noon, P.E., Vyvermann, W., Bryant, C.L., Gore, D.B., Appleby, P., Gilmour, M., Verleyen, E., Sabbe, K., Jones, V.J., Ellis-Evans, J.C., Wood, P.B., 2001. Were the Larsemann Hills ice-free through the Last Glacial Maximum? *Antarctic Science* 13: 440–454. doi:10.1017/S0954102001000608.

Howat, I.M., Porter, C., Smith, B.E., Noh, M.-J., Morin, P., 2019. The Reference Elevation Model of Antarctica. *Cryosphere* 13: 665–674. doi:10.5194/tc-13-665-2019.

Instituto Geográfico Militar de Chile, 1996. *Isla Rey Jorge – Península Fildes. Islas Shetland del Sur. XII Región de Magallanes y de la Antártica Chilena. República de Chile. Carta topográfica, escala 1 : 10 000*. Instituto Geográfico Militar de Chile.

Kakareka, S.V., Kukharchik, T.I., Salivonchik, S.V., Giginyak, Yu.G., Gaidashov, A.A., 2015. *Construction and Operation of the Belarusian Antarctic Station on the Mount Vechernyaya, Enderby Land. Final Comprehensive Assessment of the Environment*. Republican Center for Polar Research, National Academy of Sciences of Belarus, Minsk, Belarus (in Russian).

Kiernan, K., Gore, D.B., Fink, D., White, D.A., McConnell, A., Sigurdsson, I.A., 2009. Deglaciation and weathering of Larsemann Hills, East Antarctica. *Antarctic Science* 21: 373–382. doi:10.1017/S0954102009002028.

Li, Z., Zhu, Q., Gold, C., 2005. *Digital Terrain Modeling: Principles and Methodology*. CRC Press, New York, NY.

Lupachev, A.V., Abakumov, E.A., Abramov, A.A., Dobryanskiy, A.S., Dolgikh, A.V., Zazovskaya, E.P., Mergelov, N.S., Osokin, N.I., Shorkunov, I.G., Goryachkin, S.V., 2020a. Soil cover and permafrost of Antarctica: pattern and functioning. *Problems of Geography* 150: 242-285 (in Russian, with English abstract).

Lupachev, A.V., Abakumov, E.V., Goryachkin, S.V., Veremeeva, A.A., 2020b. Soil cover of the Fildes Peninsula (King George Island, West Antarctica). *Catena* 193: 104613. doi:10.1016/j.catena.2020.104613.

Lv, G., Xiong, L., Chen, M., Tang, G., Sheng, Y., Liu, X., Song, Z., Lu, Y., Yu, Z., Zhang, K., Wang, M., 2017. Chinese progress in geomorphometry. *Journal of Geographical Sciences* 27: 1389–1412. doi:10.1007/s11442-017-1442-0.

Mergelov, N.S., 2014. Soils of wet valleys in the Larsemann Hills and Vestfold Hills Oases (Princess Elizabeth Land, East Antarctica). *Eurasian Soil Science* 47: 845–862. doi:10.1134/S1064229314090099.

Minár, J., Krcho, J., Evans, I.S., 2016. Geomorphometry: quantitative land-surface analysis. In: Elias, S.A. (Ed.), *Reference Module in Earth Systems and Environmental Sciences*. Elsevier, Amsterdam, the Netherlands. doi:10.1016/B978-0-12-409548-9.10260-X.

Ministry of Merchant Marine of the USSR, 1972. *Schirmacher Oasis, Queen Maud Land, Antarctica. Topographic Map, scale 1 : 25 000*. Ministry of Merchant Marine of the USSR, Moscow, USSR.

Moore, I.D., Grayson, R.B., Ladson, A.R., 1991. Digital terrain modelling: a review of hydrological, geomorphological and biological applications. *Hydrological Processes* 5: 3–30. doi:10.1002/hyp.3360050103.

Myasnikov, O.V., 2011. Geological structure of the Vechernegorskaya territory (Western Enderby Land, Antarctica). In: Samodurov, V.P. (Ed.), *Actual Problems of Geology and Prospecting of Mineral Deposits: Proceedings of the 5th University Geological Readings, Minsk, Belarus, 8–9 April 2011*. Belarus State University, Minsk, Belarus, pp. 17–20 (in Russian).

Myasnikov, O.V., Fedotova, L.R., Vasilyonok, E.A., 2021. A digital atlas of rocks of East Antarctica (Thala Hills): methods of formation. In: Lukashev, O.V. (Ed.), *Problems of Regional Geology of the West East European Platform and Adjacent Territories: Proceedings of the 2nd International Scientific Conference, Minsk, Belarus, 16 Feb. 2021*. Belarus State University, Minsk, Belarus, pp. 290–295 (in Russian).

REMA, 2018–2022. *Reference Elevation Model of Antarctica (REMA)*. Polar Geospatial Center, University of Minnesota, Saint Paul, MN. https://www.pgc.umn.edu/data/rema/

Schmid, T., López-Martínez, J., Guillaso, S., Serrano, E., D'Hondt, O., Koch, M., Nieto, A., O'Neill, T., Mink, S., Dur'an, J.J., Maestro, A., 2017. Geomorphological mapping of ice-free areas using polarimetric RADARSAT-2 data on Fildes Peninsula and Ardley Island, Antarctica. *Geomorphology* 293: 448–459. doi:10.1016/j.geomorph.2016.09.031

Shary, P.A., 1995. Land surface in gravity points classification by a complete system of curvatures. *Mathematical Geology* 27: 373–390. doi:10.1007/BF02084608.

Shary, P.A., Sharaya, L.S., Mitusov, A.V., 2002. Fundamental quantitative methods of land surface analysis. *Geoderma* 107: 1–32. doi:10.1016/S0016-7061(01)00136-7.

Shrivastava, P.K., Roy, S.K., Srivastava, H.B., Dharwadkar, A., 2019. Estimation of paleo-ice sheet thickness and evolution of landforms in Schirmacher Oasis and adjoining area, cDML, East Antarctica. *Journal Geological Society of India* 93: 638–644. doi:10.1007/s12594-019-1242-5.

Simonov, I.A., 1971. *Oases of East Antarctica*. Hydrometeoizdat, Leningrad, USSR (in Russian).

Sokratova, I.N., 2010. *Antarctic Oases: History and Results of Investigations*. AARI, St. Petersburg, Russia (in Russian, with English abstract).

Stüwe, K., Braun, H.-M., Peer, H. 1989. Geology and structure of the Larsemann Hills area, Prydz Bay, East Antarctica. *Australian Journal of Earth Sciences* 36: 219–241. doi:10.1080/08120098908729483.

Verleyen, E., Hodgson, D.A., Sabbe, K., Vyverman, W., 2004. Late Quaternary deglaciation and climate history of the Larsemann Hills (East Antarctica). *Journal of Quaternary Science* 19: 361–375. doi:10.1002/jqs.823.

Wilson, J.P., 2018. *Environmental Applications of Digital Terrain Modeling*. Wiley-Blackwell, Chichester, UK.

Wilson, J.P., Gallant, J.C. (Eds.), 2000. *Terrain Analysis: Principles and Applications*. Wiley, New York, NY.